\documentclass[12pt, a4paper]{report}
\usepackage[english]{babel}
\usepackage[utf8]{inputenc}
\usepackage{graphicx}
\usepackage[onehalfspacing]{setspace}
\usepackage{geometry}
\usepackage{fancyhdr}
\usepackage{subcaption}
\usepackage{dcolumn, bm, amsmath, environ, mathtools, makecell, amssymb, textcomp, caption, units, textgreek, float, tabularx, array, booktabs, gensymb, amsmath}
\usepackage[T1]{fontenc}
\usepackage[flushleft]{threeparttable}
\usepackage{booktabs}
\usepackage{xcolor}
\usepackage{notoccite}

\usepackage{etoolbox}
\makeatletter
\patchcmd{\@makechapterhead}{\vspace*{50\p@}}{}{}{}
\patchcmd{\@makeschapterhead}{\vspace*{50\p@}}{}{}{}
\makeatother

\title{Masterthesis}
\author{Stephan Erdmann}
\date{\today}

\begin{document}

\begin{titlepage}
\thispagestyle{empty}
\begin{figure}[t]
	\includegraphics [width=8cm]{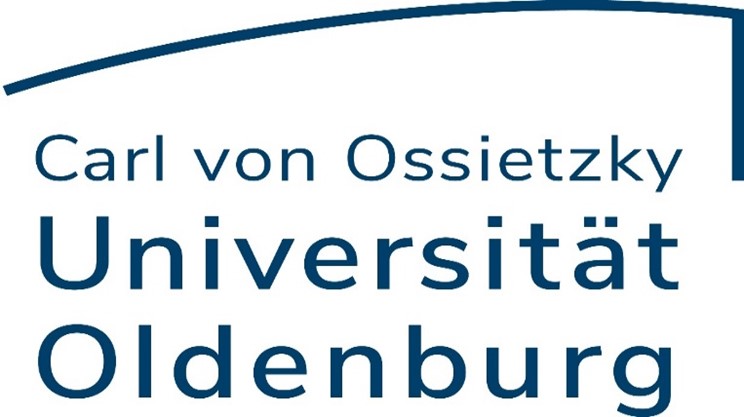}
	\centering
\end{figure}
	\centering

\vspace{1cm}

Institut für Chemie

\vspace{1.5cm}

Masterstudiengang Chemie

\vspace{1.5cm}

\textbf{{\LARGE Masterarbeit}}

\vspace{1.5cm}

\textbf{{\Large \textit{Ab Initio} Magnetic Properties of Rare-Earth Lean Nd-based Hard Magnets}}

\vspace{1cm}

vorgelegt von: Stephan Erdmann

\vspace{1cm}

Matrikelnummer: 4546697

\vfill

Betreuender Gutachter: Herr Prof. Dr. Thorsten Klüner

\vspace{0.5cm}

Zweitgutachter: Herr Dr.-Ing. Halil \.{I}brahim Sözen

\vspace{0.5cm}

Oldenburg, 20.12.2022

\end{titlepage}
\tableofcontents

\vspace{0.5cm}

\textbf{Bibliography}

\vspace{0.5cm}

\textbf{List of Abbreviations}

\vspace{0.5cm}

\textbf{Erklärung}

\newpage
\listoffigures
\listoftables
\addcontentsline{toc}{chapter}{Acknowledgements/Danksagung}
\chapter*{Acknowledgements/Danksagung}
After writing this work i want to thank Prof. Dr. Thorsten Kl\"uner who enabled me to write this thesis in his working group and for supervising this work. I also want to thank Dr. Ing. Halil \.{I}brahim S\"ozen for taking the position as the second supervisor. I also want to thank him for helping me with the program package VASP, his tutelage with the different calculations and his general support in this work. Furthermore, I want to thank my friends and especially my family for the support I got in the last years of my studies. Special thanks also go to the working group Theoretische Chemie of the Carl-von-Ossietzky university in Oldenburg who were all ready to help if it was needed. Lastly, I want to thank the team of the HPC-Cluster CARL of the university on which the calculations in this work were done.\\\\
Mein besonderer Dank gilt Herrn Prof. Dr. Thorsten Klüner dafür, dass ich diese Abschlussarbeit in seinem Arbeitskreis anfertigen konnte. Weiterhin möchte ich besonders Dr. Ing. Halil \.{I}brahim S\"{o}zen für seine Hilfe im Umgang mit dem Programmpaket VASP, seiner Unterstützung bei den Berechnungen, der generellen Betreuung durch ihn sowie der Kontrolle als Zweitprüfer in dieser Arbeit. Ebenfalls danke ich meinen Freunden und besonders meiner Familie welche mich während des Studiums stets unterstützt und mir geholfen haben.
Auch möchte ich generell der Arbeitsgruppe der Theoretischen Chemie in Oldenburg für die Bereitschaft zur Hilfe bei Fragen und für die gute Aufnahme im Arbeitskreis danken. Zuletzt möchte ich mich beim Team des HPC-Clusters CARL der Universität Oldenburg bedanken, auf dem die Berechnungen in dieser Arbeit durchgeführt wurden.


\chapter{Introduction to Permanent Magnets and Magnetism}
Magnetic materials have played a critical role in the development of electric motors and renewable energy technologies in the last years and are an essential part of many other applications like generators or actuators. Since the need and demand for these technologies are steadily increasing, more and more effort is put into the development of these materials \cite{Gutfleisch2011}. A big share of this interest is focused on the improvement of rare-earth-based hard magnetic materials that are mostly composed of rare-earth (RE) and transition metal (TM) elements. Due to resource criticality, a big focus is devoted to the reduction of the rare-earth elements in these compounds and on the development of new rare-earth lean or rare-earth free hard magnets \cite{Skokov2018,Poenaru2019}. \\

In the first chapter some fundamental and basic magnetic properties such as magnetic moment $m_{tot}$, magnetization $M_S$, the maximum energy product $|BH|_{max}$, the Curie temperature $T_C$ and the magnetocrystalline anisotropy energy (MAE) are explained to provide a robust understanding for the readers, who are unfamiliar with the topic. In the second part, the historical development of magnetic materials is reported to demonstrate the historical impact of magnets on world development. In the third part, the situation of the current global market and the main usage fields of hard magnetic materials in the last decade are described. At the end of this chapter, the motivation for this thesis is explained.


\section{Basic Properties of Magnets and Fundamentals}

Since the discovery of the first magnetic material, which were naturally magnetized pieces of iron called lodestone, its properties to attract ferrous materials has drawn the attention of many scientists and researchers to it. These properties had a great impact on the development of the world. For instance, the invention of the compass, one of the earliest magnetic devices, made the navigation on sea much easier and also enabled much larger explorations. These explorations included the discovery of America by Christopher Columbus in 1492 or the voyage around the world by Ferdinand Magellan from 1519 to 1522 \cite{Coey2009}.\\


The origin of magnetism can be attributed to the magnetic moments of the different atoms in a compound. The classical Bohr definition explains their origin in the motion of an electrical charge, in other words of electrons, and the resulting magnetic field that is generated with their orbital motion. This orbital motion generates an orbital magnetic moment and a spin magnetic moment associated with it, both of which are measured in Bohr magneton $\mu_B$ \cite{Coey2009,Levy1973,Cullity2009}. The intrinsic spin angular momentum of electrons is according to the quantum mechanical definition $s = \frac{1}{2}$. The total magnetic moment is therefore made out of the orbital (L) and the spin (S) magnetic moment that has two ways to be put together in the forms of LS and JJ couplings with J = L+S. When the electron shell of an element is filled completely the electrons pair themselves with those electrons that have an opposing intrinsic magnetic moment. This leads to an overall total magnetic moment of zero.\\

Until the 19$^{\text{th}}$ century, the shape of magnets was strongly restricted due to the shape-dependent demagnetization caused by the internal magnetic field $H_d$ that occurred in these materials ($H_d$ = - NM, with N as the shape-dependent demagnetization factor and M as the magnetization). This limited the shape of magnets to be bar and needle-like. A small improvement was the development of magnets in the shape of a horseshoe that avoided the self-demagnetization problem but still restricted the shape to a U-form. Nevertheless, these magnets at least led in combination with copper coils and iron to the first magnetic motors and generators in the 19$^{\text{th}}$ century \cite{Coey2009,Cullity2009}.\\

\begin{figure}[h]
\centering
\includegraphics[width=0.9\textwidth]{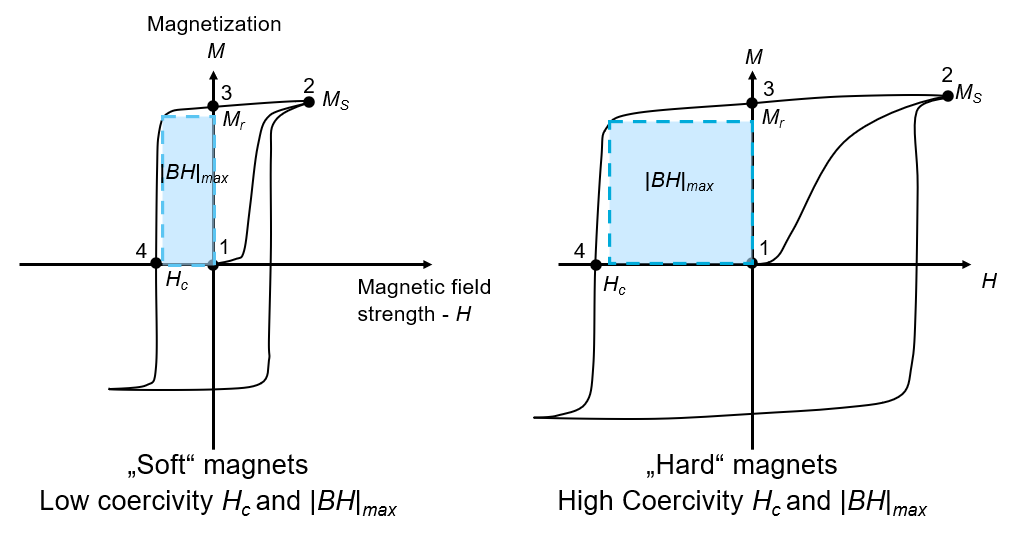}
\caption{Hysteresis loop for soft (left) and hard (hard) magnetic materials. Important magnetic properties like the magnetization saturation $M_S$ are marked and the maximum energy product |BH|$_{max}$ is is depicted as a blue square in the second quadrant.\label{hysteresis_loop}}
\end{figure}

When the first ferrimagnetic hexagonal ferrites were discovered in 1951, a better understanding of the coercivity led to further progress and enabled the manufacturing of magnets with a coercivity $H_c$ higher than their spontaneous magnetization $M_S$ and a break of the shape restrictions. At that time, it was possible to design magnets in every way possible, which caused a significant increase in usability. Ideally, a high coercivity and a high magnetization are desired for hard magnetic materials. This can best be explained with the hysteresis loop (Fig.~\ref{hysteresis_loop}), a plot of the magnetization $M$ against the magnetic field strength $H$ \cite{Coey2009,Levy1973,Cullity2009}.\\


In the hysteresis loop, the magnetization starts at zero at point 1 and increases with the magnetic field strength until the maximum at point 2 is reached. This maximum is the magnetization saturation $M_S$ and describes the maximum possible magnetization for a compound. In this process, the magnetic domains in the material, small regions with the same direction of the magnetic moments, are at first oriented in different directions to minimize the magnetostatic energy. When the material reaches $M_S$, all the domain walls move in the direction of the external field and align parallel to this direction. Once the magnetic field $H$ is reduced to zero, it consequently leads to a slight decrease in magnetization. At this point 3, the remaining magnetization is called remanence magnetization $M_r$ with most of the magnetic domains still lined up in the same direction and only a few having changed back to their former state. The remanence magnetization $M_r$ for hard magnetic materials is needed to be as large as possible as it describes how much a magnet stays magnetized without an external field applied. In the next step of the hysteresis, the magnetic field $H$ is further decreased until the magnetization reaches a value of zero. This point 4 defines the coercivity field strength $H_c$ and relates to the field strength needed to completely demagnetize the material. At this point, an equal amount of magnetic domains are aligned in two opposing directions and therefore cancel each other out. A further reduction of the magnetic field strength followed by an increase back to zero results in a mirrored version of the first two quadrants of this loop. With the hysteresis loop the maximum energy product $|BH|_{max}$, a quantitative measurement for the performance of a magnet, can be derived by the area of a square in the second quadrant of the loop. With these properties, the shape of the hysteresis loop for hard and soft magnetic materials can be defined as it is shown in Fig.~\ref{hysteresis_loop}. For soft magnetic materials, a small $|BH|_{max}$ is resulting from a narrow hysteresis loop due to a low coercivity. This means that soft magnetic materials are very easy to demagnetize. Hard magnetic materials on the other hand have a wide hysteresis loop, therefore a high coercivity and a high $|BH|_{max}$ which results in a high resistance to demagnetization.\\

\begin{figure}[h]
\centering
\begin{subfigure}{0.98\textwidth}
\includegraphics[width=\textwidth]{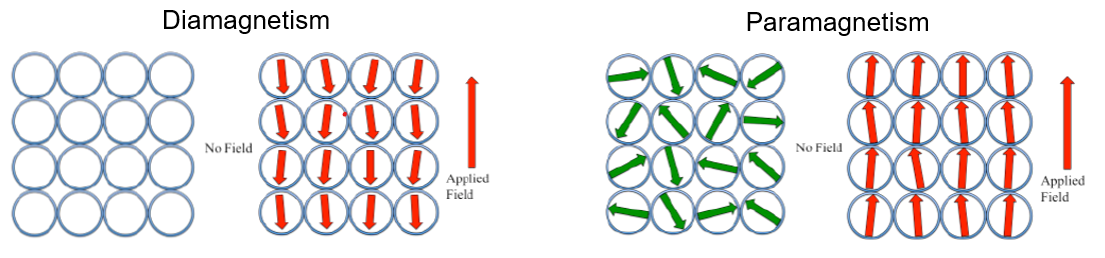}
\vspace{0.25cm}
\end{subfigure}
\begin{subfigure}{0.98\textwidth}
\includegraphics[width=\textwidth]{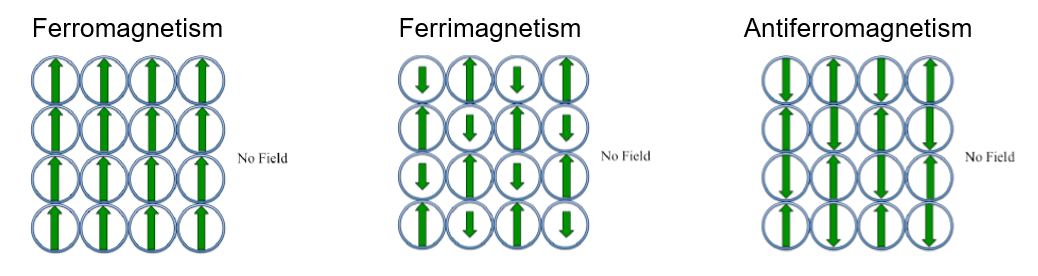}
\end{subfigure}
\caption{Different magnetic domain configurations for the different types of magnetism with and without an external field applied. The image is taken from \cite{Sozen2019}.\label{types_of_magnetism}}
\end{figure}

Since the magnetic domains can react in a variety of ways to the appliance of the external field $H$, different types of magnetism are possible. This response can be used to classify materials into one of five different classes of magnetism, namely diamagnetism, paramagnetism, ferromagnetism, ferrimagnetism and antiferromagnetism as visible in Fig.~\ref{types_of_magnetism}. Note that ferrimagnetism and antiferromagnetism are considered as subclasses of ferromagnetism \cite{Levy1973,Cullity2009}.\\

The first two types, dia- and paramagnetism, are the most common classes of magnetism and only noticeable when an external field is applied. Diamagnetism is a very weak form of magnetism and is found in most materials. Without an external field, no magnetic moments can be attributed to the material. When an external field is applied the magnetic moments are induced by the field and aligned in the opposite direction of the field. This causes a weakening of the magnetic field due to the opposite direction of the induced magnetic moments. Diamagnetism is generally attributed to materials in which all electrons are paired and hence have a total magnetic moment of zero. \\

With paramagnetism, the local magnetic moments in a material are all finite, due to unpaired electrons, and independently leading in different directions which results in an overall magnetic moment of zero. When an external field is applied, like before with diamagnetism, the magnetic moments are realigned into the same direction as the external field \cite{Levy1973,Cullity2009}.\\ 

Ferromagnetism has two additional subclasses, which are ferrimagnetism and antiferromagnetism. In a ferromagnetic material, the local magnetic moments are all linked and oriented parallel to each other. With an external field, the direction of these linked local magnetic moments changes to the same direction as the field. When the field is removed the magnetic moments stay in the induced direction which leads to the macroscopic effect of magnetism. The best-known material for this type of magnetism is, as the name already suggests, iron. \\

In the subclass of antiferromagnetism instead of one sublattice with linked magnetic moments, there are two sublattices alternating with each other. The magnetic moments of these lattices are identical in their order to each other but are aligned in opposing directions. This means that the magnetic moments cancel each other out and a total magnetic moment of zero is the result \cite{Levy1973,Cullity2009}.\\

The subclass of ferrimagnetism shows a similar case to antiferromagnetism as the magnetic moments are linked to each other in two alternating magnetic sublattices directed in opposing directions. Nevertheless, the difference from antiferromagnetism is the magnetic moments of the sublattices are not equal to each other and do not cancel each other out completely. This means that ferrimagnetic materials still have a net magnetic moment similar to ferromagnets but are considerably weaker. This subclass is often seen in materials consisting of different atoms such as ferrites where each atom species provides one of the sublattices \cite{Levy1973,Cullity2009}. \\

\begin{figure}[h]
\centering
\begin{subfigure}{0.44\textwidth}
\includegraphics[width=\textwidth]{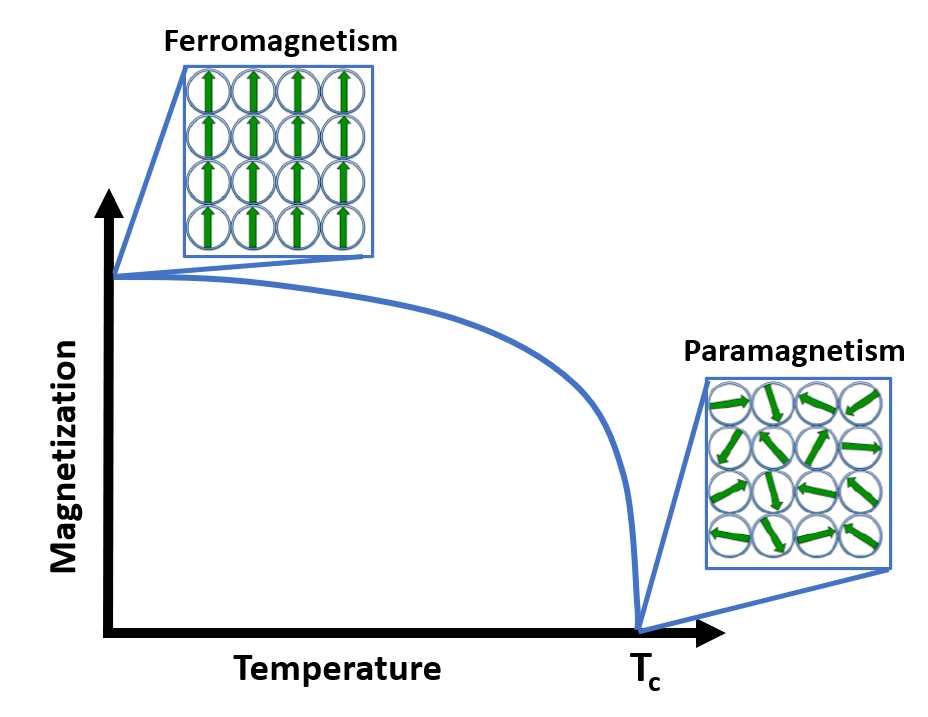}
\end{subfigure}
\begin{subfigure}{0.015\textwidth}
\flushright{\textbf{a)}}
\end{subfigure}
\begin{subfigure}{0.45\textwidth}
\includegraphics[width=\textwidth]{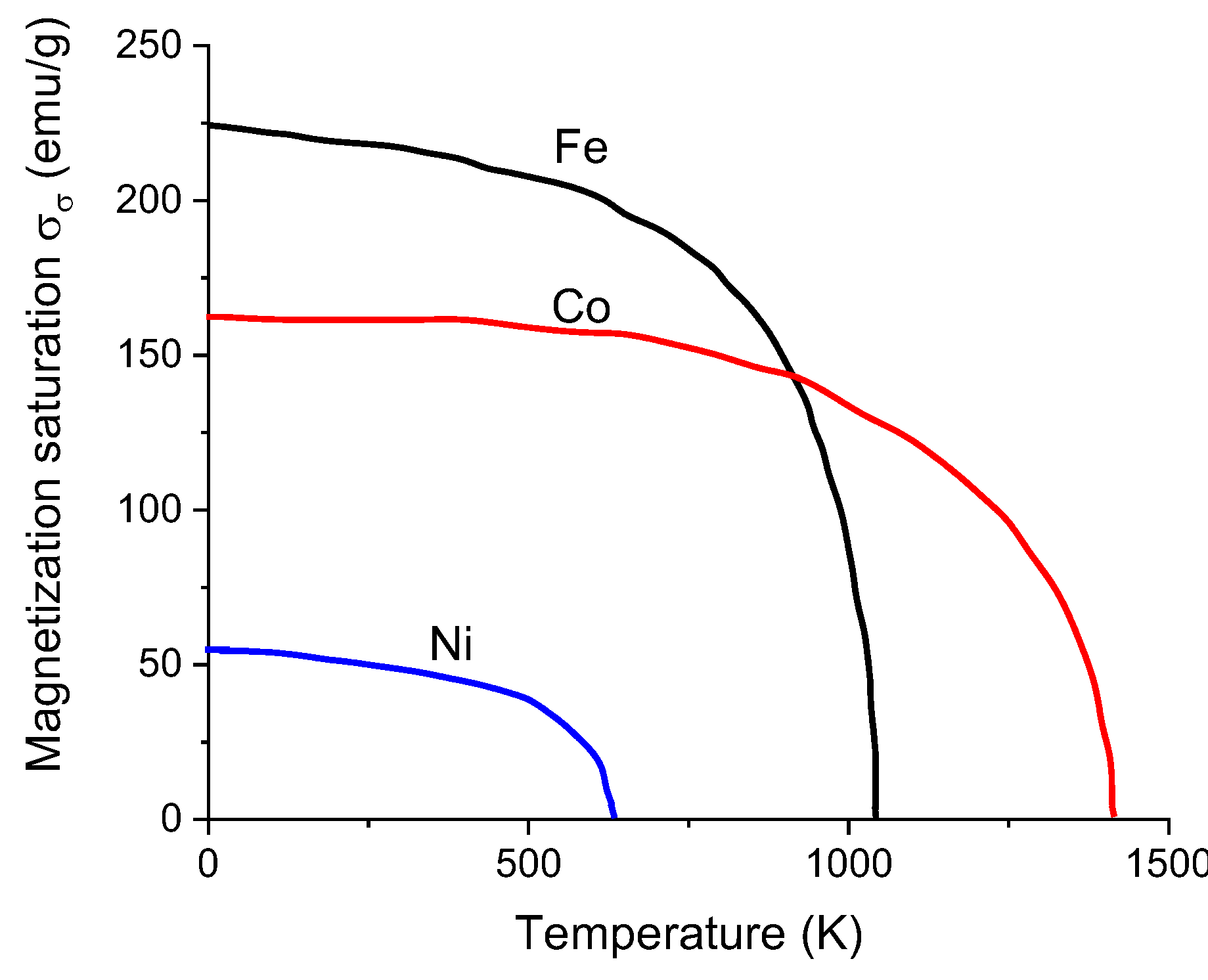}
\end{subfigure}
\begin{subfigure}{0.015\textwidth}
\flushright{\textbf{b)}}
\end{subfigure}
\caption{a) Change of the magnetic ordering from ferro- to paramagnetism with increasing temperature up to the Curie temperature $T_C$; b) Finite temperature magnetization curves for the elements Fe, Co and Ni up to their $T_C$'s of 1044, 1388 and 628 K. Image b) is based on data from \cite{Cullity2009}.\label{curie_elements}}
\end{figure}

With the knowledge of the different types of magnetism and magnetization, which is a result of the realignment of the magnetic domains, it is now possible to explain the influence of temperature on the strength of magnetization in ferromagnetic materials. At low temperatures, the magnetic domains in a ferromagnet are strictly ordered and aligned in the same direction as explained before. As magnetic ordering is a thermodynamic process an increase in the temperature causes the magnetic order to get slowly disrupted until the order is completely lost (see Fig.~\ref{curie_elements} a)) This results in a transition from a ferromagnetic material at low temperatures to a paramagnetic material above a certain temperature called the Curie temperature $T_C$. Above $T_C$ the total magnetization is zero, which means that $T_C$ is an important intrinsic magnetic property that has to be taken into account for the possible field of application. A magnet with a low $T_C$ can not be utilized in various industrial fields, which need high operation temperatures. In Fig.~\ref{curie_elements} b) the finite temperature magnetization of the metals Fe, Co and Ni are shown with their respective $T_C$'s of 1044, 1388 and 628 K \cite{Cullity2009}.\\

\begin{figure}[h]
    \centering
    \includegraphics[width=0.75\textwidth]{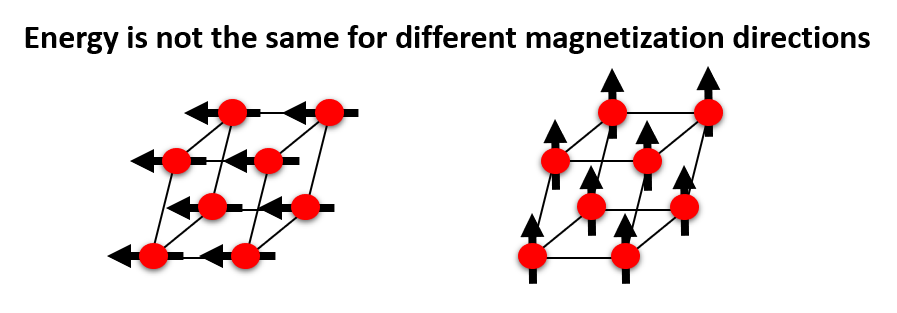}
    \caption{Unequal energies for the different magnetization directions. Image adapted from \cite{FZ-Julich}.\label{fig_MAE_energies}}
\end{figure}

As mentioned above, a better understanding of coercivity was a big milestone in the evolution of magnets as it enabled the manufacturing of magnets with a higher coercivity $H_c$ than their magnetization $M_S$ and a break of the shape restrictions. A key part of this coercivity is the magnetocrystalline anisotropy energy (MAE). In a ferromagnet, the magnetization direction of the magnetic domains usually aligns through one or more easy axes that represent the energetically most favored directions. This implies that the magnetization direction plays an important role as the energy needed to realign the domains is not the same for each direction (see Fig.~\ref{fig_MAE_energies}). With only one easy magnetization direction in a compound, the anisotropy is called uniaxial \cite{Coey2009,Cullity2009}. For the ThMn$_{12}$-type compounds, considered in this work, only one easy axis and therefore uniaxial anisotropy is reported \cite{Li2019}.\\

\begin{figure}[h]
    \centering
    \includegraphics[width=0.75\textwidth]{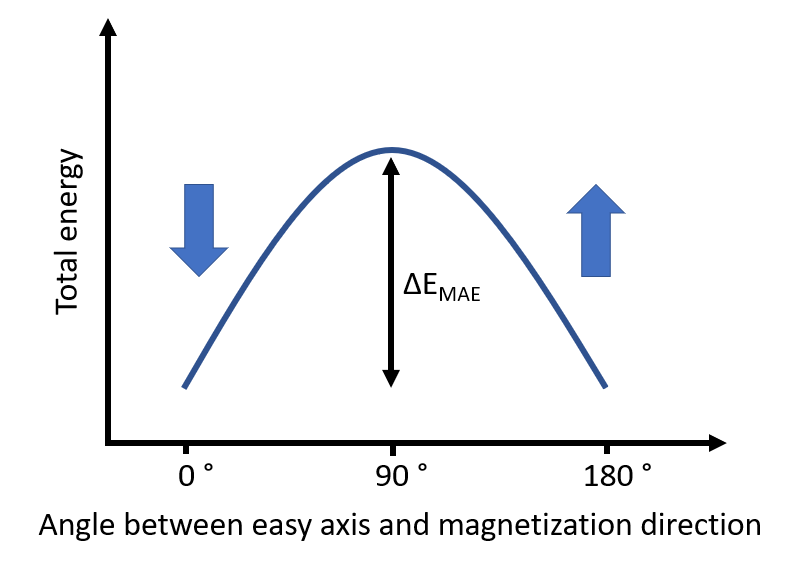}
    \caption{Change of the magnetic energy in relationship to the angle between the easy axis and the magnetization direction with a peak at an angle of 90°. \label{fig_MAE_angles_diff}}
\end{figure}

Since the origin of the magnetism lies within the motion of the electrons and therefore in the circulating electron currents, the energy for a certain magnetization distribution in the direction r $M(r)$ is the same as for the opposing direction along the same axis ($M(r) = -M(r)$). This implies that the energy is different for deviations from the easy axis with an energy peak at an angle of $ \theta = 90 ^{\circ}$ as illustrated in Fig.~\ref{fig_MAE_angles_diff}.\\

In the case of uniaxial anisotropy the energy difference $\Delta E_{MAE}$ resulting for each angle differing from the easy magnetization axis can be expressed with the following equation \cite{Coey2011}

\begin{equation}
    \Delta E_{MAE} = K_1 \cdot sin(\theta)^2,
\end{equation}

with $\theta$ as the angle between the direction of the magnetization M and the easy axis and $K_1$ as the anisotropy constant. Normally the anisotropy energy is described by more than one constant ($K_2$, $K_3$) but with uniaxial anisotropy $K_1$ alone is sufficient as all higher constants are at least 2 orders smaller than $K_1$. Both $E_{MAE}$ and $K_1$ are measured in J/m$^3$, ranging from low values below 1 kJ/m$^3$ up to more than 10 MJ/m$^3$. Most commonly both values are expressed in MJ/m$^3$ \cite{Coey2009}.

Another property influenced by the anisotropy energy, and therefore by the anisotropy constant, is the anisotropy field $H_a$. As seen in the hysteresis loop, given in Fig.~\ref{hysteresis_loop}, the anisotropy field $H_a$ has an upper limit represented by the coercivity field $H_c$ ($H_a \leq H_c$) as this point marks the complete demagnetization of the material. The connection of the anisotropy constant $K_1$ to the anisotropy field $H_a$ can be expressed as \cite{Coey2011}

\begin{equation}
    H_a = \frac{2K_1}{\mu _0 M_S},
\end{equation}
with the magnetic constant $\mu _0 = 4 \pi \times 10^{-7}$ J/(A$^2$m).


The suitability of a magnetic material to serve as a hard magnet can be determined with the anisotropy constant $K_1$ and the magnetization $M_S$ by considering the magnetic hardness factor $\kappa$ as

\begin{equation}
\kappa = \sqrt{\frac{K_1}{\mu_{0}M_{S}^{2}}}.
\end{equation}

The magnetic hardness factor $\kappa$ defines whether a magnet is accepted as a soft ($\kappa < 0.1$), a semi-hard ($ 0.1<\kappa <1.0$) or a hard magnet ($\kappa >1.0$). The most prominent magnet Nd$_2$Fe$_{14}$B is attributed with a hardness factor of 1.54 (see Tab.~\ref{tab_ref_kappa}). \\

\begin{table}[h]
\caption{Curie temperature $T_C$, magnetization $\mu _0$M$_S$, anisotropy constant $K_1$ and hardness factor $\kappa$ for common hard magnets. Data is taken from \cite{Skomski2016}. \label{tab_ref_kappa}}
\centering
\begin{tabular}{@{}ccccc@{}}
\toprule \hline
Magnet & T$_c$ (K) & $\mu _0$M$_S$ (T) & $K_1$ (MJ/m$^3$) & $\kappa$ \\ \midrule
Nd$_2$Fe$_{14}$B & 588    & 1.61                            & 4.9                            & 1.54                  \\
Sm$_2$Co$_{17}$  & 1190   & 1.21                            & 4.2                            & 1.89                  \\
SmCo$_5$    & 1020   & 1.05                            & 17.0                           & 4.40                  \\
Alnico 5 & 1210   & 1.40                            & 0.32                           & 0.45                  \\ \hline \bottomrule
\end{tabular}
\end{table}

Magnets composing of RE and TM elements, such as Nd$_2$Fe$_{14}$B, have in most cases a high hardness factor $\kappa$ due to a high anisotropy and a high magnetization. The high anisotropy can be attributed to the RE-elements with their 4\textit{f}-electrons. The high magnetization of these magnets on the other hand mostly belongs to the TM elements, foremost to elements like Fe or Co. The hardness factors $\kappa$ of a few common magnetic materials are listed in Tab.~\ref{tab_ref_kappa} in addition to some intrinsic magnetic properties such as Curie temperature $T_C$, saturation magnetization $M_S$ and magnetocrystalline anisotropy constant $K_1$.\\



\section{Historical Development of Magnets}

The first discovered magnetic materials were naturally magnetized iron pieces called lodestone. It is assumed that these iron pieces were magnetized by huge electrical currents in lightning strikes and were already known by the people in ancient Greece and China since the 4$^{\text{th}}$ to 6$^{\text{th}}$ century. First devices using these iron pieces were compasses and a tool named a "South pointer". The South pointer was used in China for geomancy and utilized a lodestone carved into a spoon that turned on a small base to align its handle with the earth's magnetic field. It was used to design the grid-like street plans of old Chinese towns. The invention of the first navigational compasses around 1088 by Shen Kua and a century later in Europe (first European reference around 1157-1217 by Alexander Neckham \cite{Levy1973}) made greater explorations at sea possible and led to the discovery of America in 1492 by Christopher Columbus or the voyage around the world by Ferdinand Magellan from 1519 to 1522 \cite{Coey2009}.\\

In the middle ages many superstitions were attributed to magnets like the possibility of a perpetuum mobile and magnetic levitation, of which the latter was actually achieved at the end of the 20$^{\text{th}}$ century. 
In 1269 the french crusader Petrus Peregrinus de Maricourt was the first in Europe to give a detailed description of a floating compass and also the first one to discover the magnetic lines and poles of a magnet. Based on this work William Gilbert extended in his monograph \textit{De Magnete} in 1600 the idea of Peregrinus and came to the conclusion that the alignment of free-rotating magnets into a particular direction is due to the earth itself being one big magnet \cite{Levy1973}.\\

After this the next noteworthy development with magnetic materials appeared in 1743, when Daniel Bernoulli invented the first horseshoe magnet that did not suffer from demagnetization due to its own internal field. Magnets in this U-shape are even today one of the most commonly applied symbols for magnetism \cite{Coey2009,Levy1973}. The next milestone in the history of magnets was the connection of electricity and magnetism found by Hans-Christian Oerstedt in 1820. On basis of his discovery Faraday found the phenomenon of magnetic induction in 1821 and also the connection of light and magnetism with the magneto-optic Faraday effect in 1845. The results of both, Oerstedt and Faraday, motivated James Clerk Maxwell in the middle of the 19$^{\text{th}}$ century to the formulation of the Maxwell equations which laid the theoretical foundation for the further development of hard magnets in the future \cite{Coey2009,Levy1973}.\\

\begin{figure}
\centering
\includegraphics[width=0.70\textwidth]{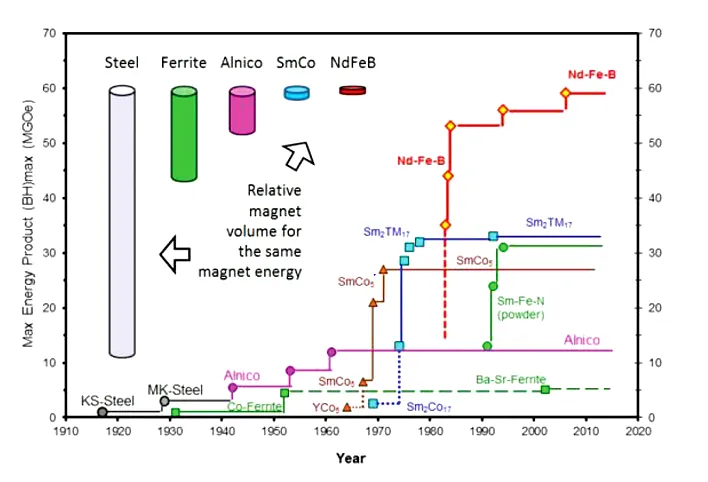}
\caption{Development of the different types of magnets in the 20$^{\text{th}}$ century and the beginning of the 21$^{\text{st}}$ century. The volume needed of each magnetic material to achieve the same magnetic strength is shown under the diagram. Image taken from \cite{Abdelbasir2019}.\label{development_magnets}}
\end{figure}

However, until the early 20$^{\text{th}}$ century almost no further improvement on permanent magnets was made due to the aforementioned problem of the demagnetization and the resulting shape restrictions. It is therefore no surprise that the scientific focus of that time was mainly on electromagnetism as it seemed far more effective and enabled inventions like the first electrical motors.
Nevertheless, the development of hard magnets was at a breaking point and each developed type of magnet was further improved with a better performing one in a short period of time. This evolution in the beginning of the 20$^{\text{th}}$ century up to the start of the 21$^{\text{st}}$ century is shown in Fig.~\ref{development_magnets}.\\

The first developed magnets in the 20$^{\text{th}}$ century were steel magnets \cite{Ellis1935} coming up around 1917 with a very low maximum energy product $|BH|_{max}$ of only 8 kJ/m³ (1 MGOe). These magnets were improved using W and Cr together with Fe-C alloys to suppress the domain wall movement by treatment under appropriate heat. Later the addition of Co led to an increase of the energy product and the focus of attention was then shifted generally to Co alloys and the resulting Alnico compounds \cite{McCurrie1982} around 1940.\\

These compounds, as the name suggests, consist of aluminum (Al) nickel (Ni) and cobalt (Co) and offered much better magnetic properties than the magnetic steels with a high Curie temperature $T_C$, a high remanent force, a good resistance against corrosion in addition with a great reduction of the material volume needed to achieve the same strength of the magnetic properties, as can be seen in Fig.~\ref{development_magnets}. A downside of these compounds is a low coercivity \cite{TremoletdeLacheisserie2005} and therefore an easy demagnetization of these magnets.\\

Another group of magnets developed around the 1930s and becoming popular in the 1950s were the ferrite magnets \cite{Sugimoto1999}, also called ceramic magnets, consisting mainly of ferric oxide $\alpha$-Fe$_2$O$_3$. The reason they gained popularity was the invention of hard hexagonal ferrites together with their extremely low production costs, due to easy production steps, cheap raw materials and a good corrosion resistance. Even today, they are still used as the second highly preferred magnet in the market \cite{Coey2020} because of their low costs although they have a comparably low energy product of 25 kJ/m$^3$, which is less than the more expensive alnico magnets have. The main drawback of the ferrite magnets are their brittleness, which makes it harder to process them into certain shapes, and a low mechanical strength.\\

In the mid of the 1960s a breakthrough took place in the historical development of permanent magnets by using RE-TM consisting magnets. The first magnet of this group was the YCo$_5$ compound \cite{Strnat1966}, and it showed that the combination of RE and TM provides superior magnetic properties such as high $T_C$ and high magnetization, attributed to the TM elements, and high anisotropy due to the RE elements. The soon after discovered SmCo$_5$ compound has an energy product of 200 kJ/m$^3$ (22.5 MGOe), which is a significant increase compared to the alnico 5 compound with 60 kJ/m$^3$ (10 MGOe). This improvement led to an further decrease by about a factor of 2 of the needed material volume to achieve a specific magnetic strength compared to the alnico magnets. These Sm-Co compounds were further refined with the Sm$_2$Co$_{17}$ phase (it is also called 2:17 phase) that enhanced the $|BH|_{max}$ to a value of 294 kJ/m³. This group of magnets is still used today as high performance magnets in motors or other complex applications due to their high $T_C$ and high $M_S$. However, the usage of these magnets is rather limited due to the high costs of the Sm and Co elements, which is also the greatest disadvantage of these magnets.\\

After the Sm-based magnets, the strongest type of RE-TM-magnets with the Nd$_2$Fe$_{14}$B compounds were invented. These compounds were independently reported in 1984 for the first time by Sagawa \textit{et al.} \cite{Sagawa1984} and Croat \textit{et al.} \cite{Croat1984} with the $|BH|_{max}$ of 286 kJ/m³ (36 MGOe). As a result of the high costs for the elements Sm and Co the Nd$_2$Fe$_{14}$B compounds were preferred for many applications that do not require a high $T_C$. Main advantage of these materials is a high energy product that was improved to a value of more than 500 kJ/m$^3$. The greatest disadvantage of the Nd$_2$Fe$_{14}$B magnets is their low $T_C$ of only 585 K, which makes it impossible to use these magnets in applications with high working temperatures. This drawback can be treated by the addition of expensive and as critically labeled elements like Co or Dy.\\

The last depicted magnet type in Fig.~\ref{development_magnets} are the Sm-Fe-N magnets which are still in focus of research today. They exhibit promising properties such as a high $T_C$, a high magnetization $M_S$ and a high resistance to demagnetization.\\

\section{World Market and Usage of Magnets}
\label{sec_world_market}
\begin{figure}[h]
    \centering
    \includegraphics[width=0.85\textwidth]{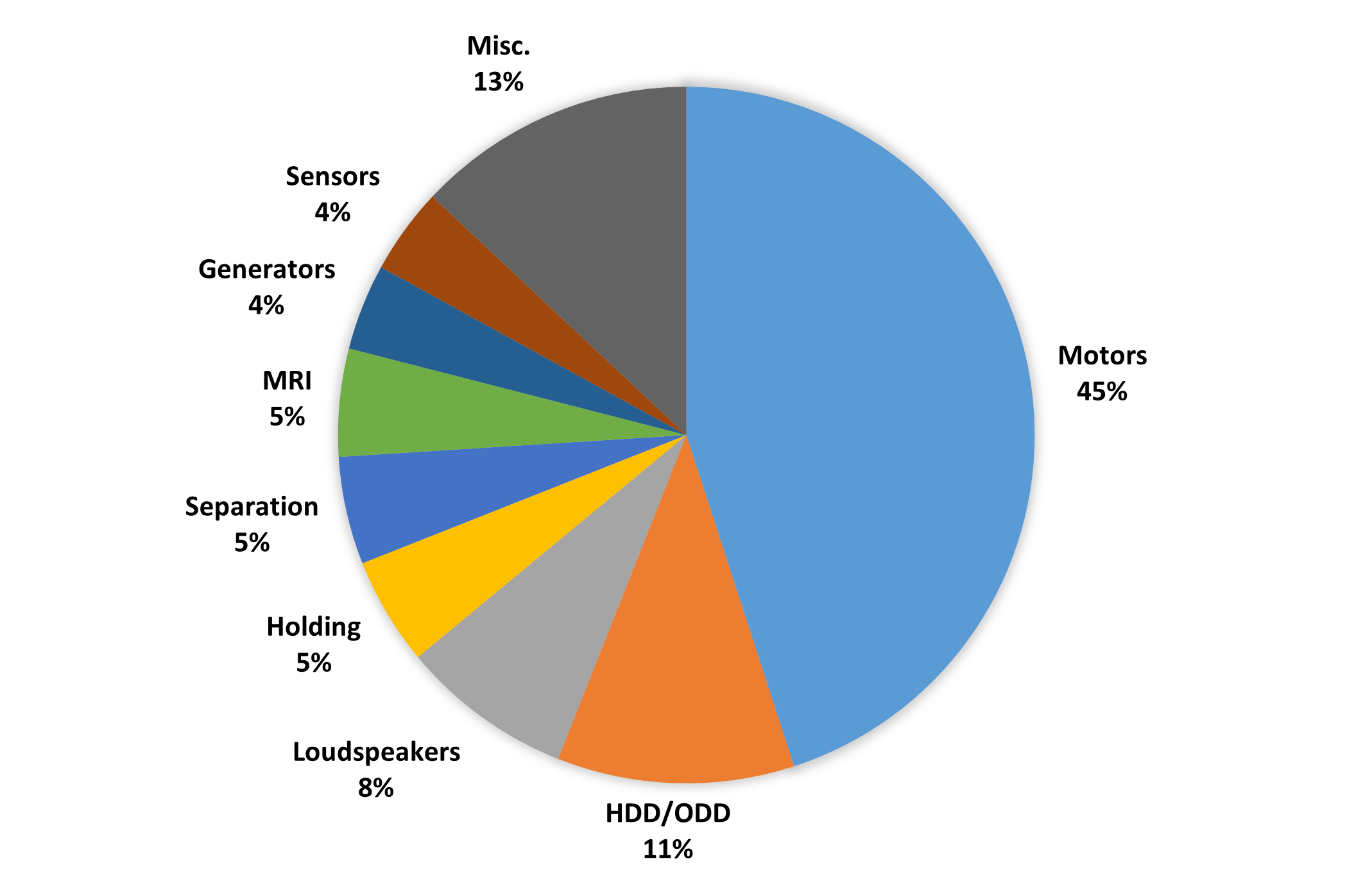}
    \caption{Application of permanent magnets by market share in 2019. The image is based on data from \cite{Cui2022}. \label{fig_application_of_magnets_in_2019}}
\end{figure}

\begin{figure}[h]
\centering
\includegraphics[width=0.95\textwidth]{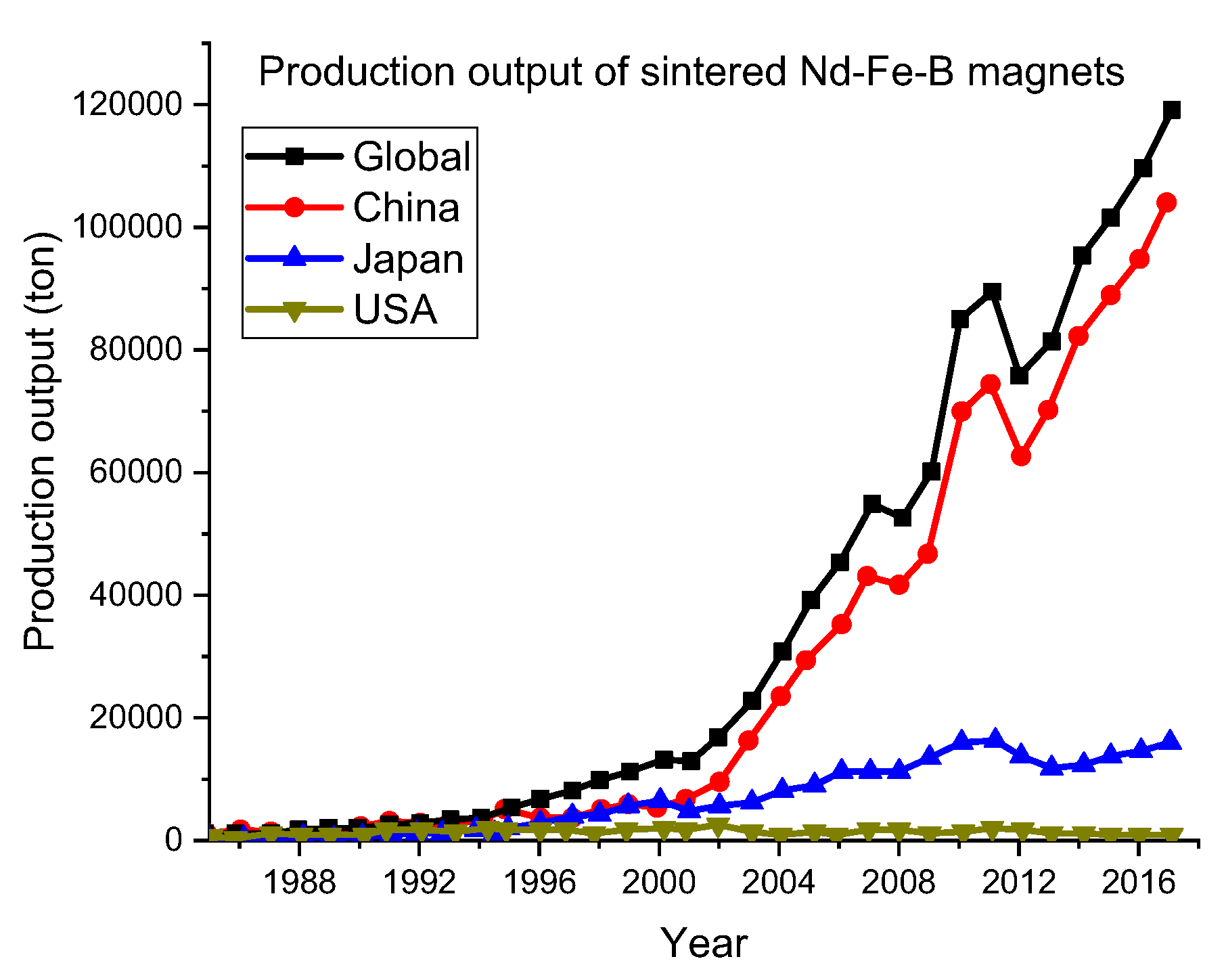}
\caption{Worldwide production of sintered Nd-Fe-B magnets from 1984 to 2017. The image is based on data from \cite{Magnets}\label{fig_world_magnet_production}.}
\end{figure}

In the last decade, the demand for hard magnets has steadily increased as magnets play an essential role in an emerging number of industrial applications \cite{Gutfleisch2011}. In Fig.~\ref{fig_application_of_magnets_in_2019} the percentual application of magnets in different technological fields is shown. The largest application field of permanent magnets is appeared as motors with 45 \%, which also yields the importance of the motor technology as well. The mentioned trend of an increase can also be seen in Fig.~\ref{fig_world_magnet_production} by the world production of sintered Nd-Fe-B magnets from 1984 to 2016 for selected countries. Since the start of the 21$^{\text{st}}$ century the production of these hard magnets has increased significantly from roughly 20000 tons in the year 2000 to around 120000 tons in 2016, thus increasing to 6 times the amount than before. Further, it  has to be noted that Fig.~\ref{fig_world_magnet_production} only shows the production of sintered Nd-Fe-B magnets and does not include other types of magnets like hard ferrites or Sm-based ones.

\begin{figure}[h]
\centering
\includegraphics[width=0.7\textwidth]{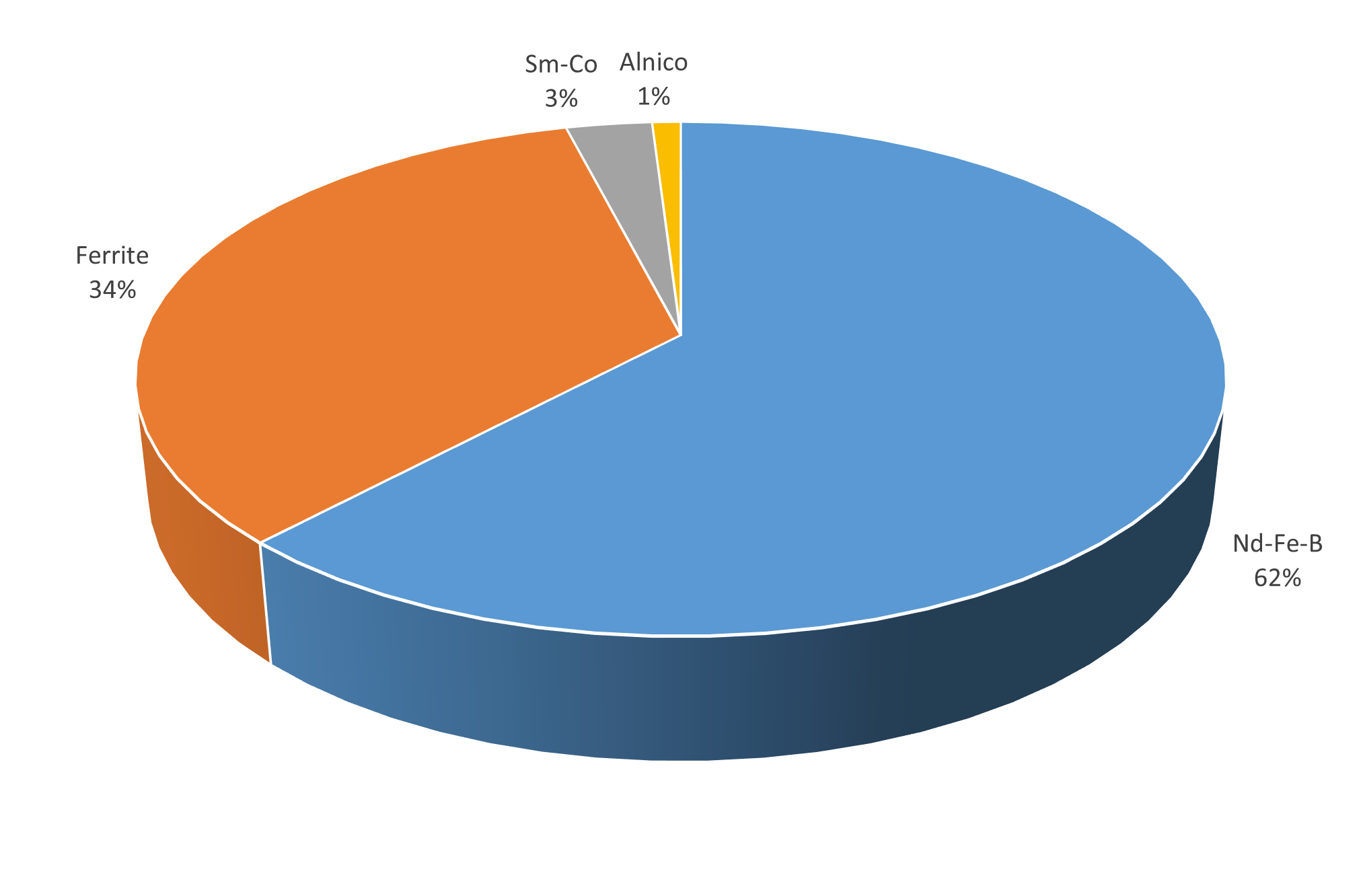}
\caption{Percentage sales of major permanent magnet alloy types in 2020 in the global market. The image is based on data from \cite{Coey2020}\label{fig_percentage_magnets}}
\end{figure}

The percentage of sales of Nd-Fe-B magnets has to be included to get a broader and complete picture of the increasing demand for hard magnets. In Fig.~\ref{fig_percentage_magnets}, the percentage sales of the different types of magnets are shown. With a majority of 62 \% the Nd-Fe-B magnet takes the biggest share followed surprisingly by the hard ferrite magnets. However, magnetic steels and Alnico magnets do not have sufficiently enough magnetic properties to reach the requirements for most industrial applications, and the strong Sm-based magnets are too expensive for a broad utilisation. This results in only small shares of these magnets on the general sales.

\begin{table}[h]
\caption{Production of RE-oxides and available reserves for different countries. Data is taken from \cite{Cordier2022}. \label{tab_production_RE}}
\footnotesize
\begin{tabular}{@{}ccccccc@{}}
\toprule \hline
Country &
  \begin{tabular}[c]{@{}l@{}}Production \\ in 2016 {[}t{]}\end{tabular} &
  \begin{tabular}[c]{@{}l@{}}Production\\ in 2017 {[}t{]}\end{tabular} &
  \begin{tabular}[c]{@{}l@{}}Production\\  in 2018 {[}t{]}\end{tabular} &
  \begin{tabular}[c]{@{}l@{}}Production\\ in 2019 {[}t{]}\end{tabular} &
  \begin{tabular}[c]{@{}l@{}}Production\\ in 2020 {[}t{]}\end{tabular} &
  \begin{tabular}[c]{@{}l@{}}Reserves\\ in 2020 {[}t{]}\end{tabular} \\ \midrule
China         & 105.000 & 105.000 & 120.000 & 132.000 & 140.000 & 44.000.000 \\
United States & 0       & 0       & 18.000  & 26.000  & 39.000  & 1.400.000  \\
India         & 1.700   & 1.500   & 2.900   & 3.000   & 2.900   & 6.900.000  \\
Australia     & 14.000  & 20.000  & 21.000  & 21.000  & 21.000  & 4.000.000  \\
Russia        & 3.000   & 3.000   & 2.700   & 2.700   & 2.700   & 21.000.000 \\
Malaysia      & 300     & 300     & 0       & 0       & 0       & n.a.       \\
Brazil        & 1.100   & 2.000   & 1.100   & 1.000   & 600     & 21.000.000 \\
Thailand      & 800     & 1.600   & 1.000   & 1.800   & 3.600   & n.a.       \\
Vietnam       & 300     & 100     & 920     & 920     & 700     & 22.000.000 \\
Myanmar       & 0       & 0        & 19.000  & 25.000  & 31.000  & n.a        \\ \hline\bottomrule
\end{tabular}
\end{table}

As demonstrated before the RE-TM magnets, in particular Nd-Fe-B magnets make up the majority of hard magnets used in industrial processes, the importance of a steady supply of the needed resources for these magnets has a fundamental importance. This marks a crucial problem associated with RE elements as most of them like Nd, Dy, Sm or Tb are labeled as critical elements for future applications \cite{Bauer2011,Pellegrini2014}. Another downside with the RE elements lies within the available global reserves as a majority of them are located in China, as visible in Tab.~\ref{tab_production_RE}. In addition to the high amount of reserves, China also produced around 60\% of the world RE elements (measured in RE-oxides) in 2020. With this monopoly in the world market, China has a significant impact on the market prices of RE elements. This was experienced in 2009, when the prices changed by a factor of up to ten due to export quotas, export tariffs and other restrictions implemented by the Chinese government. This monopoly was also used as leverage against the United States in 2019, when tensions between both countries arose due to trade restrictions by the US. As a response to China's position in the global market, some countries such as the US have started to raise their production of RE elements in the last years, as given in Tab.~\ref{tab_production_RE}.

\newpage

\chapter{Case Description and Motivation of the Work}

As mentioned in Sec.~\ref{sec_world_market} there is an increasing demand for affordable and resource-efficient hard magnets in rising application fields such as renewable energies and electrical transportation \cite{Gutfleisch2011}. Most of the currently applied hard magnets consist of RE elements together with 3\textit{d} TM elements, with Nd$_{2}$Fe$_{14}$B as the strongest magnet at the moment \cite{Croat1984,Sagawa1984}.\\

However, as mentioned before, most of the RE elements used in hard magnetic materials such as Nd, Dy, Sm and Tb are often considered to be critical elements for future applications \cite{Bauer2011,Pellegrini2014}. In addition, given the increasing demand in combination with high prices for RE-elements and the supply monopoly in the global market (Tab.~\ref{tab_production_RE}), there are considerable efforts to design new and alternative RE-lean and RE-free hard magnetic materials \cite{Skokov2018,Poenaru2019}.\\

In order to overcome this challenge, several approaches have been proposed in the literature. First, instead of using critical Dy, coercivity enhancement has been achieved by grain size reduction \cite{Sepehri-Amin2011,Hono2012}. Second, the grain boundary diffusion process has been performed, which eliminates Dy and Tb-like critical materials and enhances coercivity substantially \cite{Li2009,Sepehri-Amin2013}. Third, the critical RE component has been partially substituted by a more abundant element that offers a low-cost alternative product for low to moderate-temperature applications. As a potential candidate Ce has been investigated intensively \cite{Li2015,Jin2018}.\\

To this extent, this work is focused on the last strategy, developing RE-lean permanent magnets based on the ThMn$_{12}$ prototype structure (will be referred as the 1:12 phase). Note that the composition ratio of RE:Fe=1:12 has a lower RE content than Nd$_2$Fe$_{14}$B magnets and promising magnetic properties \cite{Mooij1988,Buschow1991,Gabay2018}. In a work by S\"ozen \textit{et al.}, \textit{ab initio}-based thermodynamic phase stabilities of such RE-lean magnets have been reported \cite{Sozen2022}. Abundant and inexpensive Ce and Y have been chosen for a potential substitution of the critical Nd and their impact on the stability of the 1:12 phase has been discussed for Nd-Fe-Ti compounds.\\

\begin{figure}[h]
\centering
\begin{subfigure}{0.32\textwidth}
\includegraphics[width=\textwidth]{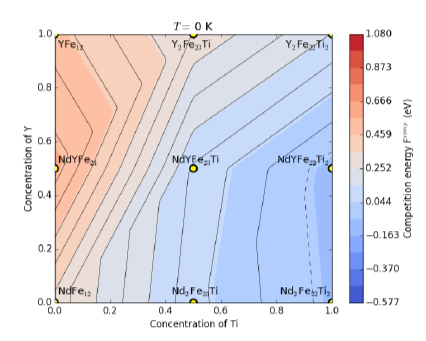}
\end{subfigure}
\begin{subfigure}{0.32\textwidth}
\includegraphics[width=\textwidth]{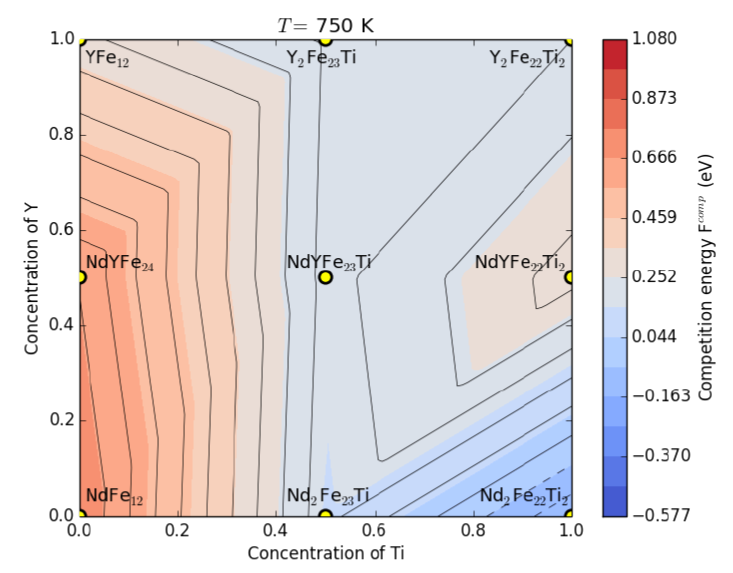}
\end{subfigure}
\begin{subfigure}{0.31\textwidth}
\includegraphics[width=\textwidth]{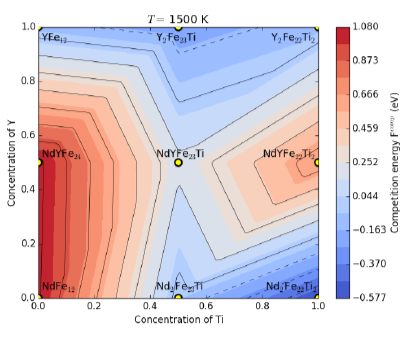}
\end{subfigure}
\begin{subfigure}{0.32\textwidth}
\includegraphics[width=\textwidth]{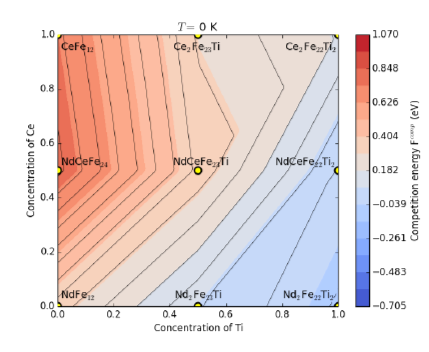}
\end{subfigure}
\begin{subfigure}{0.32\textwidth}
\includegraphics[width=\textwidth]{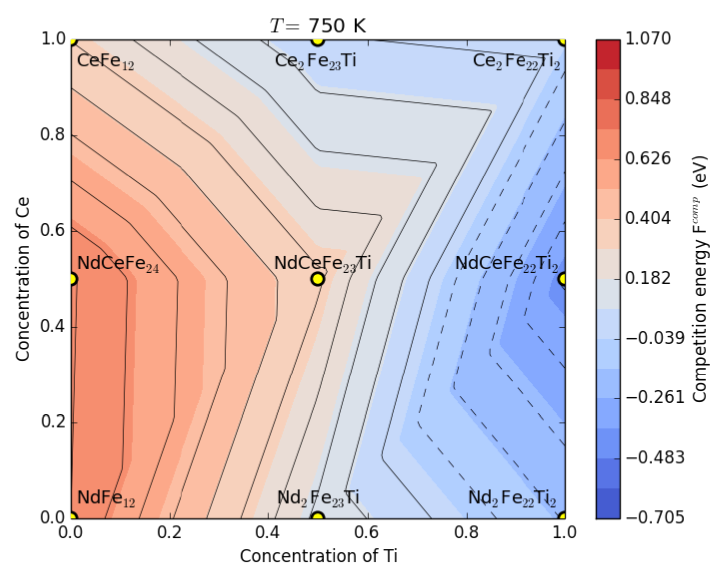}
\end{subfigure}
\begin{subfigure}{0.32\textwidth}
\includegraphics[width=\textwidth]{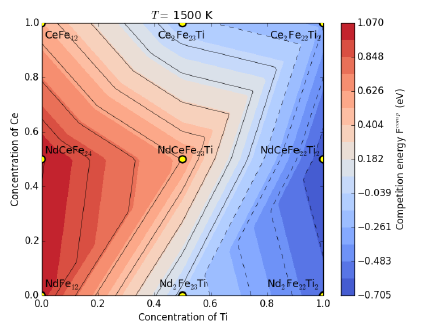}
\end{subfigure}
\caption{Calculated competition energies of NdFe$_{12}$ alloys with increasing concentrations of Ti and Ce or Y with three different temperatures 0, 750 and 1500 K. The calculations for the (Nd,X)Fe$_{12-y}$Ti$_y$ alloys (X = Y and Ce) were performed in 2 formula units with 26 atoms supercells in the limits of $0<x<0.5$ and $0<y<1$. Linear interpolation has been performed between the calculation points that correspond to intermetallic compositions. Images are taken from \cite{Sozen2022}. \label{fig_phase_stability}}
\end{figure}

In that study, state-of-the-art approaches for vibrational, electronic and magnetic entropy contributions of the Helmholtz free energy, $F$($T,V$), have been calculated for all hard magnetic and relevant competing phases. Competition energy formalisms have been developed for each ternary and quaternary system to understand the relative finite temperature phase stabilities. It has been revealed that Ce is a promising candidate to compensate Nd with a composition of (Nd,Ce)Fe$_{11}$Ti, which have appeared as stable quaternarys at all considered temperatures from 0 to 1500 K (Fig.~ \ref{fig_phase_stability}). For further details about competition energy formalisms it is advised to read Ref.~ \cite{Sozen2022} as competition energies are not a focus of this thesis.\\

Nevertheless, the effect of such substitutions on the intrinsic magnetic properties was not addressed before. In order to fill this gap and produce sustainable and reliable hard magnetic materials with reduced Nd content, detailed \textit{ab initio}-based investigations on the intrinsic magnetic properties of (Nd,X)-Fe-Ti-N alloys (X= Y and Ce) have been performed in this thesis. In order to have a systematic understanding, the intrinsic magnetic properties have been calculated starting from RFe$_{12}$ (R: Y, Ce and Nd) binaries. Since, the 1:12 phase is thermodynamically not stable for the considered RE-elements \cite{Hirayama2015a,Suzuki2017,Hirayama2017}, Ti is considered as a stabilizer in different concentrations for ternary RFe$_{12-y}$Ti$_y$ ($0.5\le y\le1$) compounds. Then as quaternary (Nd,X)Fe$_{12-y}$Ti$_y$ compounds, half of the Nd atoms have been replaced with alternative elements, which are Y and Ce, respectively. In total this makes 15 different ThMn$_{12}$-type Nd-X-Fe-Ti compounds and in addition, their nitrogenated cases have been examined as well. \\

\chapter{Theoretical Background}

In order to investigate the intrinsic magnetic properties of the considered compounds in this study, \textit{ab initio} calculations have been performed. The main goal is to achieve precise and accurate magnetic properties providing an atomistic understanding. In each calculated elastic and magnetic material, properties have been computed and compared with available experimental data.\\

In this chapter the theoretical fundamentals of the used methods are introduced. First of all, density functional theory (DFT) is explained as it is the core method in this thesis. Furthermore, fundamental points of quantum mechanics are explained, and a focus is put on the treatment of magnetism and \textit{f}-electrons in DFT. Due to the limitations of DFT regarding localized electrons,  a DFT+\textit{U} approach is also considered in this study, and its technical aspects are summarised.\\

\section{Density Functional Theory}

Density functional theory (DFT) is one of the most used quantum mechanical methods to describe the electronic structure of many-body systems in chemistry and physics. It is based on the Hohenberg-Kohn theorems \cite{Hohenberg1964} postulated in 1964 by Hohenberg and Kohn. As Hartree-Fock (HF) and post-Hartee-Fock (post-HF) methods, DFT describes the interactions of electrons based on quantum mechanical approaches using universal principles such as the Schrödinger equation. These kinds of calculations are called first-principle or \textit{ab initio} methods as they do not require additional experimental data with the selected input.\cite{Parr1989,Edition2001}\\ 

\begin{figure}[h]
\centering
\includegraphics[width=0.75\textwidth]{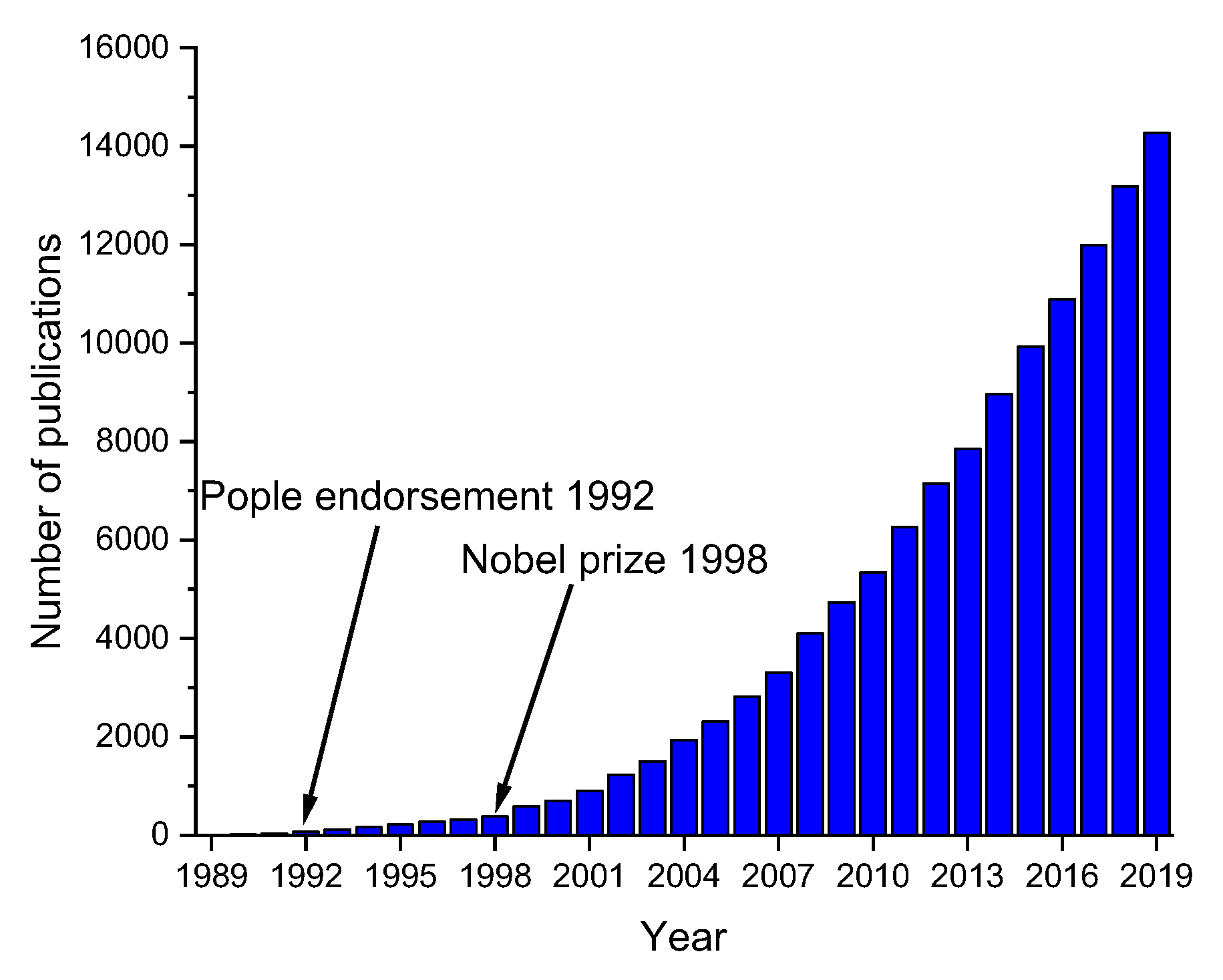}
\caption{Increasing number of publications using DFT. The Pople endorsement and the Nobel prize in chemistry are marked according to the appropriate year. The image is based on data from \cite{Dumaz2021}\label{increasing number of papers}}
\end{figure}

Contrary to the Hartree-Fock and post-Hartree-Fock methods, which use a Slater determinant, a determinant of single-electron wave functions to describe the many-electron wave function, DFT relies on the electron density of the investigated system to express the ground state properties of it. The many-body wave functions with 3\textit{N} variables are reduced to the electron density that only depends on 3 variables and is therefore much easier to handle. This approach by Hohenberg and Kohn was improved by Kohn and Sham with the Kohn-Sham equations \cite{Kohn1965} only one year later. They partially reintroduced the description via a wave function and mapped the many-body problem onto a system of non-interacting quasiparticles, thus simplifying the multi-electron problem into a problem of non-interacting electrons in an effective potential \cite{Sozen2019}. In this potential, the external potential and the Coulomb interactions between the electrons are included. This approach is the basis of current DFT methods \cite{Becke2014}. Since then, the popularity of DFT increased steadily as can be seen in Fig.~\ref{increasing number of papers}. This can mostly be attributed to the development of better and more precise functionals. A major acknowledgment of DFT was the awarded Nobel Prize in chemistry in 1998 by Walter Kohn and John Pople.\\

\section{Schrödinger's Equation}
As mentioned above, \textit{ab initio} methods are independent of experimental data and they only rely on the Schrödinger equation. This equation is the starting point of quantum mechanical understanding of a given material system \cite{Sozen2019}. The Schrödinger equation forms the basis of many \textit{ab initio} approaches and its time independent, non-relativistic state is defined as follows
\begin{equation}
\hat{H}\Psi _i (x_1,x_2,...x_N,R_1,R_2,...,R_M) = E_i \Psi _i(x_1,x_2,...x_N,R_1,R_2,...,R_M),
\end{equation}
with $\hat{H}$ as the Hamiltonian for a molecular system consisting of M nuclei and N electrons in the absence of magnetic or electric fields. This many-body Hamiltonian is a differential operator that represents the total energy with the following expression
\begin{equation}
\hat{H} = \hat{T}_e+\hat{T}_n+\hat{V}_{nn}+\hat{V}_{en}+\hat{V}_{ee}.
\end{equation}
In this equation, $\hat{T}_e$ is the kinetic energy of the electrons, and $\hat{T}_n$ is the kinetic energy of the nuclei. $\hat{V}_{nn}$, $\hat{V}_{en}$ and $\hat{V}_{ee}$ describe the electrostatic energy of attraction and repulsion for the nuclei-nuclei, electron-nuclei and electron-electron interactions. When using atomic units the Hamiltonian has the following form
\begin{equation}
\hat{H} = - \frac{1}{2} \sum_{i=1}^N \nabla_i^2 - \frac{1}{2} \sum_{A=1}^M \frac{1}{M_A} \nabla_A^2 - \sum_{i=1}^N \sum_{A=1}^M \frac{Z_A}{r_{iA}} + \sum_{i=1}^N \sum_{j>i}^N \frac{1}{r_{ij}} + \sum_{A=1}^M \sum_{B>A}^M \frac{Z_AZ_B}{R_{AB}},
\end{equation}
with $i$ and $j$ marking the $N$ electrons and $A$ and $B$ denoting the $M$ nuclei. $M_A$ represents the mass of nucleus A, and Z as the charge of the respective nucleus \cite{Edition2001}.

\section{Born-Oppenheimer Approximation}
Due to the complexity of the Schrödinger equation, an exact solution for most systems of interest is impossible. To reduce the complexity and make solutions possible, some approximations and methods were developed. The first and most prominent approximation is the Born-Oppenheimer approximation \cite{Born1927}, which separates the electronic and ionic degrees of freedom from each other. This assumption is based on the much higher velocity of the electrons due to their lower weight in comparison to the nucleus (as an example, in the case of H the nucleus weighs $\sim$ 1800 times more than its electron). This leads to an approximation in which the cores are seen as stationary. With this, the kinetic energy of the nuclei is now zero and the potential energy of the nucleus-nucleus repulsion is a constant. This reduces the complex Hamiltonian to the electronic Hamiltonian $\hat{H} _{elec}$:

\begin{equation}
    \hat{H}_{elec} = \hat{T}_e + \hat{V}_{en} + \hat{V}_{ee} = - \frac{1}{2} \sum_{i=1}^N \nabla_i^2 - \sum_{i=1}^N \sum_{A=1}^M \frac{Z_A}{r_{iA}} + \sum_{i=1}^N \sum_{j>i}^N \frac{1}{r_{ij}}.
    \label{eq_electronic_hamiltonian}
\end{equation}

Solving the Schrödinger equation with $\hat{H} _{elec}$ leads consequently to the electronic wave function $\Psi_{elec}$ and the electronic energy $E_{elec}$. Since $\Psi_{elec}$ only depends on the electronic coordinates and the coordinates of the nuclei are only entered parametrically and do not appear explicitly, the total energy $E_{tot}$ is a sum of the electronic energy $E_{elec}$ and the constant nuclear repulsion term $E_{nuc} = \sum _{A=1}^M \sum _{B>A}^M \frac{Z_A Z_B}{R_{AB}}$ \cite{Edition2001}.\\

With the total energy of the ground state of an electronic system calculated via Eq. \ref{eq_electronic_hamiltonian}, the ground state energy $E_0 = E[N,V_{ext}]$ and the ground state wave function $\psi_0$ can be determined with the variational principle for a system with $N$ electrons and a given nuclear potential $V_{ext}$. A full minimization of $E[\Psi]$ will then lead to the true ground state $\Psi_0$

\begin{equation}
    E_0 = \underset{\Psi \rightarrow N}{\text{min}} E[\Psi] = \underset{\Psi \rightarrow N}{\text{min}} \langle \Psi |\hat{T}_e + \hat{V}_{Ne} + \hat{V}_{ee} | \Psi \rangle.
\end{equation}

\section{The Hohenberg-Kohn Theorem}

Even though the Born-Oppenheimer approximation simplifies the Hamiltonian to a great amount, it is still too complicated to be solved. Especially the $V_{ee}$ term with a 3\textit{N} dimensionality turns out to be a challenging part. To solve these problems Hohenberg and Kohn postulated the Hohenberg-Kohn theorems \cite{Hohenberg1964} in 1964 and reduced the many-body wave function to the electron density $\rho (r)$, which represents the fundamental base for DFT\cite{Parr1989,Edition2001}. 

In their first theorem, they proved that since the electron density $\rho$ determines the number of electrons in a system, it also defines with $\rho(r)$ the ground state wave function $\Psi$ and every other electronic property. In a relatively simple way, they showed this by employing only the minimum-energy principle for the ground state. \\

They considered the electron density $\rho(r)$ for the nondegenerate ground state of \textit{N}-electron systems and  defined \textit{N} by quadrature. With the electron density, the external potential $\nu(r)$ and all other properties were also defined. When two external potentials $\nu$ and $\nu'$ are differing by more than a constant and each one gives the same density $\rho$ for its ground state then one could have two different Hamiltonians $\hat{H}$ and $\hat{H}'$ with the same ground state density but with different wave functions $\Psi$ and $\Psi'$. Trying to solve the $\hat{H}$ problem with $\Psi'$ as a trial function and also the opposite way with $\hat{H}'$ and $\Psi$ one could get the following expressions
\begin{equation}
E_0 < \langle \Psi'|\hat{H}| \Psi' \rangle = \langle \Psi'|\hat{H}'| \Psi' \rangle + \langle \Psi'|\hat{H}-\hat{H}'| \Psi' \rangle = E_0' + \int \rho(r)[\nu(r) - \nu'(r)]\,dr,
\end{equation}
and 
\begin{equation}
E_0' < \langle \Psi|\hat{H}'| \Psi \rangle = \langle \Psi|\hat{H}| \Psi \rangle + \langle \Psi|\hat{H}'-\hat{H}| \Psi \rangle = E_0 - \int \rho(r)[\nu(r) - \nu'(r)]\,dr,
\end{equation}
where $E_0$ and $E_0'$ are the ground state energies for $\hat{H}$ and $\hat{H}'$. 
Adding both equations the contradiction $E_0 + E_0' < E_0' + E_0$ is obtained, which indicates that there cannot be two different $\nu$ that give the same electron density for their ground states. With this they proved that the electron density determines $N$ and $\nu$ and all following ground state properties like the kinetic energy $T(\rho)$, the potential energy $V(\rho)$ and the total energy $E(\rho)$ \cite{Parr1989,Edition2001}.\\

In their second theorem, Hohenberg and Kohn introduced the variational principle in the following form
\begin{equation}
    E_0 \leq E(\tilde{\rho}) = T(\tilde{\rho}) + V_{Ne}(\tilde{\rho}) + V_{ee}(\tilde{\rho}),
\end{equation}
with $\tilde{\rho}(r)$ as any trial density that satisfies the needed boundary conditions such as $\tilde{\rho}(r) \geq 0$ and $\int \tilde{\rho}(r)\,dr = N$. This means that the minimum energy for the system can only be attained if the correct ground state density is used. Every other trial density $\tilde{\rho}(r)$ results in higher energies than the ground state energy which makes it possible to determine the ground state density by the variation of $\tilde{\rho}(r)$ until a minimum is reached \cite{Parr1989,Edition2001}. This way we arrive at the following equation

\begin{equation}
    \langle \tilde{\Psi} | \hat{H} |\tilde{\Psi} \rangle = T[\tilde{\rho}] + V_{ee}[\tilde{\rho}] + \int \tilde{\rho}(r) \nu_{ext}\,dr = E[\tilde{\rho}] \geq E_0[\tilde{\rho}_0] = \langle \Psi_0 | \hat{H} |\Psi_0 \rangle.
    \label{eq_hohenberg_kohn}    
\end{equation}

Additionally, the kinetic energy $T(\rho)$ and the Coulomb interaction energy $V_{ee}(\rho)$ can be summarized in the Hohenberg-Kohn functional $F_{HK}(\rho)$
\begin{equation}
    F_{HK}(\rho) = T(\rho) + V_{ee}(\rho).
\end{equation}
Furthermore, the Coulomb interaction energy $V_{ee}(\rho)$ in this assumption includes the classical repulsion term $J(\rho)$ and a non-classical exchange term $K_{ex}(\rho)$.

\begin{equation}
   V_{ee}(\rho) = J(\rho) + K_{ex}(\rho).
\end{equation}

\section{The Kohn-Sham Equations}
However, Eq. \ref{eq_hohenberg_kohn} is not providing an exact solution. Therefore, Kohn and Sham reformulated the current approach in 1965 \cite{Kohn1965} and introduced a new scheme by mapping the fully interacting electronic system onto a fictitious system of non-interacting quasi particles moving in an effective potential. The Kohn-Sham (KS) equations can be reformulated as

\begin{equation}
    \hat{H}_{KS} \psi _i = \epsilon_i \psi _i,
\end{equation}
with $\epsilon _i$ as the orbital energy of the corresponding KS orbital $\psi_i$ and the KS Hamiltonian defined as
\begin{equation}
    \hat{H}_{KS} = [- \frac{1}{2} \nabla ^2 + V_{eff} (r)],
    \label{KS_hamiltonian}
\end{equation}
with $V_{eff}(r)$ as an effective local potential. Therefore, the problem of finding a solution to the many-body Schrödinger equation is now replaced by solving single particle equations. Since the KS Hamiltonian is a functional of just one electron at point $r$, the electron density can be defined according to the HK theorem.
\begin{equation}
    \rho (r) = \sum _{i=1} ^N |\psi_i (r) |^2.
\end{equation}
Besides, the kinetic energy term and the classical Coulomb interaction energy of the electrons can be defined as
\begin{equation}
    T_e = - \frac{1}{2} \sum _{i=1} ^N \int \,d^3 r |\nabla \psi _i (r) | ^2, 
\end{equation}
\begin{equation}
    J(\rho) = \frac{1}{2} \int \, d^3 r d^3 r' \frac{\rho (r) \rho (r')}{|r - r'|}.
\end{equation}
Then, the Hohenberg-Kohn ground state energy can be written according to the Kohn-Sham approach

\begin{equation}
    E_{KS} = \sum _i ^N \epsilon _i - J(\rho) + E_{xc} - \int \frac{\delta E_{xc}}{\delta \rho(r)}.
    \label{kohn-sham-approach}
\end{equation}
$\epsilon _i$ are the one electron energies and are a result of the KS equations. However, they have limited physical meaning. The last term in Eq. \ref{kohn-sham-approach} is the exchange-correlation $E_{xc}$ term, which contains all many-body interactions of exchange and a substantial part of the electron interactions (besides the Hartree term)\cite{Sozen2019}. The exchange-correlation $E_{xc}$ can therefore be defined as

\begin{equation}
    E_{xc} = T_{exact}(\rho) - T_S(\rho) + V_{ee}(\rho) - J(\rho) = T_{exact}(\rho) - T_S(\rho) + K_{ex}(\rho).
\end{equation}
This means that $E_{xc}$ is the sum of the difference between the exact kinetic energy of the interacting system $T_{exact}(\rho)$ and the kinetic energy of the non-interacting system $T_{S}(\rho)$, achieved by the Kohn-Sham approach, and the non-classical exchange part $K_{ex}(\rho)$ of the Coulomb interaction energy between the electrons $V_{ee}(\rho)$. Since $E_{xc}$ includes the unknown terms of $T_{exact}(\rho)$ and $K(\rho)$ the exchange-correlation energy can not be calculated exactly\cite{Parr1989,Edition2001}.

\section{Exchange-Correlation Energy}

\begin{figure}[h]
\centering
\includegraphics[width=0.95\textwidth]{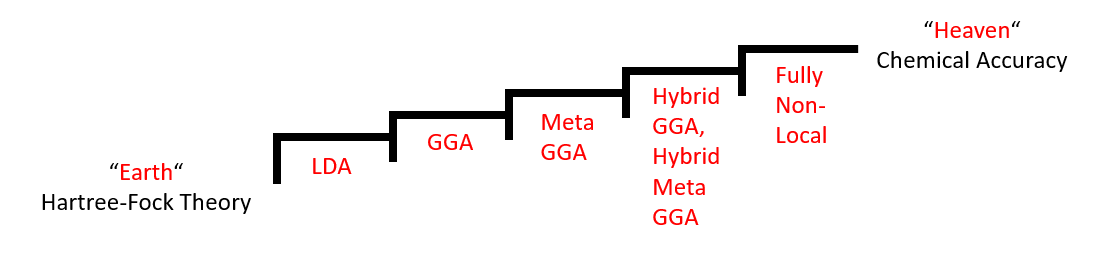}
\caption{Jacob's ladder showing the different functionals (LDA, GGA, Meta GGA, Hybrid GGA, Hybrid Meta GGA and Fully Non-Local) leading to increases of the accuracy of calculations. The image is adapted from \cite{Awoonor-Williams2018}\label{fig_jacobs_ladder}.}
\end{figure}

As explained with the Kohn-Sham equations, the exchange-correlation functional can not be calculated exactly and has to be approximated. For this approximation different methods with significantly differing quality depending on the actual system and their level of complexity have been developed. The accuracy of these functionals is often correlating with the computational cost and thus a functional that treats the exchange part correctly is also computationally very demanding. Thus, one has to find a reasonable balance between the accuracy needed and the computational demand for the calculation. In 2001 Perdew \textit{et al.} \cite{Perdew2001} introduced the \textit{Jacob's ladder}, as seen in Fig.~\ref{fig_jacobs_ladder}, which is a useful way to categorize various exchange-correlation functionals. 

Each functional on this ladder is categorized according to its level of accuracy on one of the steps. The "earth" of the ladder is representing the HF theory and the "heaven" of the ladder is the exact exchange-correlation functional. Therefore, the higher a functional is located on the ladder the more precise and computationally demanding it is. The first two steps on the ladder are the local density approximation (LDA) \cite{Vosko1980} and the generalized gradient approximation (GGA) \cite{Perdew1996}. Both of these functionals have been used in this work and are the most used functionals today.

\section{Local Density Approximation}

The first step on the \textit{Jacob's ladder} is the local density approximation (LDA). Although it is considered as a functional with a low accuracy, it is virtually the basis for all other exchange-correlation functionals. The core of this approximation is the model of a uniform electron gas with the electrons moving on a positive background charge distribution to achieve an electrically neutral system. The number of electrons \textit{N} as well as the volume of the electron gas \textit{V} are considered to approach infinity. The electron density \textit{N/V} on the other hand is assumed as finite with a constant value everywhere.

\begin{equation}
    E_{xc}^{LDA} = \int \rho (r) \epsilon _{xc} [\rho (r)] \,dr,
\end{equation}
with $\epsilon _{xc} [\rho (r)]$ as the exchange-correlation energy per particle of a uniform electron gas of density $\rho (r)$. This energy per particle is weighted with the probability $\rho (r)$ that there is in fact an electron at this point in space.

This model physically resembles an idealized metal with a perfect crystal of valence electrons and smeared out positive cores. For simple metals like sodium or systems with only a small variation of the density this assumptions is relatively good. However, for most situations in atoms and molecules, it is far from reality as the density often varies in these cases rapidly. LDA is the only functional in DFT for which the form of the exchange functional is known to a very high accuracy \cite{Edition2001,Parr1989}.

\section{Local Spin Density Approximation} 
The local density approximation can furthermore be reformulated within an unrestricted version. In this version, not the electron density $\rho(r)$ is employed but the two spin densities $\rho_{\alpha}(r)$ and $\rho_{\beta}(r)$ (with $\rho_{\alpha}(r) + \rho_{\beta}(r) = \rho(r)$). For the functional, this provides additional flexibility as it now holds two variables instead of one thus enabling more accuracy in case of an unequal number of electrons for $\alpha$ and $\beta$. This functional is known as the local-spin density approximation (LSDA) \cite{Edition2001,Parr1989}.

\begin{equation}
    E_{xc}^{LSDA}[\rho _{\alpha} \rho _{\beta}] = \int \rho (r) \epsilon _{xc} [\rho _{\alpha} (r),\rho _{\beta} (r) ] \,dr.
\end{equation}

With LSDA there are expressions known for the exchange and correlation energies for spin-compensated situations ($\rho _{\alpha} (r) = \rho _{\beta} (r) = \frac{1}{2} \rho (r)$ as well as spin-polarized cases ($\rho _{\alpha} (r) \neq \rho _{\beta} (r)$.\\

Since LDA and LSDA are only the first steps on the \textit{Jacobs ladder}, there are several drawbacks to be expected from these functionals. LDA and LSDA tend to under-predict the ground state energies as well as the ionization of systems and to over-predict the binding energies. It has also to be noted that high spin structures are often favored \cite{Edition2001}.

\section{Generalized Gradient Approximation}
Since the low accuracy of LDA and LSDA is insufficient for most applications in chemistry, these functionals are rarely used in computational chemistry. To account for the inhomogenity of the electron density, the gradient of it $\nabla \rho (r)$ is considered instead of only the charge density at a point $r$.

\begin{equation}
    E_{xc}^{GGA}[\rho] = \int \rho (r) \epsilon _{xc} (\rho (r) ,\nabla \rho)\,dr
\end{equation}

In the case of a generalized gradient approximation (GGA) functional \cite{Edition2001,Parr1989}, it takes a slowly varying electron density into account paired together with a few restrictions to restore the hole constraints lost in comparison to LDA. Generally, GGA delivers more accurate results than LDA, but in some cases, LDA still works better. Commonly used GGA functionals are the functional by Perdew and Wang (PW91) \cite{Perdew1992,Perdew1992a} and the functional by Perdew, Burke and Ernzerhof (PBE) \cite{Perdew1996}.\\

The choice of the considered functional has a huge impact on the results, therefore both the LDA (LSDA) and the GGA functional have been used in this study to achieve the most accurate results and compare the performance of functionals. As it was shown in works by S\"ozen \textit{et al.} \cite{Sozen2019,Sozen2019a,Sozen2020,Sozen2022} the lattice parameters and magnetic properties like the total magnetic moment can properly be calculated employing the PBE functional. However, in this work it has been seen that the GGA approach fails for magnetic anisotropy calculations, therefore, it has been additionally considered the LSDA functional in these cases to compare both results.

\section{Ultra-soft Pseudopotentials and the Projector-
Augmented Wave Method}
Apart from the calculation of the exchange-correlation energy $E_{xc}$ the treatment of the nuclei-electron interactions plays an important role in DFT as well. The most effective and best known methods are the projector-augmented wave method (PAW) of Blöchl \cite{Blochl1994,Kresse1999} and the ultrasoft pseudopotentials (USPP) method of Vanderbilt \cite{Vanderbilt1990}.\\

In these approaches, the effects of the motion of the core electrons is replaced by an effective potential. In the USPP method a certain core radius gets replaced by a soft nodeless pseudo wave function. This leads to a reduction of the basis set size. It is important to note that the pseudo wave function needs to have the same norm as the original wave function in the determined core radius and has to be identical to it outside of this radius. The drawback of this approach is that elements with strongly localized electrons require a large basis set for these pseudopotentials and therefore  they are expensive to calculate. Another disadvantage is that the pseudopotentials are hard to generate as many parameters have to be chosen accordingly.

\begin{figure}
    \centering
    \includegraphics[width=0.75\textwidth]{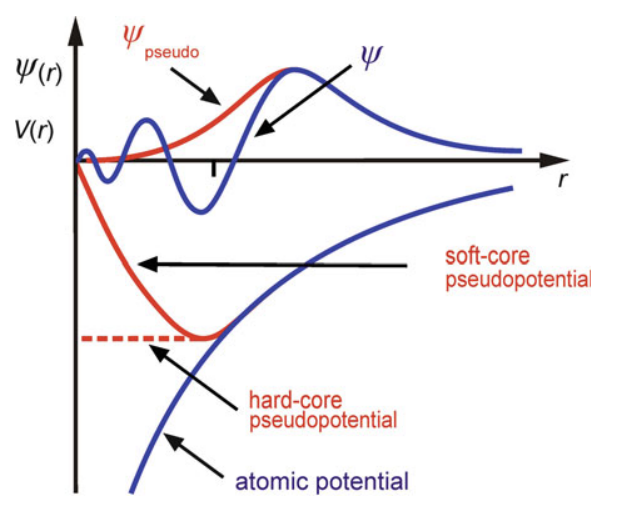}
    \caption{Radial dependence of typical pseudopotentials in comparison with the atomic Coulomb potential (blue). In the lower part a soft-core pseudopotential and a hard-core pseudopotential are shown with the solid and dashed red curve respectively. In the top part are the wave functions and the pseudo wave function resulting from the atomic (blue) and the pseudo-potential (red) depicted. It has to be noted, that both wave functions have the same values above a certain point r. The image is taken from \cite{Boer2018} \label{fig_pseudopotential}}
\end{figure}

In the PAW method, these short-comings were removed by Blöchl with the introduction of a linear transformation from the pseudo wave function to the many-electron wave function. He achieved the PAW total energy by applying this transformation to the KS functional. Furthermore, Blöchl splitted the grid of augmentation charges into a radial support grid and a regular grid. This makes it possible to use the PAW method directly with the many-electron wave function and the according many-electron potentials \cite{Sozen2019}.

\section{Magnetism in Density Functional Theory}

Since the theorems postulated by Hohenberg and Kohn \cite{Kohn1965} are only developed for spinless cases they are insufficient for the hard magnetic materials discussed in this work. The formalism was extended by Barth and Hedin \cite{VonBarth1972a} for spin polarized cases, and the generalization was done by a replacement of the scalar density with a Hermitian 2 $\times$ 2 matrix. This density matrix $n(r)$ can be defined as \cite{Bihlmayer2007}
\begin{equation}
    n_{\alpha \beta} (r) = \sum _{i=1} ^N \psi _i ^{*\alpha} (r) \psi_i ^{\beta}  (r),
\end{equation}
with $\alpha,\beta = (+), (-)$. Decomposing this matrix further, a scalar and a vectorial part corresponding to charge and magnetization densities with the Pauli matrices can be written as

\begin{equation}
    n(r) = \frac{1}{2}(n(r)I + \sigma \cdot m(r)) = \frac{1}{2}\begin{pmatrix} n(r) + m_z (r) & M_x (r) - im_y (r)\\ m_x(r) + im_y(r) & n(r) - m_z(r) \end{pmatrix}.
\end{equation}

Accordingly, the Schrödinger equation can be re-written as follows
\begin{equation}
    [(-\frac{\hbar^2}{2m}\nabla^2 + \sum _\alpha \int \frac{n_{\alpha \alpha (r')}}{|r - r'|}\,dr')I + \nu (r) + \frac{\delta E_{xc}}{\delta n(r)}] \begin{pmatrix} \psi_i ^{(+)} r \\ \psi _i ^{(-)} \end{pmatrix} = \epsilon _i \begin{pmatrix} \psi_i ^{(+)} r \\ \psi _i ^{(-)} \end{pmatrix},
\end{equation}
with I as a 2 $\times$ 2 unit matrix and the exchange-correlation matrix as well as a 2 $\times$ 2 matrix \cite{Bihlmayer2007}.

Another way to write the potential matrices are the following expressions

\begin{equation}
    \nu (r) = \nu (r) I + \mu _B \sigma \cdot B(r),
\end{equation}
and
\begin{equation}
 \nu _{XC} (r) = \nu _{XC} (r) I + \mu _B \sigma \cdot B_{xc} (r),
\end{equation}
with $B(r)$ as a magnetic field and $\mu_B = \frac{e \hbar}{2mc}$.
It is assumed that the magnetic structure is collinear. The potential matrices in this case are considered as diagonal which means that the magnetic and exchange fields are pointing in the \textit{z} direction. This leads to a decoupling of the re-written Schrödinger equation into the form of the two equations 
\begin{equation}
   (- \frac{\hbar ^2}{2m}\nabla ^2 + \nu _{Coul} (r) + \nu (r) + B_z (r) + \nu _{xc} ^{(+)} (r) ) \psi _i ^{(+)} (r) = \epsilon_i ^{(+)} \psi_i ^{(+)} (r),
\end{equation}
\begin{equation}
    (- \frac{\hbar ^2}{2m}\nabla ^2 + \nu _{Coul} (r) + \nu (r) - B_z (r) + \nu _{xc} ^{(-)} (r) ) \psi _i ^{(-)} (r) = \epsilon_i ^{(-)} \psi_i ^{(-)} (r).
\end{equation}
In these equations, $\nu_{Coul}$ marks the classical Coulomb potential and $\nu_{XC}^{+,-}$ the exchange-correlation potential in accordance to the spin up or spin down part of the diagonal density matrix. Both equations need to be solved independently \cite{Bihlmayer2007}. 

With these two equations, all kind of magnetic structures (ferromagnetic, antiferromagnetic or ferrimagnetic) can be calculated in the collinear case. The spin density and the spin moment can be obtained with the following two equations, respectively
\begin{equation}
    m(r) = -\mu_B \sum _{\alpha,\beta} \psi _\alpha ^{(+)} (r) \sigma _{\alpha \beta} \psi _\beta (r),
\end{equation}
and
\begin{equation}
    M _{spin} = \int m(r) \,dr = \int (n^{(+)}(r) - n^{(-)}(r)) \,dr.
\end{equation}

The quality of the calculated magnetic moment depends on the exchange-correlation functional used. In most cases, GGA yields better results than LSDA \cite{Sozen2019}.

\section{Treatment of \textit{f}-electrons}

Due to large self-interaction errors, strongly correlated \textit{f}-electrons cannot be represented sufficiently in DFT. The functionals of LDA and GGA are not able to fully cancel the electronic self-interaction in the Hartree term, which leads to the phenomenon that the electron sees itself. This creates false repulsion and supports electrons to artificially delocalize. This is especially a problematic case for partially filled \textit{f} states for the RE elements Pr-Eu and Tb-Yb. The exceptional cases, which are Ce and Gd, are handled reasonably well. As in the case of $\alpha$-Ce, the \textit{f}-electrons are delocalized and Gd has an half filled \textit{f}-shell. However, in case of $\gamma$-Ce, this exception is not valid anymore as the \textit{f}-electrons are strongly localized. In order to handle these elements, different methods beyond DFT can be used. Two of these methods are DFT+\textit{U} \cite{Lichtenstein1995,Dudarev1998} and the dynamical mean-field theory (DMFT) \cite{Georges1992} of which only the first one is used in this thesis.

With DFT+\textit{U}, an energy contribution called the Hubbard \textit{U} repulsion energy is added. This energy term originates in the Hubbard model, where the Hamiltonian describes moving particles on a lattice. It adds two contributions to the the Hamiltonian in form of a hopping term, that describes the probability of a particle to hop from one site to another, and a penalty energy that represents the on-site repulsion. The on-site repulsion is commonly named as \textit{U}. With the addition of \textit{U}, the artificial delocalization due to the self interaction errors are tried to be removed.

The according energy functional can be expressed as

\begin{equation}
    D^{DFT+U} [n] = E^{DFT} [n] - E^{dc} [n_i] + E_U[n_i],
\end{equation}
with $E^{dc}$ as a correction term to avoid double counting of correlation effects and $E^U$ as a repulsion term in the form of $E^U [n_i] = \frac{1}{2} U \sum _{i\neq j} n_in_j$ that is added to the occupancies of $n_i$.

The DFT+\textit{U} approach is mostly implemented in form of the method by Liechtenstein \cite{Liechtenstein1987} or by Dudarev \cite{Dudarev1998}. As only the latter one is used in this thesis, only the Dudarev approach will be described and can be formulated as

\begin{equation}
    E^{DFT+\textit{U}}_{Dudarev} = E^{DFT} + \frac{U_{eff}}{2} \sum _{I,\sigma} \sum _i n_i ^{I,\sigma} (1 - n_i ^{I,\sigma}.
\end{equation}

The effective \textit{U} parameter $U_{eff}$ is determined as the on-site Coulomb interaction parameter and is crucial for the method. To define $U_{eff}$, density of states calculations (DOS) can be performed. In case of NdFe$_{11}$Ti the experimental findings for the \textit{f} peak results in a value of 4.65 eV below the Fermi energy for $U_{eff}$.


\chapter{Computational Details}

\section{Vienna \textit{ab initio} Simulation Package Calculations}

\label{Sec_compDetails}

In this thesis, all first-principles calculations were performed in the framework of spin-polarized density functional theory (DFT) using the Vienna \textit{ab initio} Simulation Package (VASP) \cite{Kresse1996a,Kresse1996}. The projector-augmented wave method (PAW) was used as implemented in VASP. Exchange-correlation effects were treated within the generalized gradient approximation (GGA) of Perdew-Burke-Ernzerhof (PBE) \cite{Perdew1996}. In addition to this, the performance of the local spin density approximation (LSDA) \cite{Vosko1980} was examined for magnetocrystalline anisotropy (MAE) calculations in case of the clean ternary and quarternary compounds. For LSDA-MAE calculations, the optimized geometry obtained through the relaxation with the GGA-PBE functional was used as input. For clearance, calculations in this work, except for $T_C$ calculations that are referred to as DFT (DFT+\textit{U}) are treated using the GGA (GGA+\textit{U}) functional, and the treatment with LSDA is always mentioned explicitly.

The 4\textit{f}-electrons of both Ce and Nd elements were treated carefully, which is not straightforward in DFT. In the case of Ce, the single \textit{f}-electron was treated explicitly as valence state. The treatment of correlated \textit{f}-electrons in DFT requires the Hubbard \textit{U} correction. However, Ce-Fe compounds exhibit different behavior due to the hybridization of Ce-4\textit{f} with Fe-3\textit{d}-electrons. Therefore, similar to works done by S\"ozen \textit{et al.} \cite{Sozen2019a,Sozen2020}, no Hubbard \textit{U} correction scheme was applied for Ce-4\textit{f}-electrons. In the case of Nd, the \textit{f}-electrons were examined as in-core state for the DFT scheme. This treatment yielded good mechanical and elastic properties, however dramatically failed for the magnetic properties due to the missing \textit{f}-electrons. Hence, additional DFT+\textit{U} treatment via the Dudarev method \cite{Dudarev1998} was considered for the Nd case with Hubbard \textit{U}=6 eV. It has to be noted that in this work the orbital contribution of the Nd-4\textit{f} electrons to the magnetic moment are not treated and only the spin magnetic moments are considered for the calculation of the different properties. Therefore, the total magnetic moment refers only to the total spin magnetic moment.
 
The Brillouin zone was sampled with a \textit{k}-point mesh
described by a $\Gamma$-centered grid, considering
10$\times$10$\times$8 meshes for supercells
containing 26 atoms (28 for N contained cases). The cutoff
energy for the plane wave basis used was 500 eV, and the
width of the smearing parameter was 0.1 eV. The convergence criteria within the self-consistent field (SCF) scheme was set to be 10$^{-5}$ eV for all considered alloys. The values of all these input parameters provided an energy convergence with an error equal to or smaller than 1 meV/atom.

For magnetocrystalline anisotropy calculations, first a collinear self-consistent calculation with the mentioned input details was performed. The obtained geometry and magnetic configuration were used as input in non-collinear calculations that take the spin-orbit coupling into account. In each non-collinear run the magnetic moment was aligned along one of the crystallographic directions [001],[100],[010] and [110] to obtain the different energies for each direction.

\section{Korringa-Kohn-Rostoker Calculations}
 
 In order to calculate the Curie temperature $T_\text{C}$, the exchange interaction energies $J_{ij}$ for converged ferromagnetic (FM) and local-moment-disorder (LMD) states of the 1:12 phases by the Liechtenstein method were computed \cite{Liechtenstein1987}. The magnetic couplings were calculated using density functional theory following the Korringa-Kohn-Rostoker (KKR) \cite{Korringa1947,Kohn1954} Green's function method based AkaiKKR \cite{AkaiKKR} code -also known as MACHIKANEYAMA- with the implemented atomic sphere approximation (ASA) incorporating coherent potential approximation (CPA) \cite{Shiba1971,Akai1977}. Continuous concentration changes for both RE and TM sublattices were considered on the basis of KKR-CPA. All KKR calculations were based on the local density approximation (LDA) \cite{Hohenberg1964,Kohn1965}, taking the exchange-correlation function as parametrized by Moruzzi, Janak and Williams (MJW) \cite{V.L.MoruzziJ.F.Janak1978}. In this work, the scattering was considered up to \textit{d}-scattering ($l_{max} = 2$) in the systems and all \textit{f}-electrons were put in the valence state on basis of the open-core approximation \cite{Jensen1991,Richter1998}.


\chapter{Results and Analysis}
\section{Determination of the Hubbard \textit{U} Parameter by Density of States Calculations}

\begin{figure}
\begin{subfigure}{0.50\textwidth}
\includegraphics[width=0.95\textwidth]{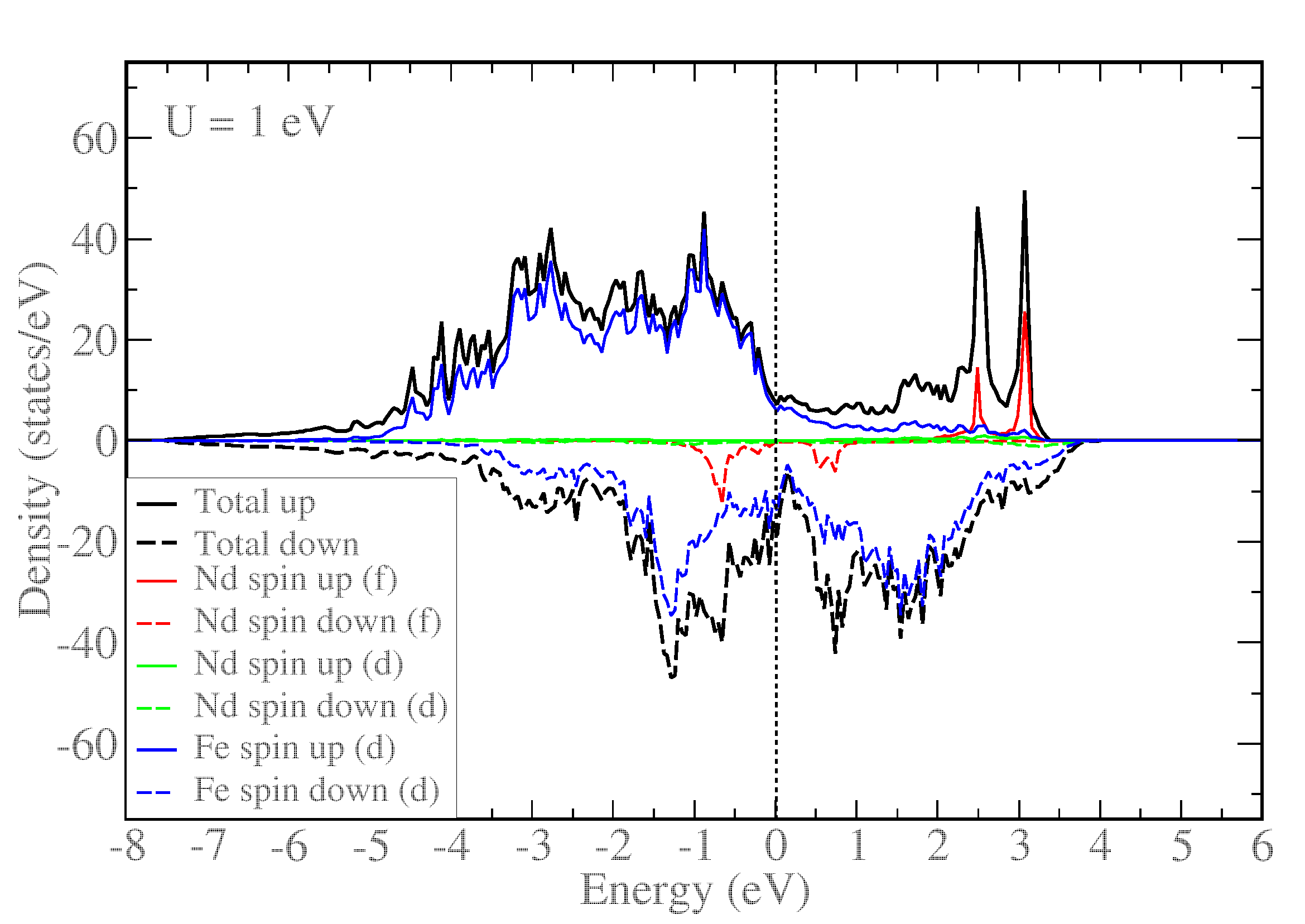}
\end{subfigure}
\begin{subfigure}{0.50\textwidth}
\includegraphics[width=0.95\textwidth]{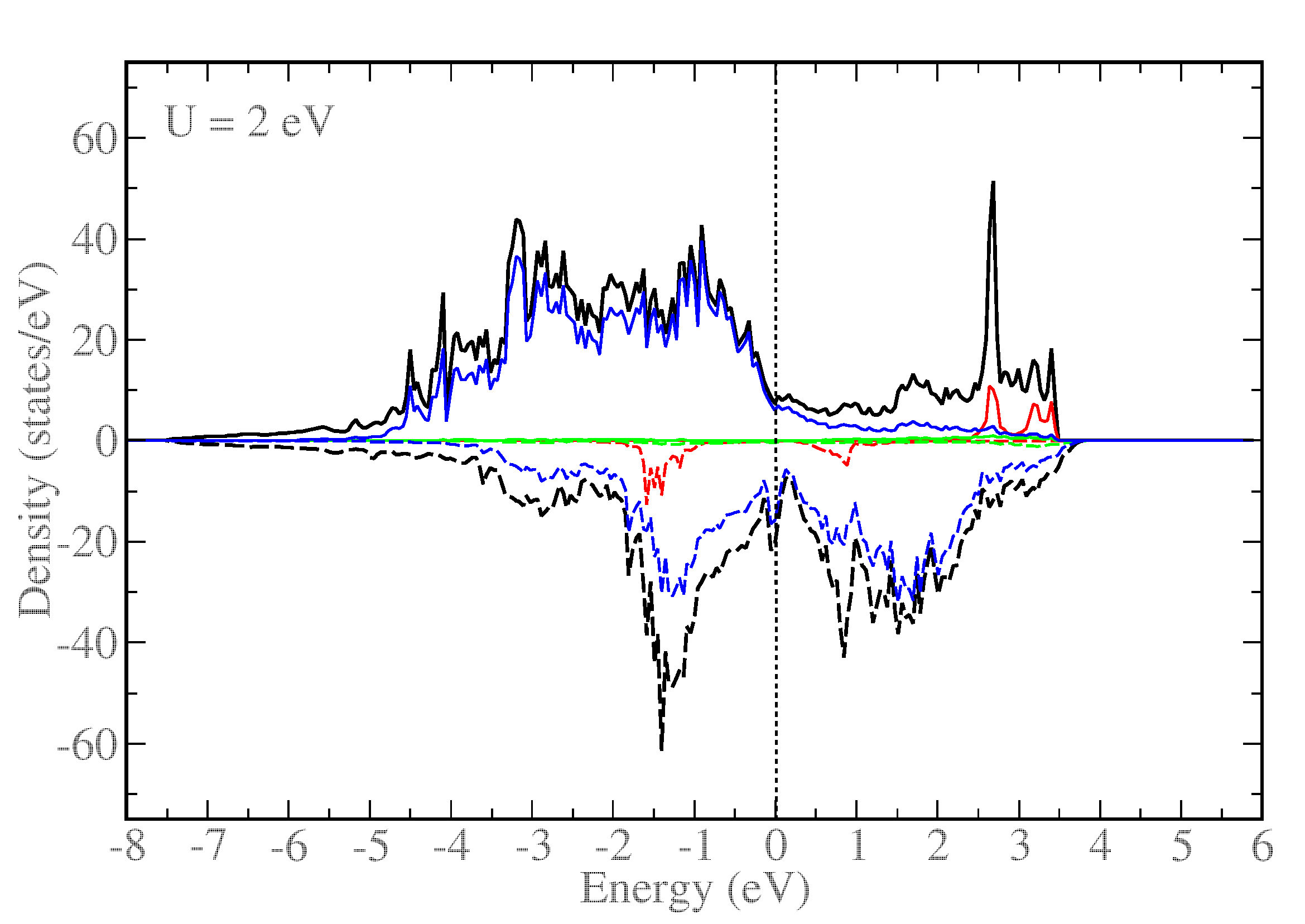}
\end{subfigure}
\begin{subfigure}{0.50\textwidth}
\includegraphics[width=0.95\textwidth]{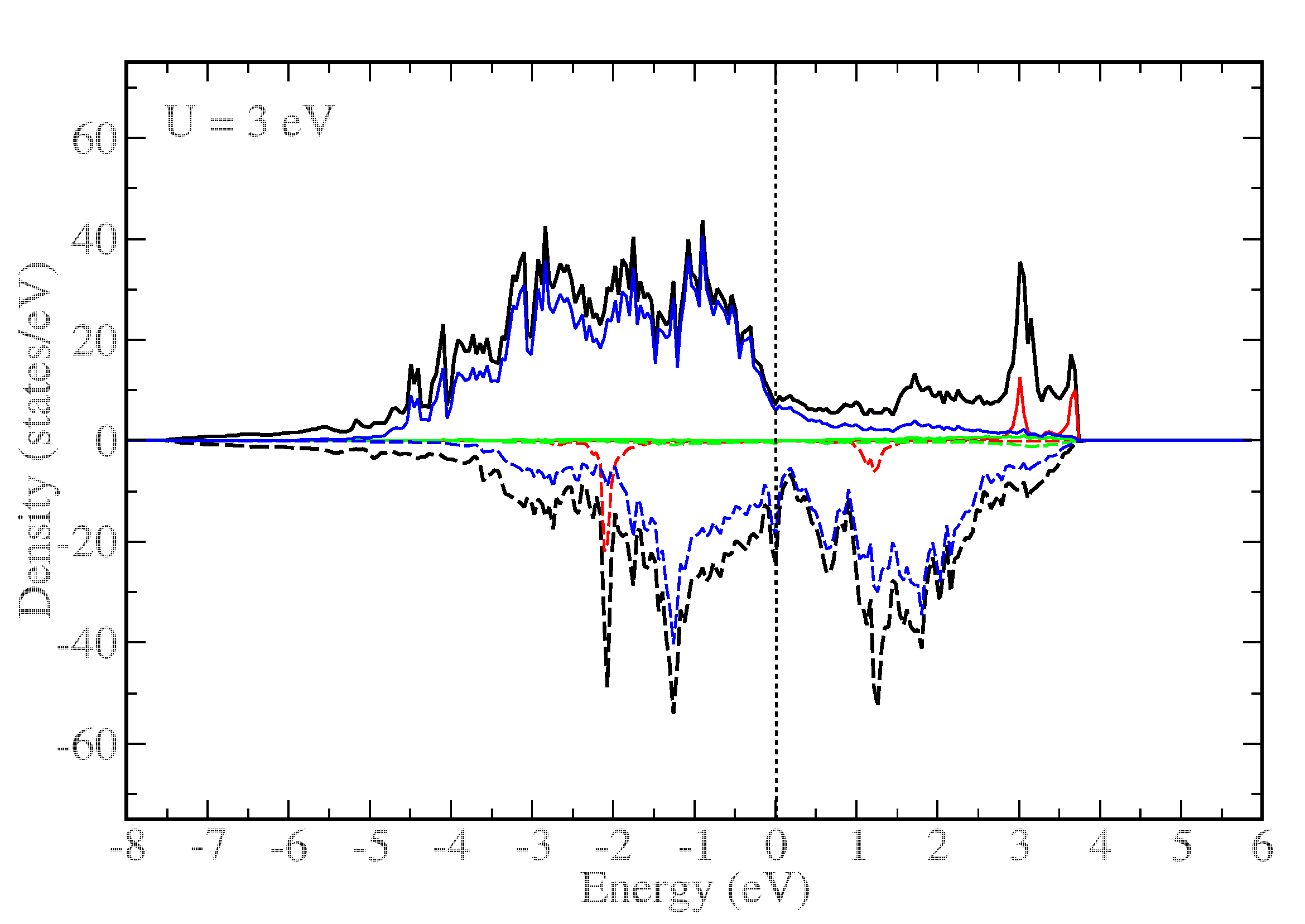}
\end{subfigure}
\begin{subfigure}{0.50\textwidth}
\includegraphics[width=0.95\textwidth]{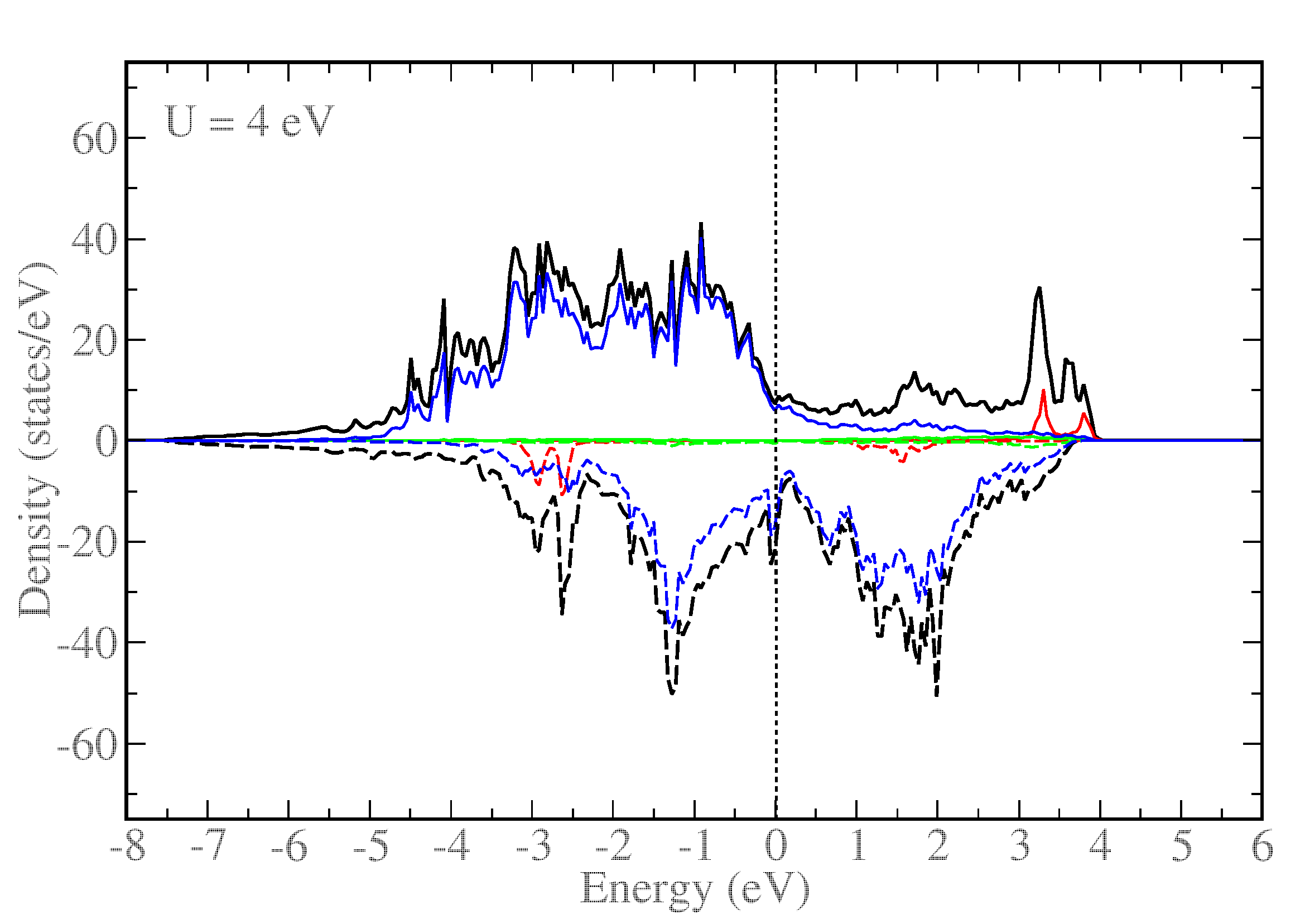}
\end{subfigure}
\begin{subfigure}{0.50\textwidth}
\includegraphics[width=0.95\textwidth]{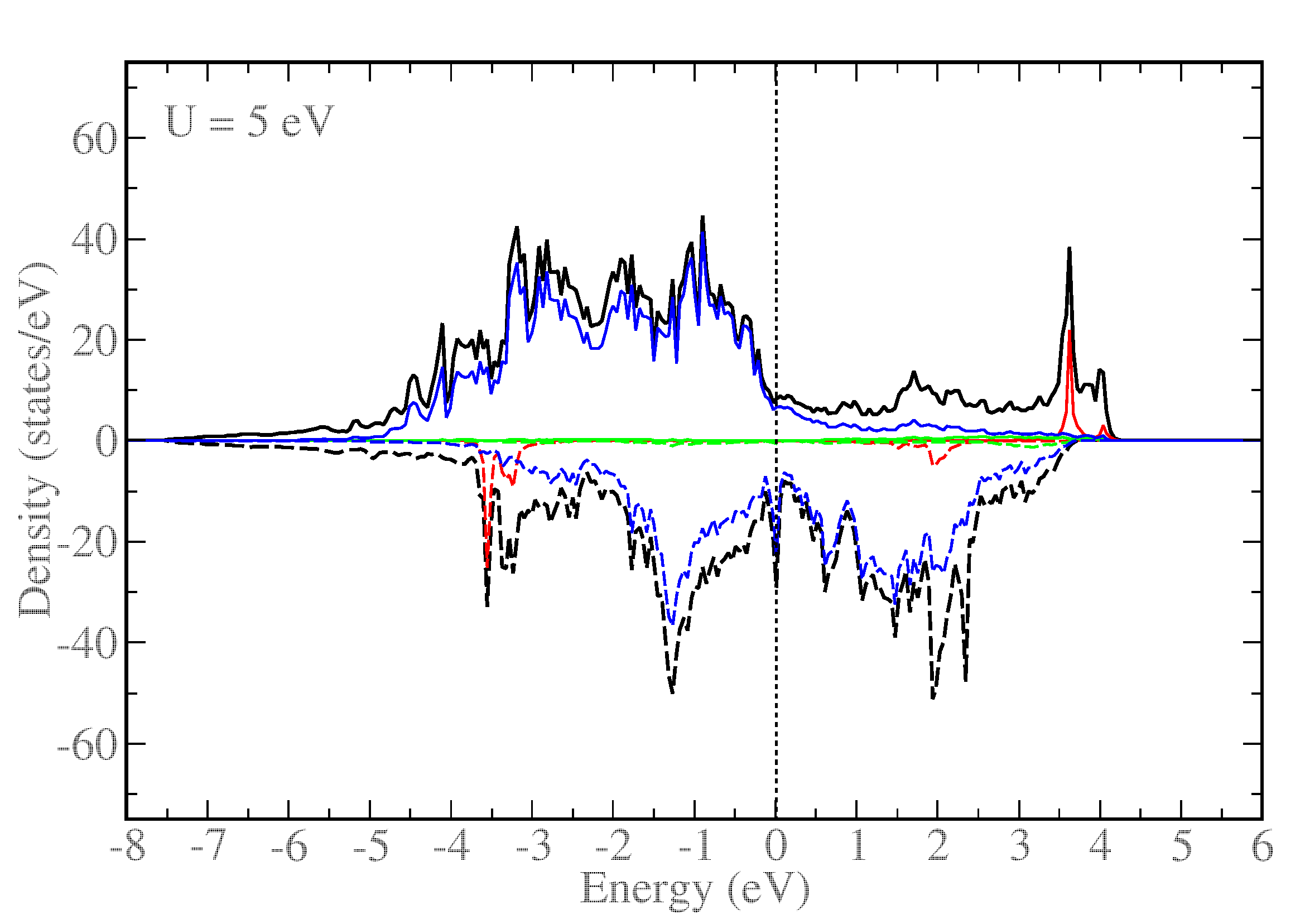}
\end{subfigure}
\begin{subfigure}{0.50\textwidth}
\includegraphics[width=0.95\textwidth]{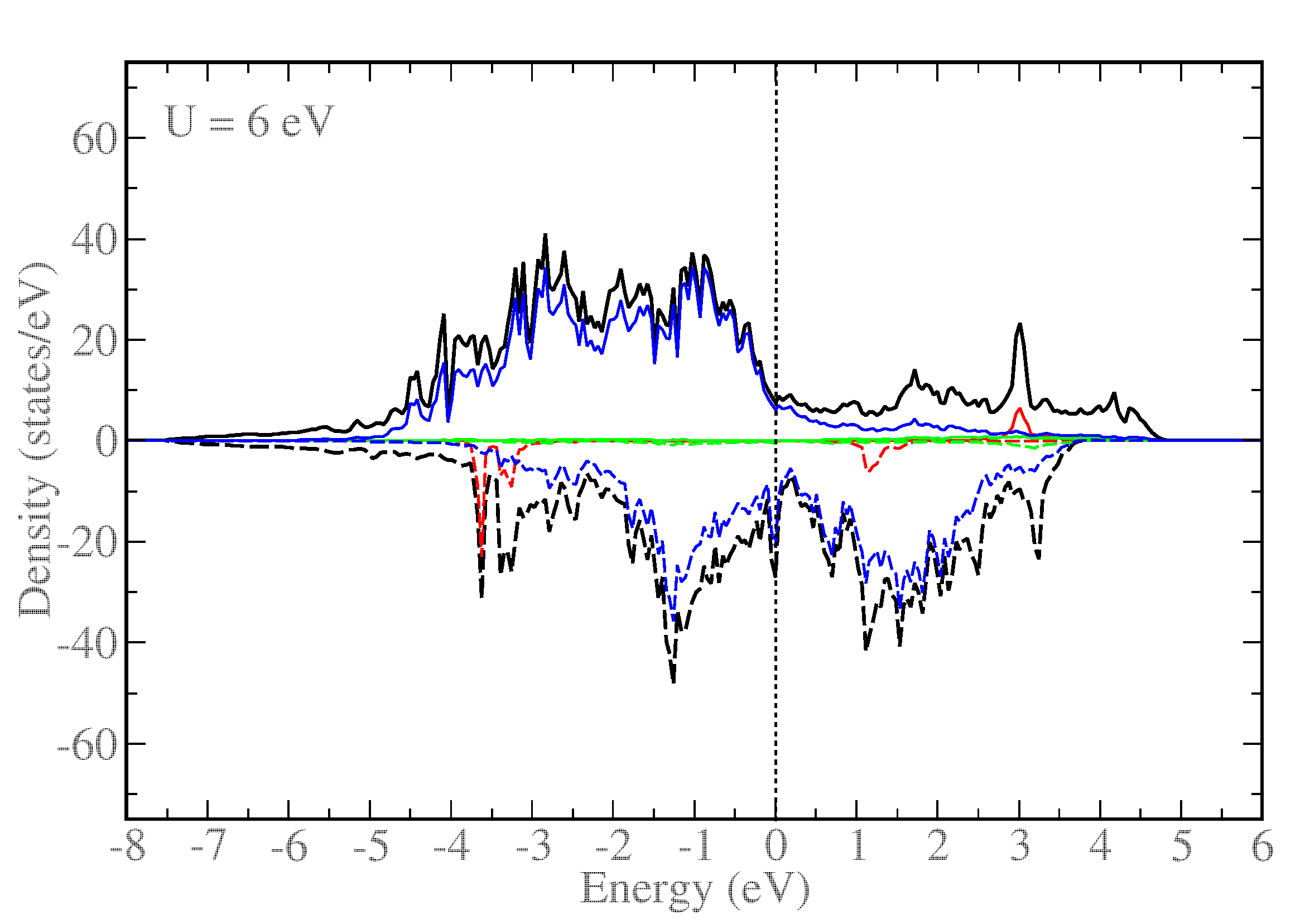}
\end{subfigure}
\begin{subfigure}{0.50\textwidth}
\includegraphics[width=0.95\textwidth]{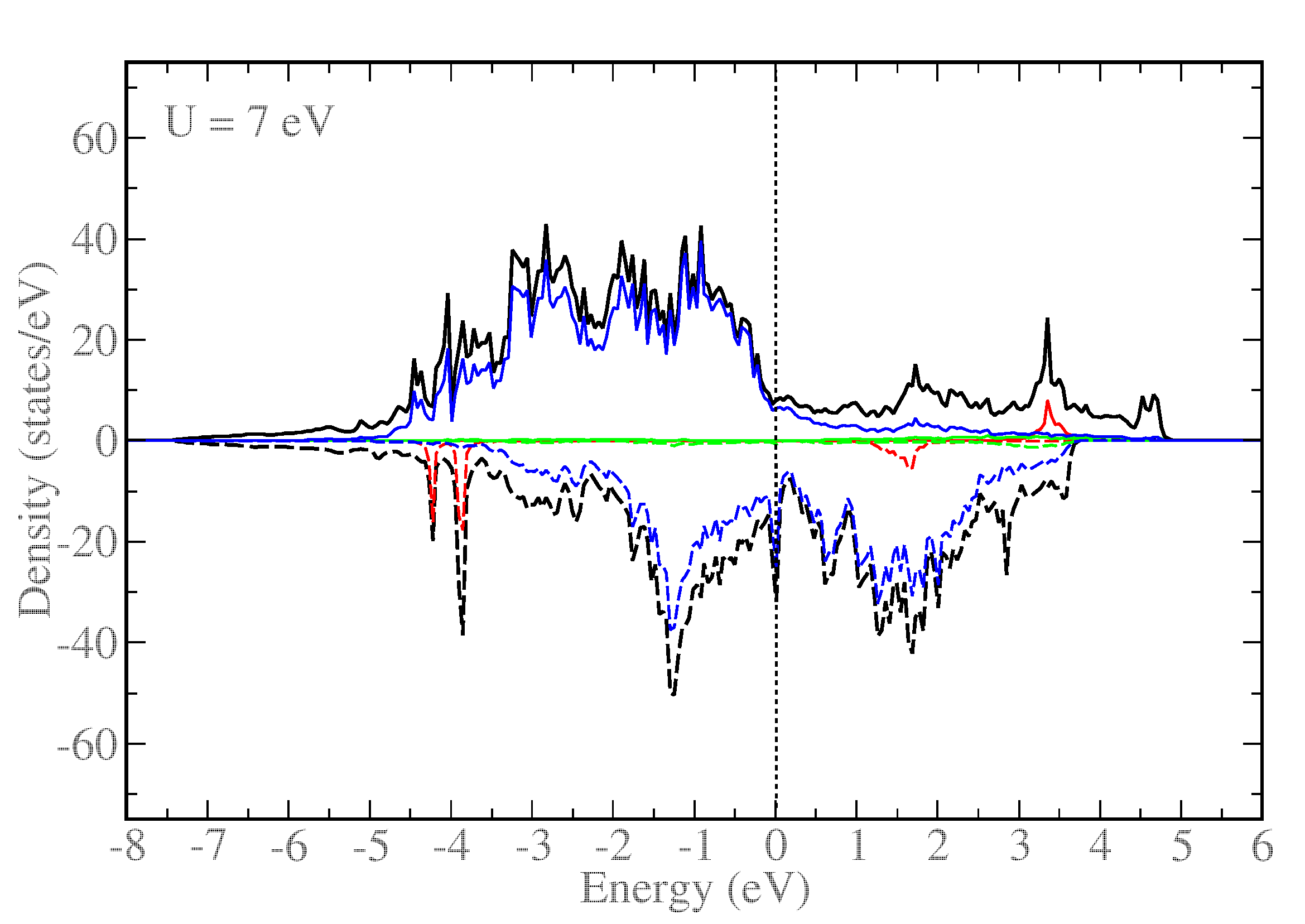}
\end{subfigure}
\begin{subfigure}{0.50\textwidth}
\includegraphics[width=0.95\textwidth]{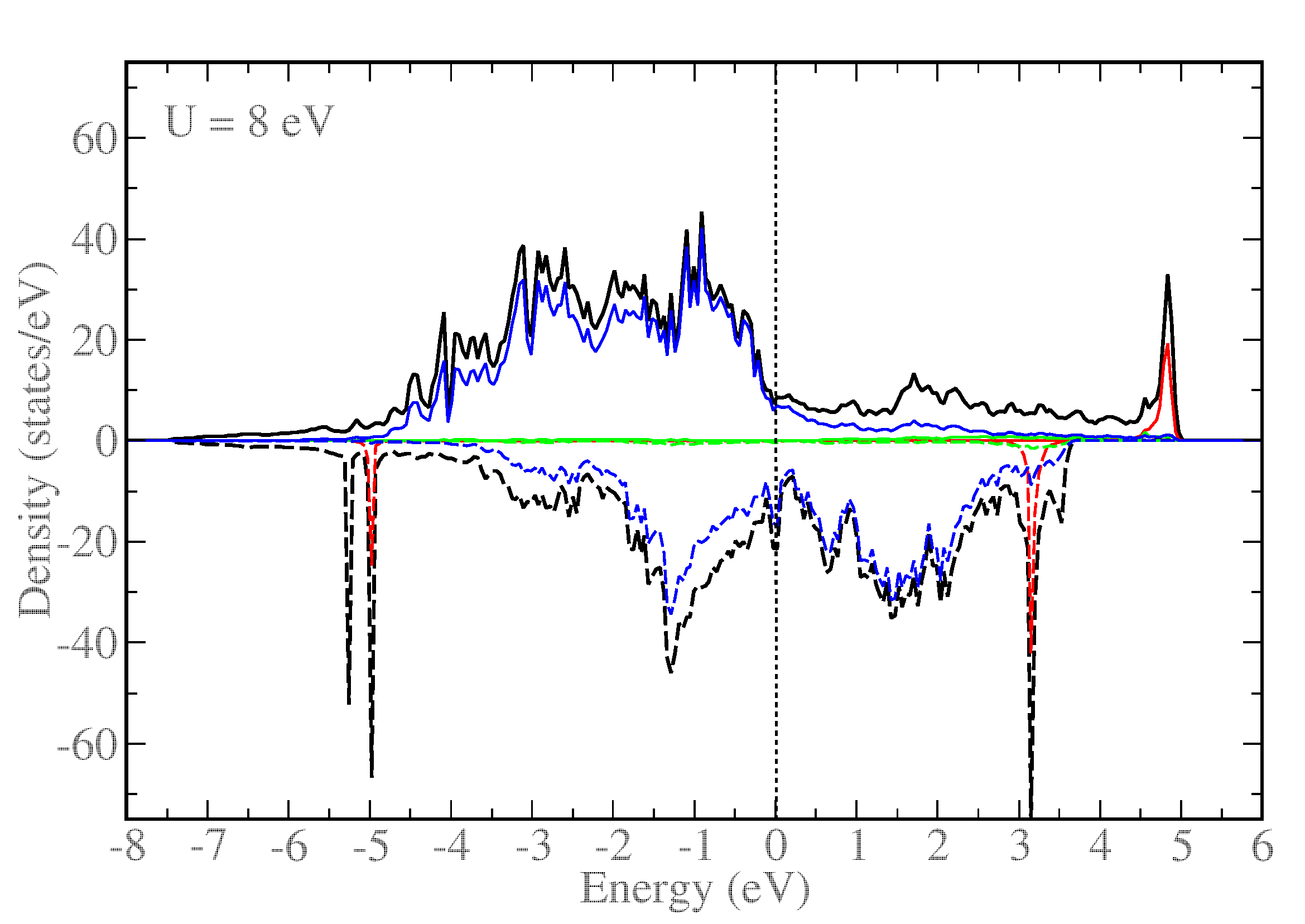}
\end{subfigure}
\caption{Calculated density of states (DOS) for the NdFe$_{11}$Ti compound. The Dudarev approach is used with the \textit{U} parameter ranging from 1 to 8 eV. Vertical dotted line represents for the Fermi level. The 4\textit{f} peaks (red lines) change according to the \textit{U} parameter and are located for small \textit{U} near the Fermi energy. \label{fig_dos_calc}}
\end{figure}

Due to the strong localization of the 4\textit{f}-electrons in Nd atoms, the DFT+\textit{U} approach is implemented in this work for all Nd containing compounds. To determine the proper value for the Hubbard \textit{U} correction, the density of states (DOS) investigations were performed for the NdFe$_{11}$Ti compound from \textit{U=}1 to 8 eV, which is given in Fig.~\ref{fig_dos_calc}.\\

As shown, the applied Hubbard \textit{U} parameter only changes the Nd-4\textit{f} states and has no considerable impact on the Fe-3\textit{d} states. It is also noticeable that small \textit{U} parameters, such as \textit{U}=1-4 are not capable to reproduce localized 4\textit{f} states, except for \textit{U}= 3 eV. The localization of 4\textit{f}-electrons only becomes visible once \textit{U}$\geq$5 eV. Once \textit{U}=5 eV, the localized 4\textit{f}-electrons peak is located $\sim$ 4 eV lower than the Fermi energy. In addition to this, increasing the \textit{U} parameter from 5 to 7 eV also does not change the location of the 4\textit{f} peak significantly. \\

The spectroscopic properties of NdFe$_{11}$Ti are reported by Lang et al. \cite{Lang1981}, and it is noted that the 4\textit{f} peak locates around 4.65 eV below the Fermi energy. In addition, the calculated magnetic moments of Nd atoms can be compared with the experimentally reported ones. In the case of \textit{U}=6 eV, Nd atoms carry -3.31 $\mu_{\text{B}}$ magnetic moment. Experimentally, it is reported to be 3.47 $\mu_{\text{B}}$ \cite{Herper2022}. In theoretical calculations, an increasing \textit{U} parameter does not modify the magnetic moments significantly, and already a promising agreement is obtained for \textit{U=}6 eV for both spectroscopic and magnetic properties. Therefore, we considered 6 eV as the Hubbard correction in case of this study. In addition, Herper \textit{et al.} \cite{Herper2022} have shown that the most accurate description of magnetic properties is achieved with a Hubbard \textit{U} value of \textit{U} between 5 and 6 eV for NdFe$_{11}$Ti. In this case, considered \textit{U=}5 eV, is also applied for binary NdFe$_{12}$ and quaternary compounds such as (Nd,Y)Fe$_{11}$Ti as well. Since all the considered compounds belong to the ThMn$_{12}$ structure, special DOS calculations for each compound have not been performed.

\section{Calculation of Lattice Constants and Cell Volume}

\begin{figure}[h]
\begin{subfigure}{0.28\textwidth}
\includegraphics[width=1.0\textwidth]{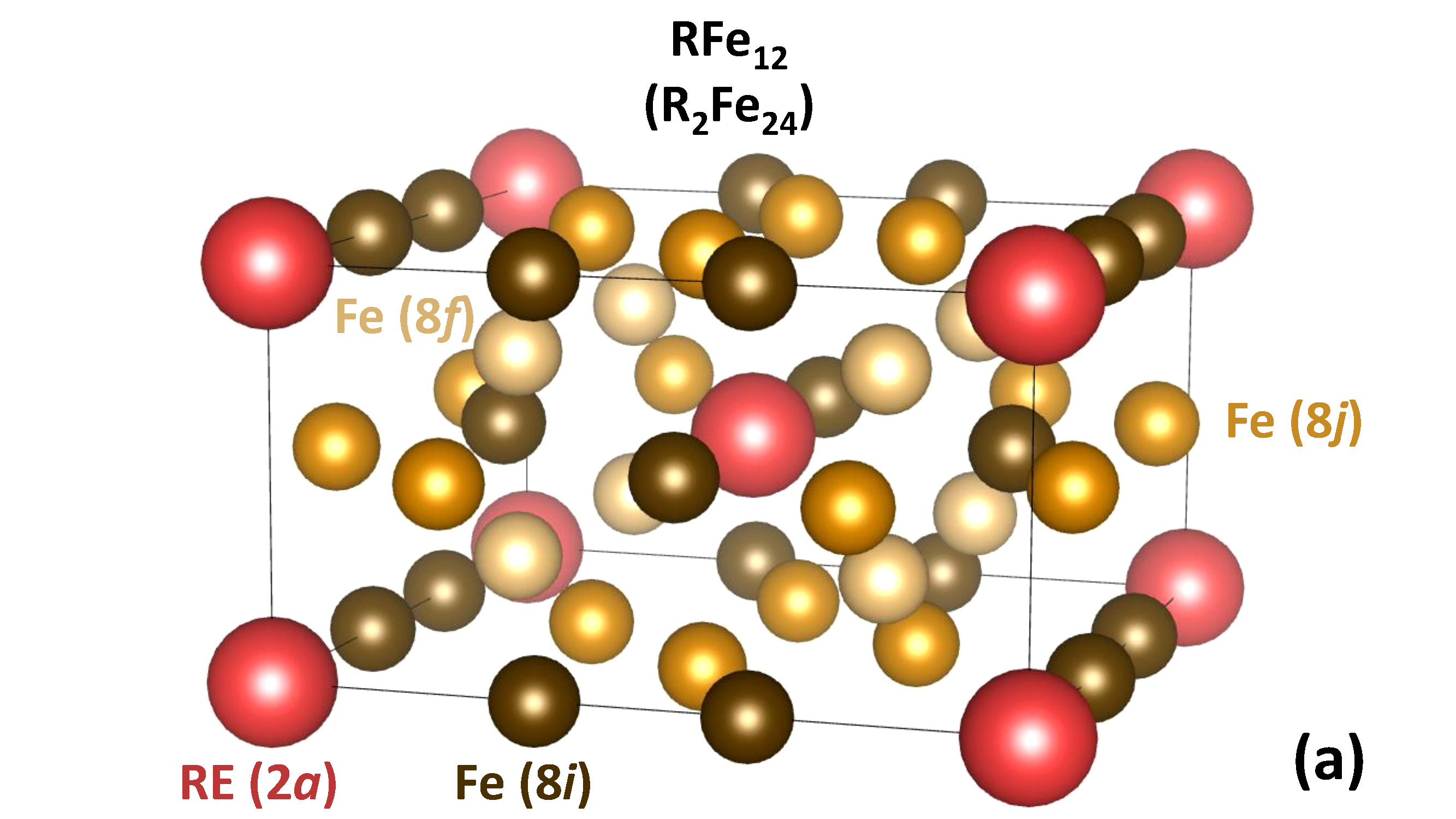}
\end{subfigure}
\begin{subfigure}{0.28\textwidth}
\includegraphics[width=1.0\textwidth]{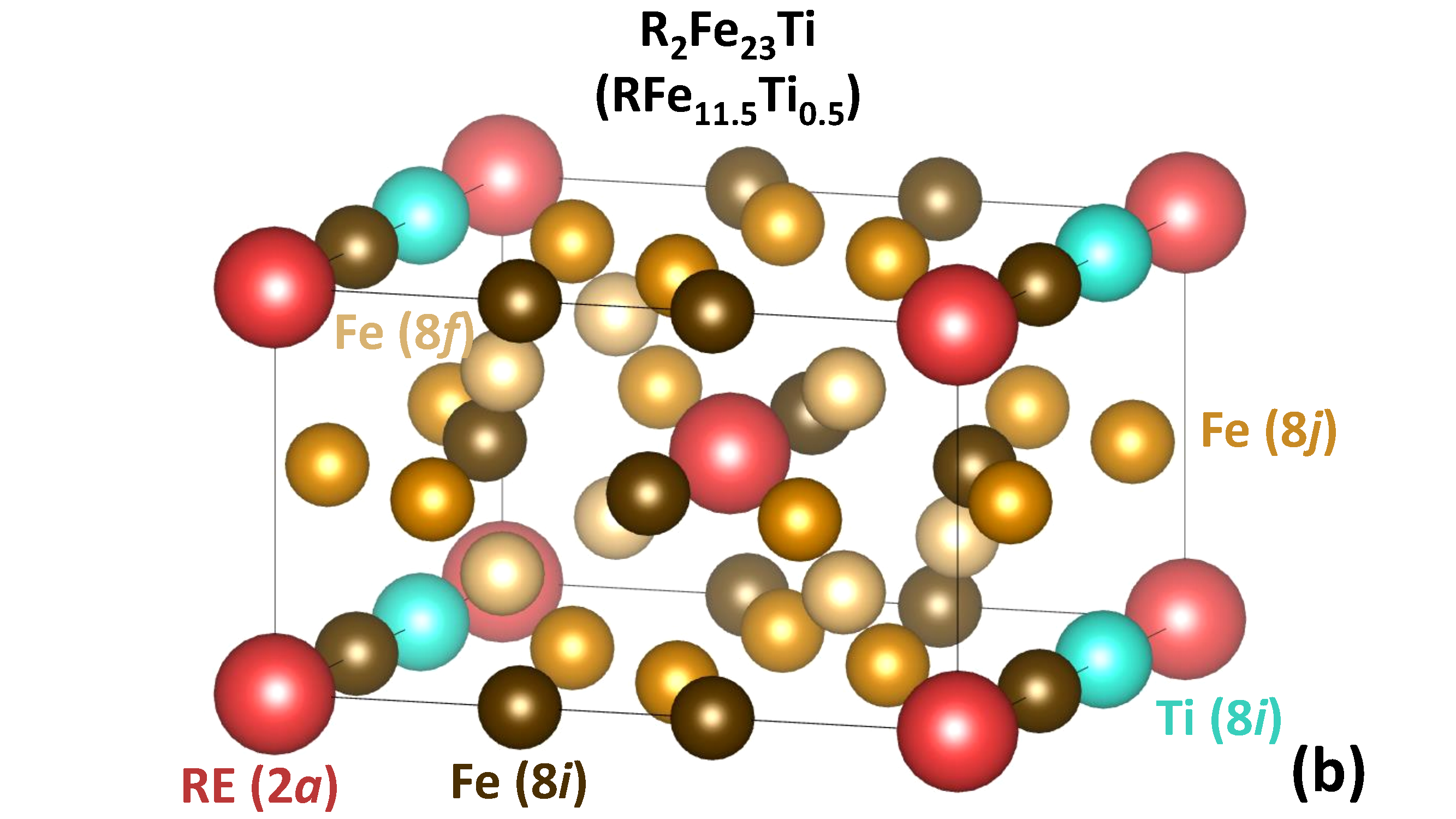}
\end{subfigure}
\begin{subfigure}{0.28\textwidth}
\includegraphics[width=1.0\textwidth]{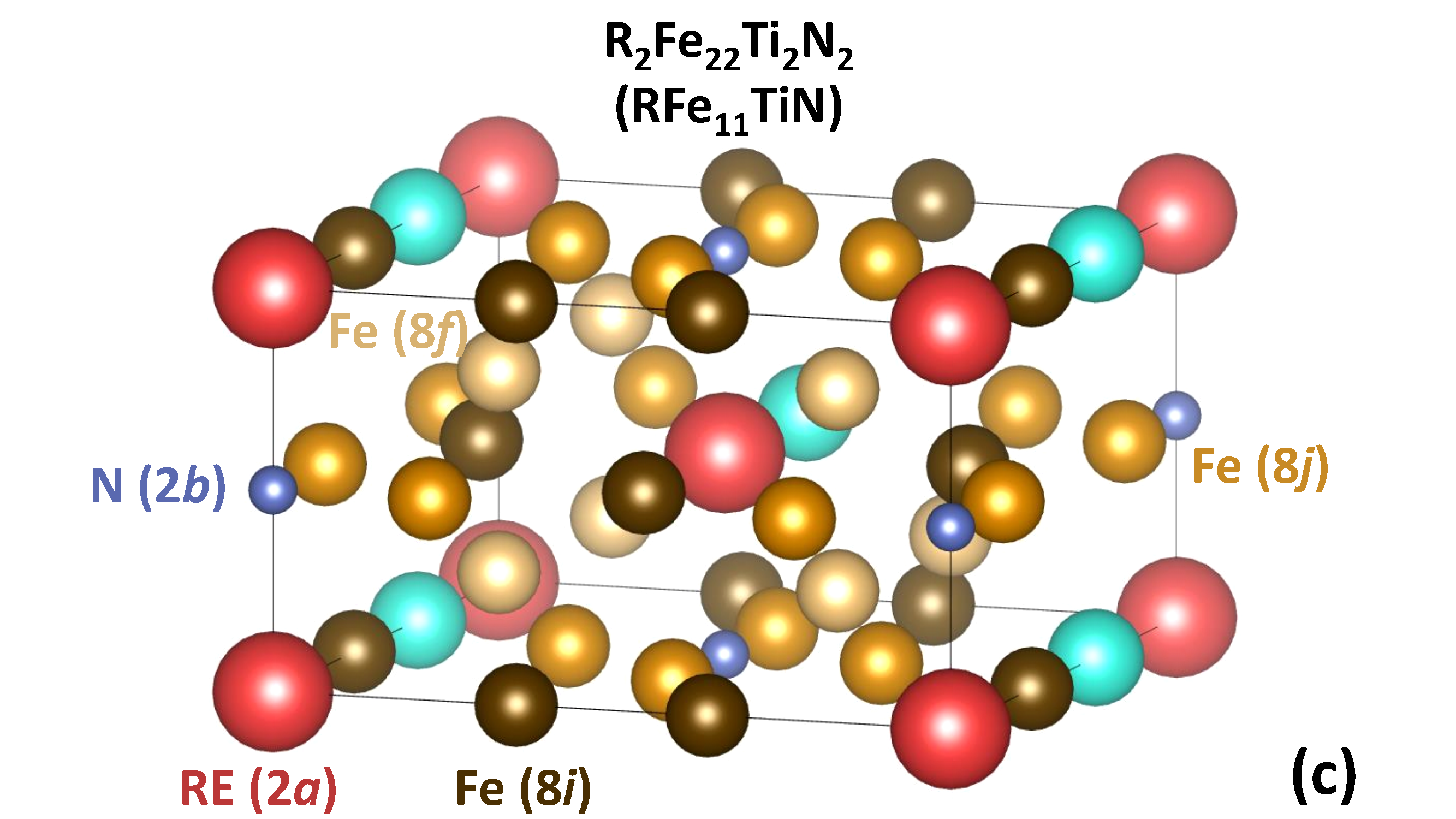}
\end{subfigure}
\begin{subfigure}{0.14\textwidth}
\includegraphics[width=1.0\textwidth]{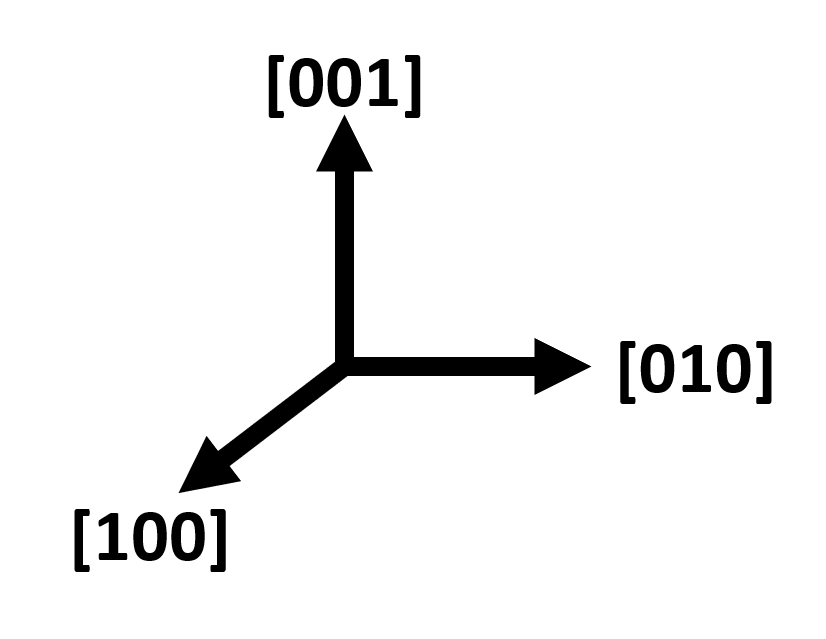}
\end{subfigure}
\caption{Schematic representation of the 2 formula unit (f.u.) (26 atoms) body-center-tetragonal ThMn$_{12}$ (I4/\textit{mmm}, space group number 139) 1:12 phases. (a) Unstable RFe$_{12}$, (b) 1 Ti atom substituted (3.8 at.\%) R$_2$Fe$_{23}$Ti (RFe$_{11.5}$Ti$_{0.5}$) and (c) 2 Ti atom substituted (7.7 at.\%) and nitrogenated supercell RFe$_{11}$TiN (R$_2$Fe$_{22}$Ti$_2$N$_2$). Ti atoms substituted in energetically favorable 8\textit{i} sites and 2 N atom considered to fully occupy interstitial 2\textit{b} site.\label{fig_unit_cells}}
\end{figure}

Besides the prototype and most well-known compound of Nd$_2$Fe$_{14}$B \cite{Drebov2013,Korner2016}, another class of typical compounds that fulfills the technological requirements for hard-magnetic applications is of the form RFe$_{12-y}$TM$_y$. The schematic representation of the RFe$_{12}$ compound is given in Fig.~\ref{fig_unit_cells}(a) which is thermodynamically not stable. However, alloyed transition metal elements, for instance, a small amount of Ti with \textit{y}$\geq$0.7, can stabilize these compounds \cite {Buschow1991,Dirba2020}. Structural parameters of CeFe$_{11}$Ti are obtained from neutron powder diffraction by Isnard \textit{et al.} \cite{Isnard1998}. Ce atoms occupy the 2\textit{a} site and Fe atoms are distributed over the three sublattices 8\textit{i}, 8\textit{j}, and 8\textit{f}, which is similar for NdFe$_{11}$Ti and YFe$_{11}$Ti as well.\\

For a theoretical study of the Ti solubility in the RFe$_{12}$ phase (R=Y, Ce and Nd), it is expected that there will be no antisites at the R(2\textit{a}) sites and that the addition of Ti will correspond to the composition RFe$_{12-y}$Ti$_y$. Starting by substituting a single Ti atom in a 26-atom supercell (2 f.u.), which corresponds to 3.8 at.\% Ti, its substitution in different sites, namely, 8\textit{i}, 8\textit{j}, and 8\textit{f} is evaluated. S\"ozen \textit{et al.} \cite{Sozen2019a} found that the lowest energy is found for Ti at the 8\textit{i} site for all the considered RFe$_{12}$ phases (see Fig.~\ref{fig_unit_cells}(b)) and it is followed by 8\textit{j} and 8\textit{f}. The solution enthalpy for the 8\textit{j} and 8\textit{f} sites are by 19 and 29 meV/atom higher for YFe$_{12-y}$Ti$_y$, respectively. This is also very similar for Ce and Nd-based 1:12 compounds, where Ti substitution for 8\textit{j} site is 20 and 18 meV/atom and for 8\textit{f} site is 29  and 27 meV/atom higher, respectively. In the case of the second Ti atom substitution, corresponding to 7.7 at.\%, all possible sites and configurations by keeping the first Ti atom at the 8\textit{i} site are checked. Again, for the second Ti atom energetical preference appeares to be the 8\textit{i} site, see Fig.~\ref{fig_unit_cells}(c). The calculated theoretical Ti preferences agree with the experimental works \cite{Isnard1998,Li2019,Tereshina2003,Akayama1994,Buschow1994} which revealed that Ti atoms are exclusively located at 8\textit{i} sites. This strong 8\textit{i} site preference of Ti atoms is mainly due to the larger Wigner-Seitz radius of the 8\textit{i} site compared to 8\textit{j} and 8\textit{f} sites \cite{Isnard1993}. The 8\textit{j} and 8\textit{f} sites have similar energies, due to their similar local coordination \cite{Wang2001}.\\

\begin{table}[h]
\centering
\scriptsize
\caption{Calculated lattice constant, cell volume and bulk modulus of the considered alloys and nitrogenated cases using DFT. In the case of Nd containing stable compounds DFT+\textit{U} has been considered only for Nd 4\textit{f}-electrons with \textit{U}=6 eV and is given in parenthesis. N is assumed to occupy all available N(2\textit{b}) sites in the 2 f.u. supercell with 2 atoms (see Fig.~\ref{fig_unit_cells} c)). For instance, Y-based 1:12 phases have the following compositions after nitrogenation: YFe$_{12}$N, Y$_2$Fe$_{23}$Ti$_2$N$_{2}$ and YFe$_{11}$TiN.
\label{Tab_lattConstANDbulkModul}}
\tiny
\begin{tabular}{@{}cccccccccc@{}}
\toprule \hline
Alloy & \multicolumn{3}{c}{\begin{tabular}[c]{@{}c@{}}Lattice constants (\AA)\\ \textit{a} \hspace{20pt} \textit{b} \hspace{20pt} \textit{c}\end{tabular}} & \begin{tabular}[c]{@{}c@{}}Cell volume\\ clean\\ (\AA$^3$)\end{tabular} & \begin{tabular}[c]{@{}c@{}}Experiment\\ cell volume\\ clean\\ (\AA$^3$)\end{tabular} & \begin{tabular}[c]{@{}c@{}}Cell volume\\ with N \\ (\AA$^3$)\end{tabular} &  \begin{tabular}[c]{@{}c@{}}Experiment\\ cell volume\\ with N\\ (\AA$^3$)\end{tabular} & \begin{tabular}[c]{@{}c@{}}Bulk modulus\\ clean\\ (GPa)\end{tabular} & \begin{tabular}[c]{@{}c@{}}Bulk modulus\\ with N\\ (GPa)\end{tabular} \\ \midrule

\begin{tabular}[c]{@{}c@{}}Y$_2$Fe$_{24}$\\ (YFe$_{12}$)\end{tabular} & 8.452 & 8.452 & 4.683 & 167.26 & 167.47* \cite{Ke2016} & 173.55 &  & 133.14 & 146.19 \\

\begin{tabular}[c]{@{}c@{}}Y$_2$Fe$_{23}$Ti\\ (YFe$_{11.5}$Ti$_{0.5}$)\end{tabular} & 8.471 & 8.457 & 4.702 & 168.43 &  & 174.78 &  & 130.61 & 145.60 \\

\begin{tabular}[c]{@{}c@{}}Y$_2$Fe$_{22}$Ti$_2$\\ (YFe$_{11}$Ti)\end{tabular} & 8.501 & 8.452 & 4.731 & 169.94 & 173.5$^a$ \cite{Yang1991} & 175.98 & 178.75$^a$ \cite{Yang1991} & 128.88 & 144.73 \\ \midrule

\begin{tabular}[c]{@{}c@{}}Ce$_2$Fe$_{24}$\\ (CeFe$_{12}$)\end{tabular} & 8.504 & 8.504 & 4.653 & 168.25 & 168.06* \cite{Ke2016} & 174.65 &  & 128.48 & 142.73 \\

\begin{tabular}[c]{@{}c@{}}Ce$_2$Fe$_{23}$Ti\\ (CeFe$_{11.5}$Ti$_{0.5}$)\end{tabular} & 8.512 & 8.522 & 4.669 & 169.33 &  & 175.73 &  & 126.68 & 143.71 \\

\begin{tabular}[c]{@{}c@{}}Ce$_2$Fe$_{22}$Ti$_2$\\ (CeFe$_{11}$Ti)\end{tabular} & 8.527 & 8.530 & 4.690 & 170.57 & 174.35$^a$ \cite{Pan1994} & 176.78 & 179.1$^a$ \cite{Pan1994} & 126.16 & 144.30 \\ \midrule

\begin{tabular}[c]{@{}c@{}}Nd$_2$Fe$_{24}$\\ (NdFe$_{12}$)\end{tabular} & 8.549 & 8.549 & 4.670 & 170.65 & 180.37* \cite{Hirayama2015} & 177.60 & 183.36* \cite{Hirayama2015} & 127.96 & 145.55 \\

\begin{tabular}[c]{@{}c@{}}Nd$_2$Fe$_{23}$Ti\\ (NdFe$_{11.5}$Ti$_{0.5}$)\end{tabular} & \begin{tabular}[c]{@{}c@{}}8.552\\ (8.561)\end{tabular} & \begin{tabular}[c]{@{}c@{}}8.570\\ (8.543)\end{tabular} & \begin{tabular}[c]{@{}c@{}}4.686\\ (4.692)\end{tabular} & \begin{tabular}[c]{@{}c@{}}171.73\\ (171.81)\end{tabular} &  & \begin{tabular}[c]{@{}c@{}}178.81\\ (178.26)\end{tabular} &  & \begin{tabular}[c]{@{}c@{}}127.17\end{tabular} & \begin{tabular}[c]{@{}c@{}}144.94\end{tabular} \\

\begin{tabular}[c]{@{}c@{}}Nd$_2$Fe$_{22}$Ti$_2$\\(NdFe$_{11}$Ti)\end{tabular} & \begin{tabular}[c]{@{}c@{}}8.561\\ (8.517)\end{tabular} & \begin{tabular}[c]{@{}c@{}}8.584\\ (8.619)\end{tabular} & \begin{tabular}[c]{@{}c@{}}4.708\\ (4.706)\end{tabular} & \begin{tabular}[c]{@{}c@{}}173.01\\ (172.92)\end{tabular} & \begin{tabular}[c]{@{}c@{}} 176.2$^a$ \cite{Piquer2004}\\ 180.35$^a$ \cite{Yang1991} \end{tabular} & \begin{tabular}[c]{@{}c@{}}179.95\\ (179.42)\end{tabular} & 183.35$^a$ \cite{Yang1991} & \begin{tabular}[c]{@{}c@{}}127.20\end{tabular} & \begin{tabular}[c]{@{}c@{}}143.48\end{tabular} \\ \midrule

\begin{tabular}[c]{@{}c@{}}NdYFe$_{24}$\end{tabular} & 8.501 & 8.501 & 4.676 & 168.97 &  & 175.64 &  & 130.88 & 145.67 \\

\begin{tabular}[c]{@{}c@{}}NdYFe$_{23}$Ti\end{tabular} & \begin{tabular}[c]{@{}c@{}}8.511\\ (8.534)\end{tabular} & \begin{tabular}[c]{@{}c@{}}8.512\\ (8.527)\end{tabular} & \begin{tabular}[c]{@{}c@{}}4.695\\ (4.719)\end{tabular} & \begin{tabular}[c]{@{}c@{}}170.06\\ (170.24)\end{tabular} &  & \begin{tabular}[c]{@{}c@{}}176.83\\ (176.47)\end{tabular} &  & \begin{tabular}[c]{@{}c@{}}128.77\end{tabular} & \begin{tabular}[c]{@{}c@{}}145.02\end{tabular} \\

\begin{tabular}[c]{@{}c@{}}NdYFe$_{22}$Ti$_2$\end{tabular} & \begin{tabular}[c]{@{}c@{}}8.528\\ (8.511)\end{tabular} & \begin{tabular}[c]{@{}c@{}}8.524\\ (8.499)\end{tabular} & \begin{tabular}[c]{@{}c@{}}4.719\\ (4.704)\end{tabular} & \begin{tabular}[c]{@{}c@{}}171.54\\ (171.75)\end{tabular} &  & \begin{tabular}[c]{@{}c@{}}178.01\\ (177.65)\end{tabular} &  & \begin{tabular}[c]{@{}c@{}}127.94\end{tabular} & \begin{tabular}[c]{@{}c@{}}143.88\end{tabular} \\ \midrule

\begin{tabular}[c]{@{}c@{}}NdCeFe$_{24}$\end{tabular} & 8.525 & 8.525 & 4.662 & 169.44 &  & 176.25 &  & 129.11 & 142.92 \\

\begin{tabular}[c]{@{}c@{}}NdCeFe$_{23}$Ti\end{tabular} & \begin{tabular}[c]{@{}c@{}}8.538\\ (8.542)\end{tabular} & \begin{tabular}[c]{@{}c@{}}8.540\\ (8.534)\end{tabular} & \begin{tabular}[c]{@{}c@{}}4.678\\ (4.679)\end{tabular} & \begin{tabular}[c]{@{}c@{}}170.55\\ (170.66)\end{tabular} &  & \begin{tabular}[c]{@{}c@{}}177.37\\ (176.90)\end{tabular} &  & \begin{tabular}[c]{@{}c@{}}127.40\end{tabular} & \begin{tabular}[c]{@{}c@{}}143.23\end{tabular} \\

\begin{tabular}[c]{@{}c@{}}NdCeFe$_{22}$Ti$_2$\end{tabular} & \begin{tabular}[c]{@{}c@{}}8.544\\ (8.535)\end{tabular} & \begin{tabular}[c]{@{}c@{}}8.558\\ (8.559)\end{tabular} & \begin{tabular}[c]{@{}c@{}}4.700\\ (4.695)\end{tabular} & \begin{tabular}[c]{@{}c@{}}171.83\\ (171.59)\end{tabular} &  & \begin{tabular}[c]{@{}c@{}}178.41\\ (177.85)\end{tabular} &  & \begin{tabular}[c]{@{}c@{}}127.31\end{tabular} & \begin{tabular}[c]{@{}c@{}}143.02\end{tabular} \\\hline \bottomrule
\end{tabular}
\begin{tablenotes}
   \item Theoretical references are represented by *.
   \item $^a$ X-ray diffraction (XRD) analysis at 300 K.
\end{tablenotes}
\end{table}

\begin{figure}
\centering
\includegraphics[width=0.75\textwidth]{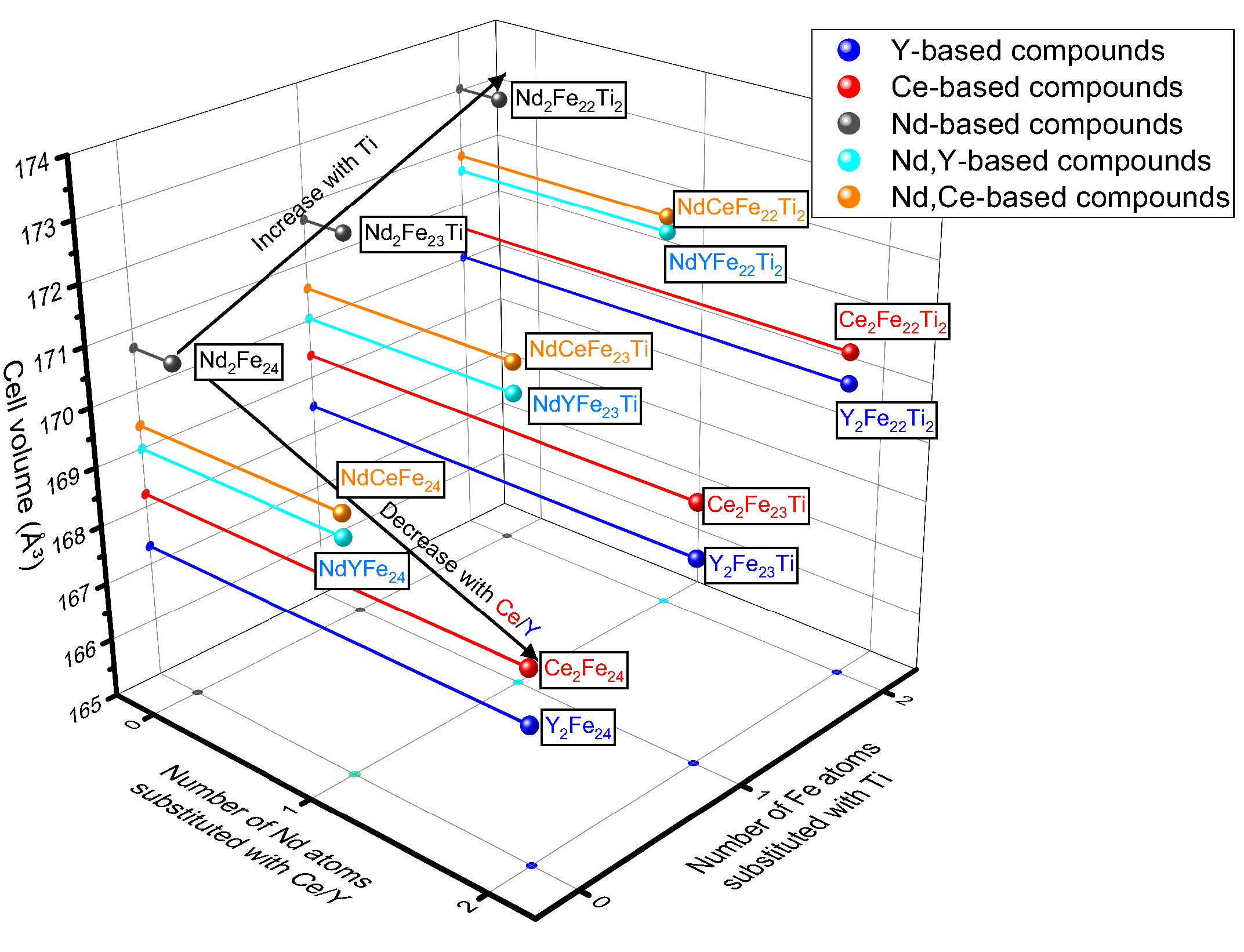}
\caption{Schematic representation of the cell volume trends for two different substitutions, which are Ti substitution on Fe sublattice and Ce and Y substitutions in Nd sublattice, via GGA calculations. Purely Nd-containing compounds are depicted in black while compounds only including Y and Ce are shown as blue and red, respectively. The quarternary compounds containing Nd,Y and Nd,Ce are illustrated as cyan and orange, respectively. The resulting trend for each substitution is further clarified by arrows.\label{fig_cell_volume}}
\end{figure}

Knowing the structure of Ti contained ternary 1:12 phases, one can calculate physical properties such as lattice constants and bulk moduli as given in Tab.~\ref{Tab_lattConstANDbulkModul}.
It has been captured an increasing lattice constant and cell volume trend as Nd$_2$Fe$_{24-y}$Ti$_y$ > Ce$_2$Fe$_{24-y}$Ti$_y$ > Y$_2$Fe$_{24-y}$Ti$_y$ for 0$\leq y\leq$2 in agreement with experimental data \cite{Suzuki2014,Isnard1998,Nikitin1998,Akayama1994}. As expected, Ce doped quaternary compounds, NdCeFe$_{24-y}$Ti$_y$, have a larger volume than Y doped cases, NdYFe$_{24-y}$Ti$_y$. The influence of each substitution can be seen in Fig.~\ref{fig_cell_volume}, where the resulting trends are shown for GGA calculated values.
Note that Y and Ce substitution has been considered only for the RE(2\textit{a}) site. The effect of nitrogen has been investigated with 2 N atoms in a 2 f.u. 28 atom supercell, see Fig.~\ref{fig_unit_cells}(c). N atoms are doped into the midpoint between RE atoms along the \textit{c} axis. The interstitial N expands the volume by $\sim$3.7 \% for Y$_2$Fe$_{24-y}$Ti$_y$ and Ce$_2$Fe$_{24-y}$Ti$_y$ and $\sim$4.1 \% for the rest of the compounds, which agrees with experimental data \cite{Akayama1994,Yang1991b,Pan1994,Suzuki2014}. As mentioned above, DFT+\textit{U} treatment is performed for Nd-contained compounds to include the influence of the 4-\textit{f}-electrons (calculated values are reported in parenthesis in  Tab.~\ref{Tab_lattConstANDbulkModul}), and a similar volume trend compared to the in-core treatment in DFT is observed as well. In addition, it has been captured that increasing Ti concentration in all considered compounds causes a reduction in bulk modulus. The softening effect of Ti addition is also the case for most of the nitrogenated compounds, except Ce$_2$Fe$_{24-y}$Ti$_y$N$_2$. \\

\newpage

\section{Total Magnetic Moment and Magnetization}

\label{Sec_Results_totalMagneticMoments}

As a first intrinsic magnetic property, the magnetization of the considered compounds has been calculated. It has to be distinguished between the total magnetic moment $m_{tot}$ and saturation magnetization $M_{S}$. The former one is calculated via the summation of each atom's magnetic moment and is expressed in units of the Bohr magneton $\mu_{\text{B}}$ per formula unit, while the latter one is in units of Tesla (T) which is estimated from the density of the compound $\rho$, the Avogadro constant N$_A$ and the molecular weight $M_{\text{alloy}}$ via

\begin{equation}
\label{equation_MS}
\mu_{0}M_{S} = m_{tot}  \mu_{B} \frac{\rho  \text{N$_A$}}{M_{\text{alloy}}}.
\end{equation}

Theoretical total magnetic moments and saturation magnetization with a comparison of available experimental data are given in Tab.~\ref{Tab_magnetization}. Knowing that the considered PBE functional tends to overestimate the magnetic moments \cite{Vishina2021}, the calculated values are within the expected range. It is important to note that having accurately relaxed structures (check Tab.~\ref{Tab_lattConstANDbulkModul}) does not guarantee correct reproduction of magnetic properties. For instance, in the case of YFe$_{11}$Ti an experimental magnetic moment is reported as 18.4 $\mu_{\text{B}}/$f.u. \cite{Suzuki2017}, whereas the calculated value in this work is 21.96 $\mu_{\text{B}}/$f.u meaning $\sim$19 \% overestimation. Similar deviation is found for CeFe$_{11}$Ti with $\sim$21\% overestimation. Experimentally, the Ce-based 1:12 phase has a lower $m_{tot}$ than the Y-based 1:12 phase. This is due to the magnetic moments carried by RE elements. In both cases, Fe and Ti atoms have rather similar magnetic moments, but Ce has -0.76 $\mu_{\text{B}}$ where Y carries -0.31 $\mu_{\text{B}}$ magnetic moment.\\

In the case of Nd$_2$Fe$_{22-y}$Ti$_y$ compounds, DFT calculations with \textit{f}-electrons in core treatment yield rather good $m_{tot}$ values. As an example, for NdFe$_{11}$Ti it is calculated to be 21.84 $\mu_{\text{B}}$, and an experimental value is measured as 21.27 $\mu_{\text{B}}$ \cite{Yang1991}. However, in this treatment 4\textit{f}-electrons are assumed as frozen in core. Treating the Nd 4\textit{f}-electrons as fully localized with DFT+\textit{U} calculations yields 18.92 $\mu_{\text{B}}$ total magnetic moment and 1.28 T saturation magnetization, which is experimentally measured as 1.38 T. With the DFT+\textit{U} approximation a reduced magnetization is seen because the contribution from the Nd ion is about -3.31 $\mu_{\text{B}}$ (mainly from \textit{f}-electrons) and it is -0.28 $\mu_{\text{B}}$ for simple DFT due to the missing \textit{f} orbital. Nevertheless, for both treatments, acceptable values compared to experimental data are achieved. \\

Y and Ce substitution to the Nd-based 1:12 phase decreases the total magnetic moments for DFT treatment because both Y (-0.31 $\mu_{\text{B}}$) and Ce (-0.76 $\mu_{\text{B}}$) have lower magnetic moments than the Nd (-0.28 $\mu_{\text{B}}$) ion. However, the situation is reversed in the case of DFT+\textit{U} treatment, since the inclusion of localized \textit{f}-electrons yields more negative magnetic moments for the Nd ion by -3.31 $\mu_{\text{B}}$. The calculated magnetic moments of Nd atoms are not influenced considerably, once the \textit{f}-electrons are localized with Hubbard \textit{U}$\geq$ 5 eV. Therefore, having  larger \textit{U} values is not changing the calculated magnetic moments of the RE site. For instance, in case of NdFe$_{11}$Ti a value of \textit{U}=5 eV results in a total magnetic moment of 18.93 $\mu_{\text{B}}/$f.u., while a value of \textit{U}=8 eV leads to 18.92 $\mu_{\text{B}}/$f.u. In addition, for each considered compound increasing Ti concentration reduces the total magnetization due to the antiferromagnetic alignment of Ti atoms against Fe atoms with an average of -1.1 $\mu_{\text{B}}$.\\

\begin{table*}[h]
\caption{Calculated magnetic moments at rare-earth site $m^{2a}_{tot}$, total magnetic moments $m_{tot}$ ($\mu_{\text{B}}/f.u$) and saturation magnetization $\mu_0M_S$ (T) with comparison against available experimental data. In the case of Nd contained compounds DFT+\textit{U} with \textit{U=} 6 eV is applied to Nd 4\textit{f}-electrons only and results are given in parenthesis. 2 N atoms have been considered in the N(2\textit{b}) site in the 2 f.u. supercell for the nitrogenated cases.  \label{Tab_magnetization}}
\tiny
\begin{tabular}{@{}cccccccccc@{}}
\toprule \hline

Alloy & \begin{tabular}[c]{@{}l@{}}$m_{tot}^{2a}$\\ clean\\ ($\mu_{\text{B}}/f.u.$)\end{tabular} & \begin{tabular}[c]{@{}l@{}}$m_{tot}$\\ clean\\ ($\mu_{\text{B}}/f.u.$)\end{tabular} & \begin{tabular}[c]{@{}l@{}} $m_{tot}$\\ experiment \\ clean \\($\mu_{\text{B}}/f.u.$)\end{tabular} & \begin{tabular}[c]{@{}l@{}}$\mu_0 M_S$\\ clean\\ (T)\end{tabular} & \begin{tabular}[c]{@{}l@{}} $\mu_0 M_S$\\experiment\\ clean \\ (T)\end{tabular} & \begin{tabular}[c]{@{}l@{}}$m_{tot}$\\ with N\\ ($\mu_{\text{B}}/f.u.$)\end{tabular} & \begin{tabular}[c]{@{}l@{}} $m_{tot}$\\experiment\\ with N\\ ($\mu_{\text{B}}/f.u.$)\end{tabular} & \begin{tabular}[c]{@{}l@{}}$\mu_0 M_S$\\ with N\\ (T)\end{tabular} &  \begin{tabular}[c]{@{}l@{}} $\mu_0 M_S$\\experiment\\ with N\\ (T)\end{tabular} \\ \midrule

Y$_2$Fe$_{24}$ & -0.30 & 26.23 & 24.30$^a$ \cite{Suzuki2017} & 1.83 & 1.66$^a$ \cite{Suzuki2017} & 28.21 & & 1.89 &  \\

Y$_2$Fe$_{23}$Ti & -0.31 & 23.95 & & 1.66 & & 25.83 & & 1.72 &  \\

Y$_2$Fe$_{22}$Ti$_2$ & -0.31 & 21.96 & \begin{tabular}[c]{@{}l@{}}18.3$^b$ \cite{Obbade1997} \\ 18.4$^a$ \cite{Suzuki2017} \\ 18.57$^c$ \cite{Yang1991} \\ 19.0$^f$ \cite{Qi1992} \\ 19.97$^e$ \cite{Herper2022} \\ 20.3$^c$ \cite{Mao1998} \end{tabular} & 1.50 & \begin{tabular}[c]{@{}l@{}} 1.24$^b$ \cite{Obbade1997} \\ 1.26$^a$ \cite{Suzuki2017} \\ 1.26$^c$ \cite{Yang1991} \\ 1.29$^f$ \cite{Qi1992} \\  1.36$^e$ \cite{Herper2022} \\ 1.38$^c$ \cite{Mao1998}  \end{tabular} & 23.38 & 21.75$^c$ \cite{Yang1991} & 1.55 & 1.42$^c$ \cite{Yang1991} \\ \midrule

Ce$_2$Fe$_{24}$ & -0.79 & 25.75 & 26.55$^{*}$ \cite{Snarski-Adamski2022} & 1.78 & 1.89$^{*}$ \cite{Snarski-Adamski2022} & 27.73 & 29.66$^{*}$ \cite{Korner2016} & 1.85 & 1.98$^{*}$ \cite{Korner2016} \\

Ce$_2$Fe$_{23}$Ti & -0.75 & 23.35 & 24.68$^{*}$ \cite{Snarski-Adamski2022} & 1.61 & 1.70$^{*}$ \cite{Snarski-Adamski2022} & 25.24 & & 1.67 &  \\

Ce$_2$Fe$_{22}$Ti$_2$ & -0.76 & 21.15 & \begin{tabular}[c]{@{}l@{}}17.4$^d$ \cite{Isnard1998} \\ 18.62$^c$ \cite{Pan1994} \\ 22.25$^f$ \cite{Akayama1994} \end{tabular} & 1.45 & \begin{tabular}[c]{@{}l@{}}1.19$^d$ \cite{Isnard1998} \\ 1.27$^c$ \cite{Pan1994} \\ 1.55$^f$ \cite{Akayama1994} \end{tabular} & 22.69 & \begin{tabular}[c]{@{}l@{}}21.69$^c$ \cite{Pan1994} \\ 25.19$^{*}$ \cite{Korner2016} \end{tabular} & 1.50 & \begin{tabular}[c]{@{}l@{}} 1.43$^c$ \cite{Pan1994} \\ 1.66$^{*}$ \cite{Korner2016} \end{tabular} \\ \midrule

Nd$_2$Fe$_{24}$ & -0.27 & 26.43 & \begin{tabular}[c]{@{}l@{}}27.2$^{*}$ \cite{Fukazawa2022} \\ 29.15$^{*}$ \cite{Korner2016} \\ 31.20$^{*}$ \cite{Hirayama2015} \end{tabular} & 1.80 & \begin{tabular}[c]{@{}l@{}}1.73$^{*}$ \cite{Fukazawa2022} \\ 1.99$^{*}$ \cite{Korner2016} \\ 2.01$^{*}$ \cite{Hirayama2015} \end{tabular} & 28.73 & \begin{tabular}[c]{@{}l@{}}31.40$^{*}$ \cite{Korner2016} \\ 32.40$^{*}$ \cite{Hirayama2015} \end{tabular} & 1.88 & \begin{tabular}[c]{@{}l@{}} 2.06$^{*}$ \cite{Korner2016} \\ 2.06$^{*}$ \cite{Hirayama2015} \end{tabular} \\

Nd$_2$Fe$_{23}$Ti & \begin{tabular}[c]{@{}l@{}}-0.28\\ (-3.30)\end{tabular} & \begin{tabular}[c]{@{}l@{}}24.04\\ (21.35)\end{tabular} & &  \begin{tabular}[c]{@{}l@{}}1.63\\ (1.45)\end{tabular} & & \begin{tabular}[c]{@{}l@{}}26.35\\ (23.14)\end{tabular} & & \begin{tabular}[c]{@{}l@{}}1.72\\ (1.51)\end{tabular} &  \\

Nd$_2$Fe$_{22}$Ti$_2$ & \begin{tabular}[c]{@{}l@{}}-0.28\\ (-3.31)\end{tabular} & \begin{tabular}[c]{@{}l@{}}21.84\\ (18.92)\end{tabular} &  \begin{tabular}[c]{@{}l@{}}21.27$^c$ \cite{Yang1991} \\ 21.90$^b$ \cite{Piquer2004} \\ 23.43$^e$ \cite{Herper2022} \\ 24.10$^{*}$ \cite{Harashima2015} \\ 24.50$^{*}$ \cite{Korner2016} \\ 25.24$^f$ \cite{Akayama1994} \\ 26.30$^{*}$ \cite{Hirayama2015} \end{tabular} & \begin{tabular}[c]{@{}l@{}}1.47\\ (1.28)\end{tabular} & \begin{tabular}[c]{@{}l@{}}1.38$^c$ \cite{Yang1991} \\ 1.48$^b$ \cite{Piquer2004} \\ 1.58$^e$ \cite{Herper2022} \\ 1.63$^{*}$ \cite{Harashima2015} \\ 1.65$^{*}$ \cite{Korner2016} \\ 1.70$^f$ \cite{Akayama1994} \\ 1.70$^{*}$ \cite{Hirayama2015}  \end{tabular} & \begin{tabular}[c]{@{}l@{}}23.87\\ (20.86)\end{tabular} & \begin{tabular}[c]{@{}l@{}}23.22$^c$ \cite{Yang1991} \\ 26.84$^{*}$ \cite{Harashima2015} \\ 26.86$^{*}$ \cite{Korner2016} \\ 27.2$^{*}$ \cite{Hirayama2015} \end{tabular} & \begin{tabular}[c]{@{}l@{}}1.55\\ (1.35)\end{tabular} & \begin{tabular}[c]{@{}l@{}}1.56$^c$ \cite{Yang1991} \\ 1.74$^{*}$ \cite{Harashima2015} \\ 1.74$^{*}$ \cite{Korner2016} \\ 1.73$^{*}$ \cite{Hirayama2015} \end{tabular} \\ \midrule

NdYFe$_{24}$ & -0.27 (Nd), -0.30 (Y) & 26.33 & & 1.82 & & 28.49 & & 1.89 &  \\

NdYFe$_{23}$Ti & \begin{tabular}[c]{@{}l@{}}-0.28 (Nd), -0.30 (Y)\\ (-3.31 (Nd), -0.30 (Y))\end{tabular} & \begin{tabular}[c]{@{}l@{}}23.99\\ (22.78)\end{tabular} & & \begin{tabular}[c]{@{}l@{}}1.64\\ (1.56)\end{tabular} & & \begin{tabular}[c]{@{}l@{}}26.11\\ (24.64)\end{tabular} & & \begin{tabular}[c]{@{}l@{}}1.72\\ (1.63)\end{tabular} &  \\

NdYFe$_{22}$Ti$_2$ & \begin{tabular}[c]{@{}l@{}}-0.28 (Nd), -0.31 (Y)\\ (-3.53 (Nd), -0.31 (Y))\end{tabular} & \begin{tabular}[c]{@{}l@{}}21.91\\ (20.47)\end{tabular} & & \begin{tabular}[c]{@{}l@{}}1.49\\ (1.39)\end{tabular} & & \begin{tabular}[c]{@{}l@{}}23.66\\ (22.10)\end{tabular} & & \begin{tabular}[c]{@{}l@{}}1.55\\ (1.45)\end{tabular} & \\ \midrule

NdCeFe$_{24}$ & -0.27 (Nd), -0.76 (Ce) & 26.09 & & 1.80 & & 28.31 & & 1.87 &  \\

NdCeFe$_{23}$Ti & \begin{tabular}[c]{@{}l@{}}-0.28 (Nd), -0.74 (Ce)\\ (-3.31 (Nd), -0.75 (Ce))\end{tabular} & \begin{tabular}[c]{@{}l@{}}23.74\\ (22.39)\end{tabular} & & \begin{tabular}[c]{@{}l@{}}1.62\\ (1.53)\end{tabular} & & \begin{tabular}[c]{@{}l@{}}25.86\\ (24.24)\end{tabular} & & \begin{tabular}[c]{@{}l@{}}1.70\\ (1.60)\end{tabular} &   \\

NdCeFe$_{22}$Ti$_2$ & \begin{tabular}[c]{@{}l@{}}-0.28 (Nd), -0.77 (Ce)\\ (-3.31 (Nd), -0.77 (Ce))\end{tabular} & \begin{tabular}[c]{@{}l@{}}21.52\\ (19.93)\end{tabular} & & \begin{tabular}[c]{@{}l@{}}1.46\\ (1.35)\end{tabular} & & \begin{tabular}[c]{@{}l@{}}23.28\\ (21.57)\end{tabular} & & \begin{tabular}[c]{@{}l@{}}1.52\\ (1.41)\end{tabular} &  \\ \hline \bottomrule
\end{tabular}
\begin{tablenotes}
   \item Theoretical references are represented by *
        \item $^a$M\"ossbauer measurement at 77 K,
                $^b$M\"ossbauer measurement at 4.2 K
                $^c$Extraction sample magnetometer (ESM) analysis at 1.5 K,
                $^d$Extraction sample magnetometer (ESM) analysis at 5 K,
                $^e$Single crystal, physical property measurement system (PPMS) measured at 10 K,
                $^f$Vibrating sample magnetometer (VSM) analysis at 4.2 K.
\end{tablenotes}
\end{table*}

The effect of interstitial nitrogenation on the magnetic properties, such as magnetization, Curie temperature and magnetocrystalline anisotropy energy of REFe$_{12-x}$Ti$_x$ (RE: Y, Ce, Nd and Sm) compounds has been studied experimentally \cite{Yang1991,Pan1994}, and a dramatic impact is commonly reported. 

In the case of magnetization, an improvement of $\sim$8\% for the magnetic moments of all the considered compounds is calculated and given in Tab.~\ref{Tab_magnetization}. For the Y-based 1:12 phase, YFe$_{11}$TiN, the magnetic moment is increased by 1.42 $\mu_{\text{B}}$ compared to the N-free compound. This is followed by 1.54 and 2.03 $\mu_{\text{B}}$ increments for CeFe$_{11}$TiN and NdFe$_{11}$TiN, respectively. Experimentally, the impact of nitrogenation on YFe$_{11}$Ti magnetisation is reported to be a 3.2 $\mu_{\text{B}}$ increase by Yang \textit{et al.} \cite{Yang1991}. A similar improvement of 3.1 $\mu_{\text{B}}/f.u.$ is measured by Pan \textit{et al.} \cite{Pan1994} for CeFe$_{11}$Ti and 2.0 $\mu_{\text{B}}/f.u.$ increase for NdFe$_{11}$Ti is reported by Yang \textit{et al.} \cite{Yang1991}. Even though, the experimental improvement of $\sim$17\% for YFe$_{11}$Ti and CeFe$_{11}$Ti is higher than our theoretical improvement, the increase due to nitrogenation is properly shown and still in good agreement. For the experimental improvement of NdFe$_{11}$Ti of $\sim$9\%, a very good agreement with the calculated value in this work can be seen.\\

The effect of nitrogenation on Y and Ce substituted quaternaries are rather similar. According to the DFT (DFT+\textit{U}) calculations, the magnetic moment increase for both (Nd,Y)Fe$_{11}$TiN and (Nd,Ce)Fe$_{11}$TiN is $\sim$1.75 (1.63) $\mu_{\text{B}}$. It is also noticeable that increasing Ti concentration reduces the impact of N. For instance, the N effect on Ti-free YFe$_{12}$N is 1.98 $\mu_{\text{B}}$, once a Ti atom is added (Y$_2$Fe$_{23}$TiN$_{2}$), the impact reduces to 1.88 $\mu_{\text{B}}$, and it is further reduced to 1.42 $\mu_{\text{B}}$ once Ti concentration increases (YFe$_{11}$TiN). 

\begin{figure}[h!]
\centering
\includegraphics[width=0.9\textwidth]{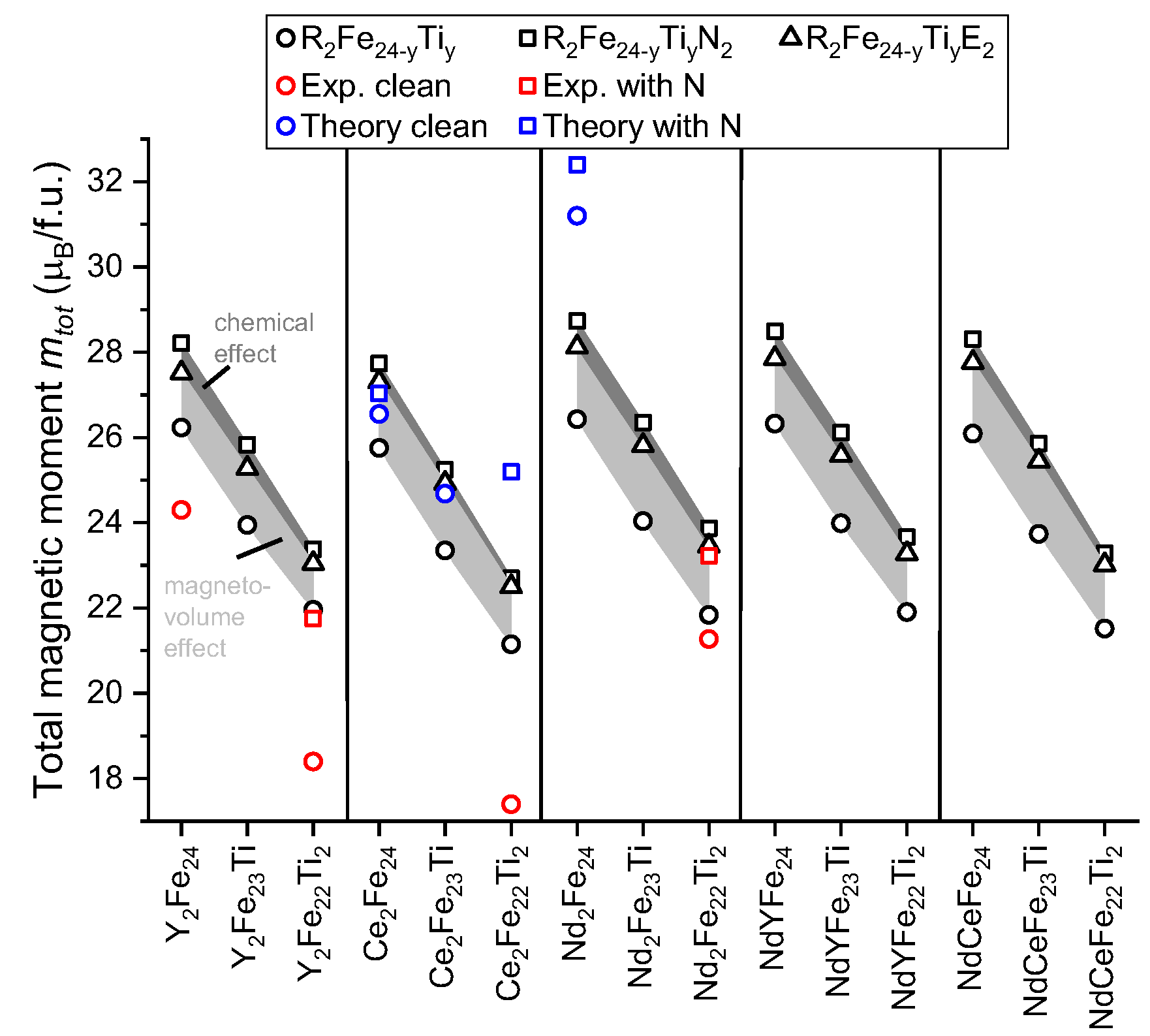}
\caption{Total magnetic moments of the clean (RFe$_{24-y}$Ti$_y$), N doped (RFe$_{24-y}$Ti$_y$N$_2$) and N removed but volume kept fixed, and atoms are allowed to relax cases (RFe$_{24-y}$Ti$_y$E$_2$). 
Since DFT+\textit{U} calculations do not change the results significantly, only DFT results are given. The light(dark) grey background coloring shows the magnetovolume (chemical) effect. The blue circles show theoretical data \cite{Snarski-Adamski2022,Hirayama2015} for the clean alloys from others, the red circles show experimental data \cite{Suzuki2017,Mao1998,Akayama1994,Yang1991b} for the clean alloys, the blue squares show theoretical nitrogenated cases \cite{Snarski-Adamski2022,Korner2016,Hirayama2015}, and the red squares show the experimental nitrogenated cases \cite{Yang1991b}.
\label{Fig_N_effect}}
\end{figure}

As given in Tab.~\ref{Tab_lattConstANDbulkModul}, nitrogenation expands the volume of all the considered compounds, and as mentioned above increases the magnetic moments. In order to understand the underlying reason, if it is based on the magnetovolume effect or a purely chemical effect, we have performed a set of simulations as given in Fig.~\ref{Fig_N_effect}. Already calculated N-free optimized R$_2$Fe$_{24-y}$Ti$_y$ compounds are given in black circles, and optimized and nitrogenated R$_2$Fe$_{24-y}$Ti$_y$N$_2$ are given in black squares. Then, the latter optimized structure is kept and the N atoms are removed (R$_2$Fe$_{24-y}$Ti$_y$E$_2$). In this set of calculations, expanded volumes are considered but N is removed and further cell relaxation in self-consistent calculations is restricted, given as black triangles. The change in the magnetic moment from the circles to the triangles corresponds to the magnetovolume effect which is associated with the expanded volume of R$_2$Fe$_{24-y}$Ti$_y$E$_2$ compared to R$_2$Fe$_{24-y}$Ti$_y$ by $\sim$4\%. This is also colored light grey. On the other hand, the change of magnetic moments from triangles to squares corresponds to the chemical effect which is colored by dark grey. As shown for the considered compounds, the magnetovolume effect has a major role in the improvement of magnetic moments by nitrogenation. \\

\section{Maximum Energy Product}
One of the performance measures of permanent magnets is the maximum energy product |BH|$_{max}$. It is an important figure-of-merit for the strength of permanent magnetic materials, and a magnet must be shaped for the most efficient use in such a way that its operating point is close to the |BH|$_{max}$ point. It is the absolute upper limit of magnetostatic energy stored in free space by a permanent magnet of unit volume \cite{Skomski2008}. By calculating the saturation magnetization $M_S$, one can calculate the |BH|$_{max}$ for an ideal square hysteresis loop as given by Ref. \cite{Coey2011}

\begin{equation}
\label{eq_bhmax}
|BH|_{max} = \frac{\mu_{0} M_{S}^{2}}{4},
\end{equation}
where $\mu_0$ is the vacuum permeability ($\mu_0=4\pi\times10^{-7}$NA$^{-2}$). 

\begin{figure}[h]
\includegraphics[width=1.0\textwidth]{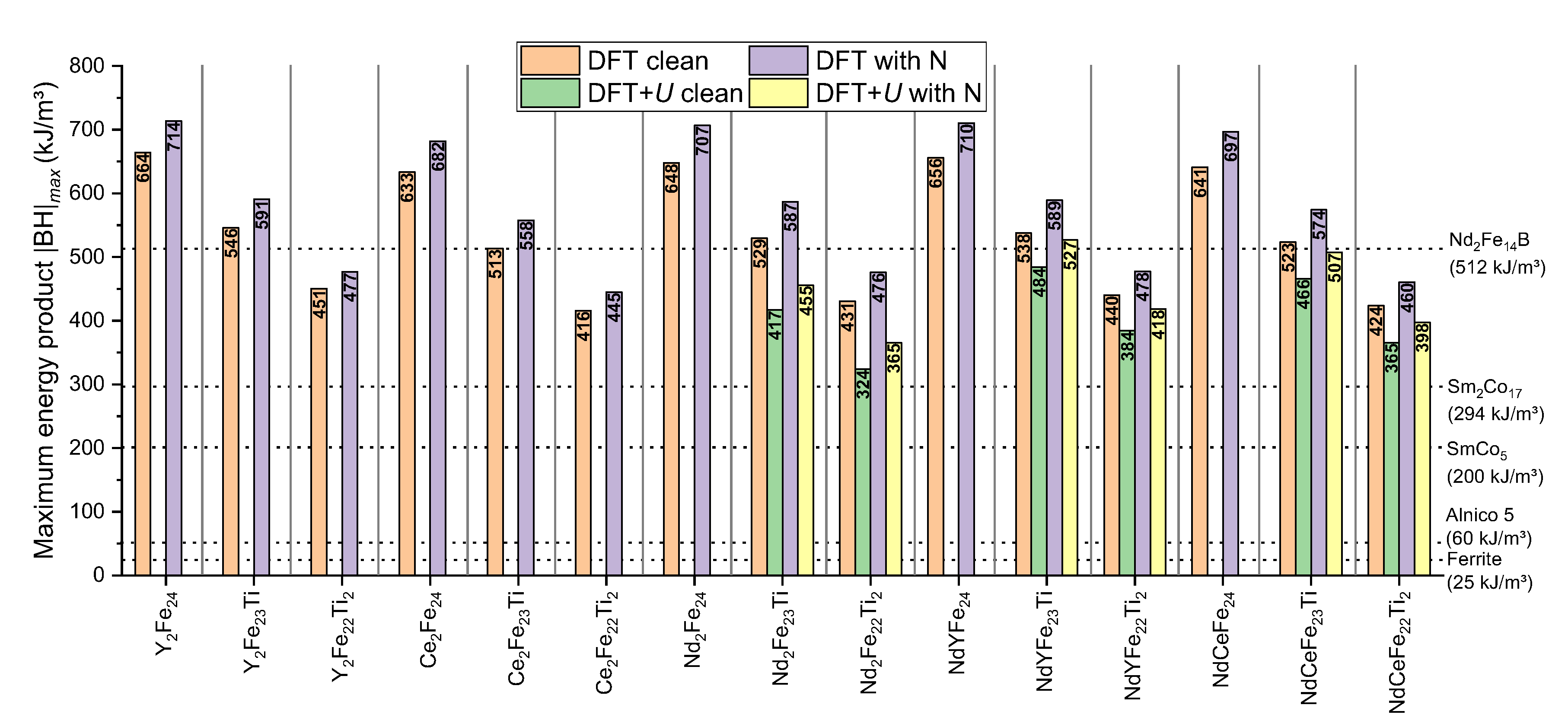}
\caption{Theoretical maximum energy product |BH|$_{max}$ values for considered compounds and nitrogenated cases. In addition to conventional DFT, DFT+\textit{U}  with \textit{U=}6 eV has been considered for stable Nd-contained alloys. Experimental values \cite{Coey2011,Fidler2004} of most common hard magnets are given as horizontal dotted lines. \label{Fig_BHmax_all}}
\end{figure}

Calculated |BH|$_{max}$ values of the considered compounds for clean and nitrogenated cases are given in Fig.~\ref{Fig_BHmax_all}. As a search criteria for a promising hard magnetic compound a |BH|$_{max}$ higher than 400 kJ/m$^3$ is important. As shown, the highest |BH|$_{max}$ values are generally obtained for Ti-free compounds, RFe$_{12}$ (RE: Y, Ce and Nd). However, they are unstable and Ti is needed for thermodynamic stabilization, as mentioned above. Although it is needed to stabilize the 1:12 phases, Ti substitution sacrifices a considerable amount of |BH|$_{max}$. An average of 3.8 at.\% Ti concentration (1 Ti atom in 2 f.u. supercell) reduces the maximum energy product by $\sim$120 kJ/m$^3$ and the loss is $\sim$100 kJ/m$^3$ once the Ti concentration is increased to 7.7 at.\% (2 Ti atom in 2 f.u. supercell).\\

Due to the missing experimental |BH|$_{max}$ measurements, a comparison for the theoretical results can not be made. Nevertheless, K\"orner \textit{et al.} \cite{Korner2016} calculated the |BH|$_{max}$ for various 1:12 phases by using tight-binding linear-muffin-tin-orbital atomic-sphere approximation (TB-LMTO-ASA). According to their findings, the |BH|$_{max}$ of CeFe$_{11}$Ti is 396 kJ/m$^3$ and NdFe$_{11}$Ti is 438 kJ/m$^3$, while in case of this work they are 416 and 431 kJ/m$^3$, respectively. This is a good theoretical agreement. In the case of DFT+\textit{U} treatment, 324 kJ/m$^3$ is found for NdFe$_{11}$Ti with \textit{U}=6 eV. Since, DFT+\textit{U} treatment yields localized 4\textit{f}-electrons with total -3.3 $\mu_{\text{B}}$ magnetic moment, it reduces $m_{tot}$ and |BH|$_{max}$ compared to DFT, which treats \textit{f}-electrons as in core. \\

Quaternary Y and Ce alloying to Nd-based 1:12 magnets have promising |BH|$_{max}$ values. In the case of Y substitution, the maximum energy product increases from 324 to 384 kJ/m$^3$ for (Nd,Y)Fe$_{11}$Ti, while it increases to 365 kJ/m$^3$ for (Nd,Ce)Fe$_{11}$Ti. As mentioned before, nitrogenation increases the total magnetization, and this similarly affects |BH|$_{max}$. For each considered compound nitrogenation increases |BH|$_{max}$ by $\sim$7-12\%. As an example, nitrogenation improves |BH|$_{max}$ for NdFe$_{11}$Ti $\sim$45 (41) kJ/m$^3$ for DFT (DFT+\textit{U} with \textit{U=} 6 eV) in the calculations of this thesis and the effect is calculated to be 49 kJ/m$^3$ by K\"orner \textit{et al.} \cite{Korner2016}.\\

Fig.~\ref{Fig_BHmax_all} also includes the |BH|$_{max}$ values of the most well known hard magnets. The highest energy product is achieved by Nd$_2$Fe$_{14}$B with 512 kJ/m$^3$. In the case of DFT+\textit{U} calculations the most promising quaternary compound is calculated to be NdYFe$_{23}$Ti with 484 kJ/m$^3$, which is 28 kJ/m$^3$ lower than |BH|$_{max}$ of Nd$_2$Fe$_{14}$B. In the case of NdCeFe$_{23}$Ti it is 465 kJ/m$^3$. As mentioned earlier, quarternaries containing a higher concentration of Ti have a lower |BH|$_{max}$, such as 384 and 365 kJ/m$^3$ for NdYFe$_{22}$Ti$_2$ and NdCeFe$_{22}$Ti$_2$, respectively. Note that nitrogenation of these compounds further increases the calculated values by $\sim$40 kJ/m$^3$. Therefore, achieving such high energy products with about 50\% less Nd would be technologically and economically very valuable.\\

\section{Curie Temperature}

\label{Sec_CurieTemp}

In addition to a strong magnetization, a high Curie temperature $T_\text{C}$ is another desired intrinsic magnetic property. The well-known Nd$_2$Fe$_{14}$B magnet is famous for its strong saturation magnetization at room temperature, however, it is also known to have a relatively low $T_\text{C}$ (588 K \cite{Coey2011}) compared to more traditional RE-based permanent magnets such as Sm$_2$Co$_{17}$ (838 K \cite{Coey2011}). Finding an Nd-lean RE magnet with a better balance between saturation magnetization and Curie temperature thus deserves extensive efforts.\\

In order to calculate Curie temperatures of the considered compounds, an effective spin Heisenberg model is employed and solved by the mean-field approximation (MFA). In the Heisenberg model, the Hamiltonian for the spin-spin interaction can be written as $\mathcal{H} =-\sum_{i \neq j}J_{ij}\vec{S_{i}}\vec{S_{j}}$, where $\vec{S_{i}}$($\vec{S_{j}}$) is the spin at site $i (j)$ and $J_{ij}$ is the exchange coupling constant between sites $i$ and $j$, which is calculated by AkaiKKR \cite{AkaiKKR}. The absolute value of the magnetic ordering temperature, $T_\text{C}$, derived from the MFA is $k_{\text{B}}T_{\text{C}}= \frac{3}{2}J_0$, where $k_{\text{B}}$ is the Boltzmann factor and $J_0$ is the total effective exchange that is the sum of $J_{ij}$ connecting a given lattice site to all remaining ones.\\

In the MFA, the systematic overestimation of $T_\text{C}$ is a well-known issue, therefore, in addition to the FM state, the local moment disorder (LMD) approach was considered as a second magnetic state as well. In the framework of the KKR-CPA method, the concept of LMD state is conveniently used to describe the paramagnetic state of a ferromagnet above $T_\text{C}$ \cite{Akai1993} (note that LMD is also called disorder local moment (DLM) \cite{Gyorffy1985}). Nevertheless, magnetic moments survive and do not vanish. As an advantage of the KKR based on Green's function theory, the exchange interaction energies for such situations and the Curie temperature for both FM and LMD states can be calculated.\\

\begin{figure}[h]
\includegraphics[width=1.0\textwidth]{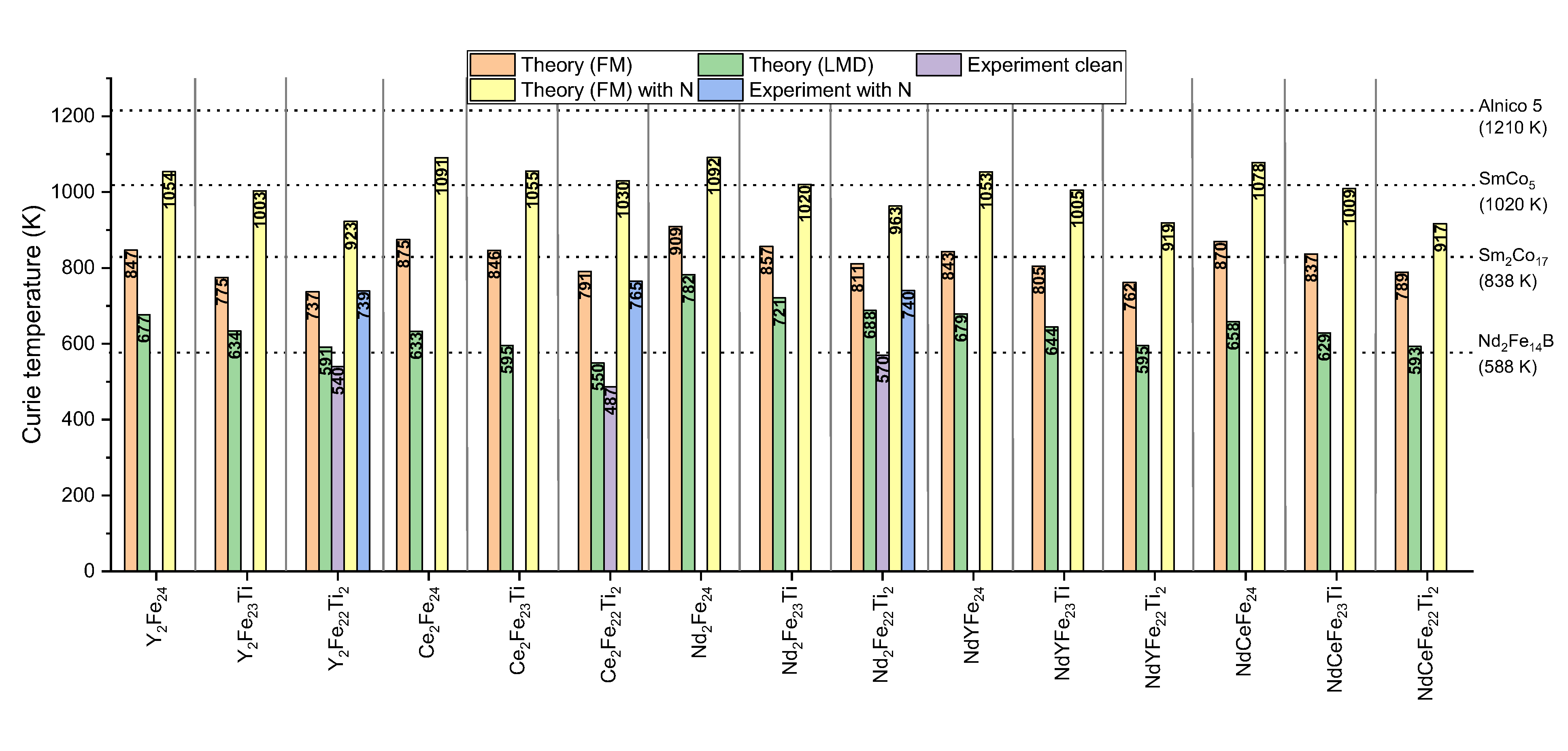}
\caption{Calculated Curie temperatures $T_{\text{C}}$ for the clean and nitrogenated 1:12 phases. Exchange interaction energies have been calculated with both ordered ferromagnetic (FM) and disordered local moment disorder (LMD) states, and the input lattice information is taken from our theoretically relaxed calculations (check Tab.~\ref{Tab_lattConstANDbulkModul}). Experimental data is taken from \cite{Yang1991,Maccari2021}.
\label{Fig_curie_temp}}
\end{figure}

The calculated Curie temperatures are given in Fig.~\ref{Fig_curie_temp} in comparison with available experimental data for clean and nitrogenated compounds and well-known permanent magnets. As shown, the calculated FM ground state approximation in this work overestimates about 40\% for most of the cases as expected. However, calculating the $T_\text{C}$ by considering the LMD as the ground state, the agreement improves significantly. The underlying reasons for the overestimation of $T_\text{C}$ for the FM ground state are two-fold: $i)$ the problem in describing the delocalized electronic state in the magnetism of intermetallics on the basis of localized degrees of freedom. $ii)$  exchange couplings between the local moments are calculated for the ground state, and this assumption does not need to be held for high temperatures near the $T_\text{C}$.\\

Nevertheless, Miyake \textit{et al.} \cite{Miyake2021} calculated the $T_\text{C}$ for various RE-based compounds for the 2:14, 2:17 and 1:12 phases (namely RE$_2$Fe$_{14}$B, RE$_2$Co$_{14}$B, RE$_2$Fe$_{17}$, RE$_2$Co$_{17}$ and RE$_2$Fe$_{22}$Ti) and reported that the performance of the LMD treatment is not always better than the FM case. For instance, in the case of RE$_2$Fe$_{14}$B, the $T_\text{C}$ values are systematically overestimated for FM treatment and agreement appeared to be better for the LMD case. However, in RE$_2$Co$_{14}$B, the $T_\text{C}$ results from the FM ground state are much better than the underestimated LMD values. In the case of the RE$_2$Fe$_{17}$ phase, both FM and LMD treatments fail to reproduce $T_\text{C}$ by substantial overestimation, while in the case of RE$_2$Co$_{17}$, the $T_\text{C}$ is well produced for the FM treatment. In agreement with this work, they also reported an overestimation of $T_\text{C}$ for the FM treatment in RE$_2$Fe$_{22}$Ti$_2$ compounds, while the LMD treatment gives much better agreement with respect to experiment. As reported by Miyake \textit{et al.} \cite{Miyake2021}, the FM ground state yields generally overestimation (or good agreement as in the case of RE$_2$Co$_{14}$B), but there is a complicated dependence of the LMD performance on the crystal structure and the transition metal element. This may originate from a difference in spin fluctuations close to the magnetic transition and needs to be further analyzed.\\

As in the case of $m_{tot}$ and |BH|$_{max}$, increasing Ti concentration reduces $T_\text{C}$ for all considered compounds shown in Fig.~\ref{Fig_curie_temp}. The average decrease of $T_\text{C}$ due to the Ti addition (1 Ti atom=3.8 at.\%) is calculated to be $\sim$45 K. Experimentally it is reported that among the ternaries, CeFe$_{11}$Ti has the lowest $T_\text{C}$ with 487 K \cite{Maccari2021}. This is followed by YFe$_{11}$Ti with 540 K \cite{Yang1991} and NdFe$_{11}$Ti with 570 K \cite{Yang1991}. The theoretical LMD results in this work perfectly reproduce the experimental trend with the calculated $T_\text{C}$ values of 550, 591, and 688 K, for CeFe$_{11}$Ti, YFe$_{11}$Ti, and NdFe$_{11}$Ti, respectively.\\

Y and Ce substitution on Nd-based 1:12 phase cause a decreasing $T_\text{C}$. According to our LMD calculations, the Curie temperature of (Nd,Y)Fe$_{11}$Ti is 595 K, and it is 593 K for (Nd,Ce)Fe$_{11}$Ti, which means $\sim$95 K reduction compared to NdFe$_{11}$Ti. However, it must be noted that the experimental $T_\text{C}$ of well-known magnets are also shown in Fig.~\ref{Fig_curie_temp}, and both Y and Ce doped Nd-based quaternary compounds still have  higher $T_\text{C}$ than Nd$_2$Fe$_{14}$B. Therefore, one can conclude that Nd-lean 1:12 phases have slightly lower but still comparable Curie temperatures as compared to Nd$_2$Fe$_{14}$B.\\

The effect of N doping on the considered compounds has been investigated as well. Although the FM treatment overestimates the calculated $T_\text{C}$, the difference between clean and nitrogenated compounds can be compared ($\Delta T_{\text{C}}^{\text{N}}=T_{\text{C}}^{\text{RFe$_{12-y}$Ti$_y$N}}-T_{\text{C}}^{\text{RFe$_{12-y}$Ti$_y$}}$). As shown in Fig.~\ref{Fig_curie_temp}, it is experimentally reported that N increases the $T_{\text{C}}$ for all ternaries RFe$_{11}$Ti (RE: Y, Ce and Nd). This improvement is measured to be 199 \cite{Yang1991}, 278 \cite{Maccari2021} and 170 \cite{Yang1991} K for YFe$_{11}$Ti, CeFe$_{11}$Ti and NdFe$_{11}$Ti, respectively. Our FM treatment also approves such an increase after nitrogenation with 186, 239 and 152 K for Y, Ce and Nd-based 1:12 phases, respectively. Besides a very promising quantitative agreement between our theoretical work and experimental data, the order of N impact is calculated correctly as well. According to the experiment, N has the highest impact on the Ce-based 1:12 phase, then it is followed by Y and Nd-based 1:12 phases which is correctly reproduced in our calculations. 


\section{Finite Temperature Magnetization}

\begin{figure}[h]
\begin{subfigure}{0.32\textwidth}
\includegraphics[width=\textwidth]{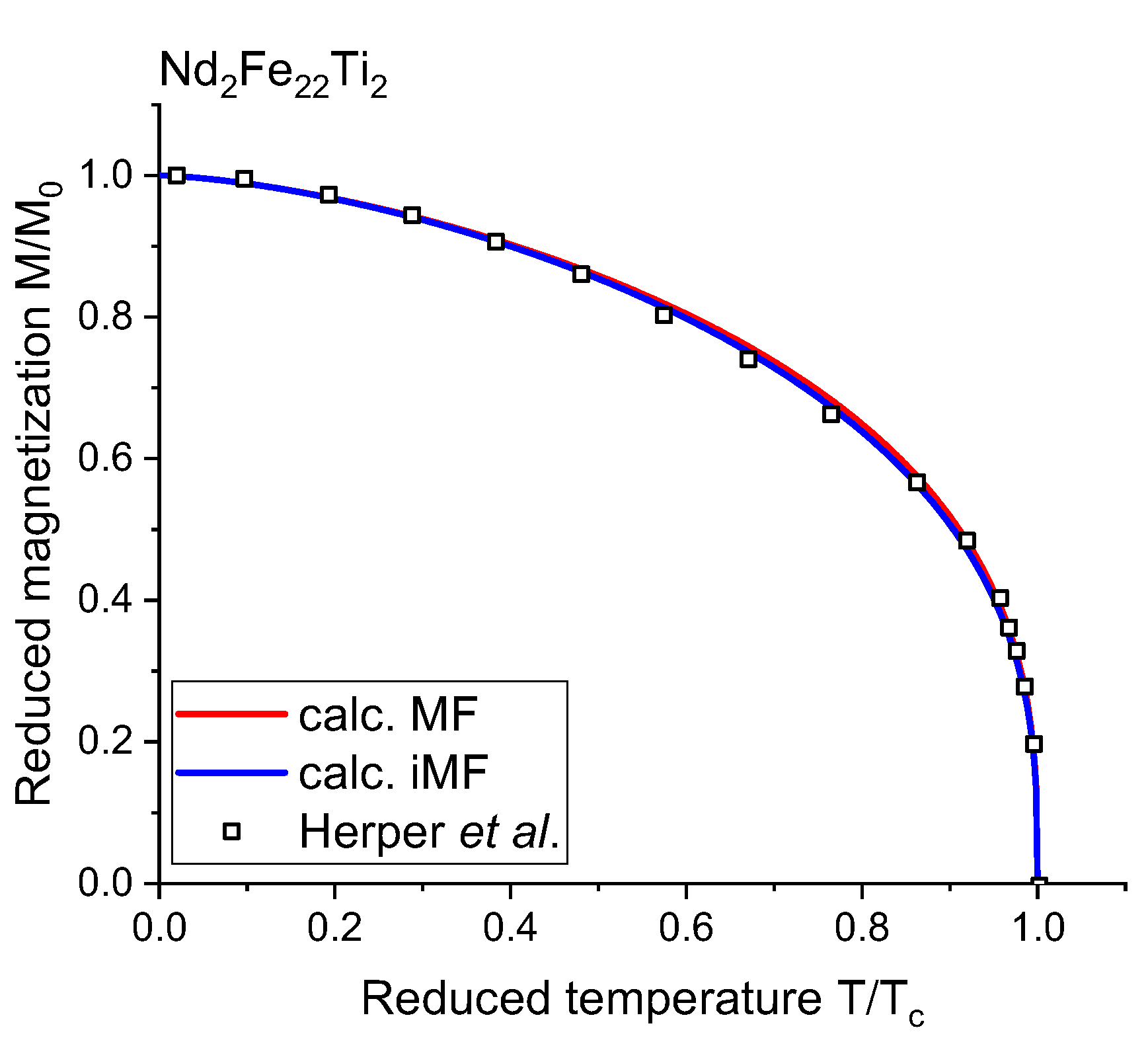}
\end{subfigure}
\begin{subfigure}{0.32\textwidth}
\includegraphics[width=\textwidth]{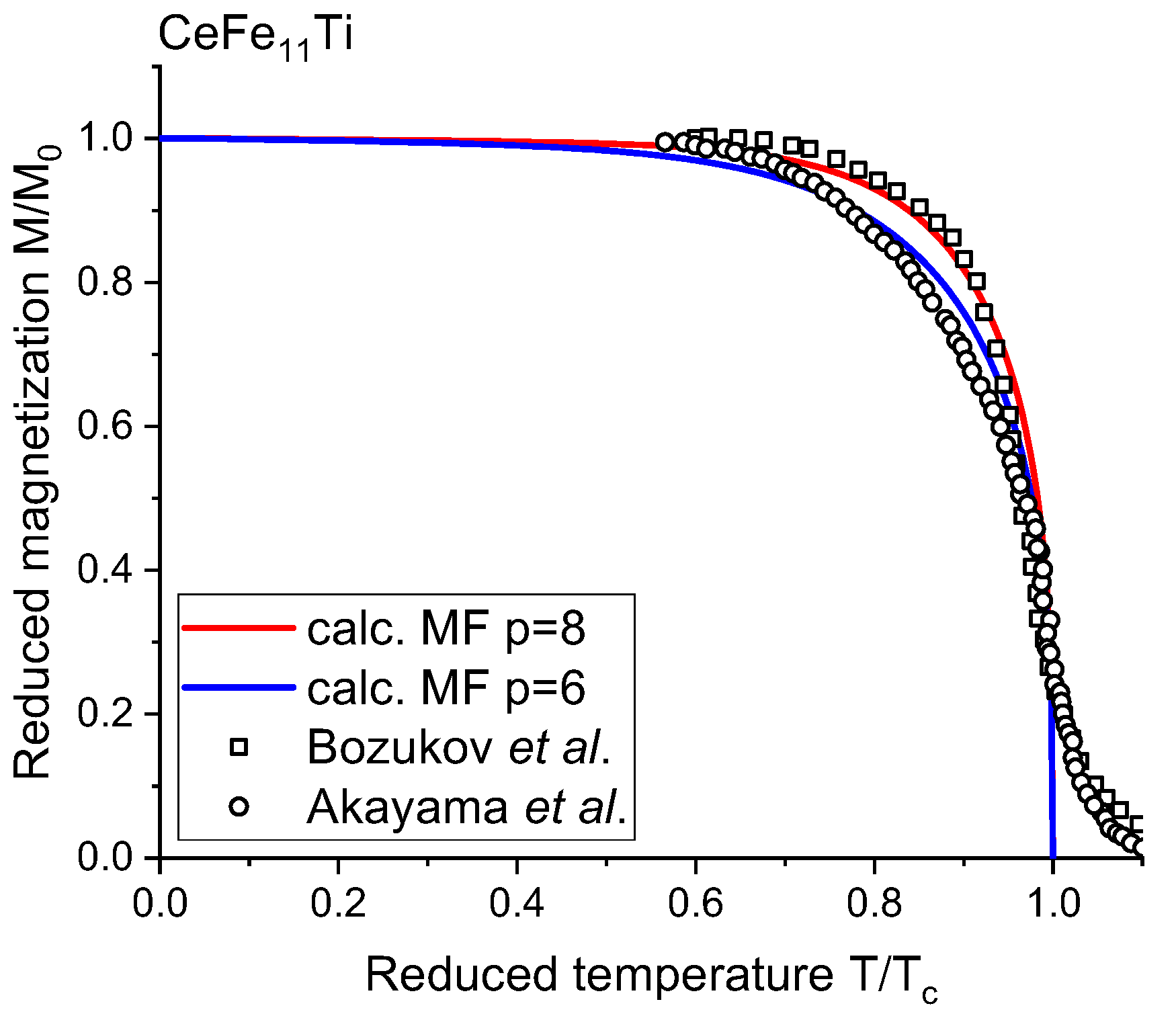}
\end{subfigure}
\begin{subfigure}{0.32\textwidth}
\includegraphics[width=\textwidth]{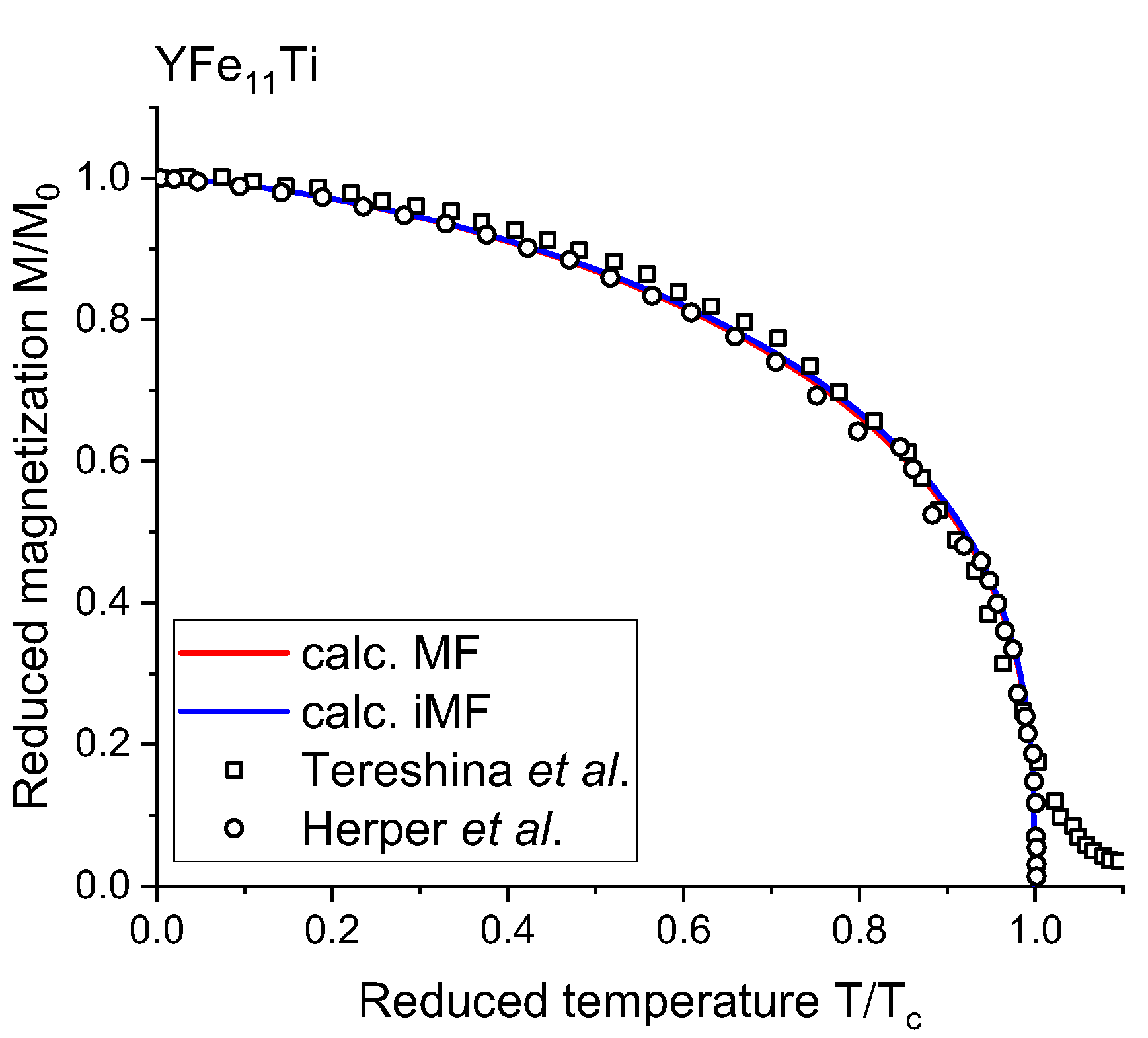}
\end{subfigure}
\caption{Finite temperature magnetization calculated on basis of the K\"ormann and Kuzmin approaches. For comparison the experimental data by Herper \textit{et al.} \cite{Herper2022}, Bozukov \textit{et al.} \cite{Bozukov1991}, Akayama \textit{et al.} \cite{Akayama1994} and Tereshina \textit{et al.} \cite{Tereshina2003} are shown. For the NdFe$_{11}$Ti and YFe$_{11}$Ti compounds the Kuzmin and the K\"ormann approaches are used. For the CeFe$_{11}$Ti compound the Kuzmin approach with $p$=8 (red) and $p$=6 (blue) are depicted.  \label{kuzmin_plot}}
\end{figure}

The temperature-dependent stability of the magnetic phases has been addressed in this thesis as well. Based on the developed computational schemes by Kuzmin \textit{et al.} \cite{Kuzmin2005} and K\"ormann \textit{et al.} \cite{Kormann2008}, one can calculate the finite temperature magnetization curves of various elements and compounds by using total magnetic moments and Curie temperature as input parameters. 
For the Kuzmin plot (labeled as MF in the diagrams), the following equation is used to determine the finite temperature magnetization.
\begin{equation}
m(\frac{M(T)}{M_0}) = [1-s(\frac{T}{T_C})^{\frac{3}{2}} - (1-s)(\frac{T}{T_C})^{p}]^{\frac{1}{3}}.
\end{equation}
The finite temperature magnetization $M(T)$ is depicted as the reduced magnetization $m(\frac{M(T)}{M_0})$ and is calculated with the temperature $T$, the Curie temperature $T_C$, the magnetization at 0 K $M_0$, and the two variables $s$ and $p$. The variable $p$ is most of the times set to be $\frac{5}{2}$ like for body centered cubic (bcc) Fe \cite{Kuzmin2005}. The variable $s$ can then be a value between 0 and $\frac{5}{2}$ and is varied until a good fit to the experimental data is achieved.
In the improved mean field approach by K\"ormann \textit{et al.} \cite{Kormann2008} on the example of bcc Fe, the variable $p$ is set to a constant value of 4. This results in a better agreement for their considered example of bcc Fe as well as for the considered compounds of NdFe$_{11}$Ti and YFe$_{11}$Ti in Fig.~\ref{kuzmin_plot}. The resulting mean-field plots, as shown in Fig.~\ref{kuzmin_plot}, show a good agreement with the experimental data by Herper \textit{et al.} \cite{Herper2022}, Bozukov \textit{et al.} \cite{Bozukov1991}, and Tereshina \textit{et al.} \cite{Tereshina2003}. In case of the NdFe$_{11}$Ti and YFe$_{11}$Ti compounds, the variable $p$ is set as described before to $\frac{5}{2}$, and the variable $s$ leads to good results with a value of about 1.0. In both cases, the difference between the data calculated with the Kuzmin and the K\"ormann approach is only marginal and both approaches reveal a good agreement with the experimental data. For the CeFe$_{11}$Ti alloy, the variable $p$ is chosen much bigger with a value of 8 in case of the data of Bozukov \textit{et al.} \cite{Bozukov1991}, and a value of 6 in case of the data by Akayama \textit{et al.} \cite{Akayama1994}. This change was applied only to the Kuzmin approach because changing the value of 4 in the equation of K\"ormann leads to the same equation as used by Kuzmin. In this case, a very small value of $s=0.05$ is chosen to get the best possible results. With these variables and changes, all of the calculated finite temperature plots are in good accordance with the experimental data.

\section{Magnetocrystalline Anisotropy}

\label{Sec_Results_MAE}

Magnetocrystalline anisotropy (MCA) is a critical magnetic property and must be addressed for a material once it has been tended to be used as a permanent magnet. Therefore, as the last intrinsic property, the magnetocrystalline anisotropy, which has a key importance for the performance of hard magnets is investigated. The MCA can be calculated by employing the magnetic force theorem \cite{Springford1980} and by total energy differences. In this study, the latter one has been considered and magnetocrystalline anisotropy energy (MAE) is defined as

\begin{equation}
\label{Eq_MAE}
\Delta E_{\text{MAE}} = E_{[110]} - E_{[001]},
\end{equation}

where $E_{[110]}$ and $E_{[001]}$ are the total energies of the considered supercells for the magnetization being oriented along the [110] and [001] directions, respectively.\\

\begin{figure}[h]
\centering
\begin{subfigure}{0.48\textwidth}
\includegraphics[width=\textwidth]{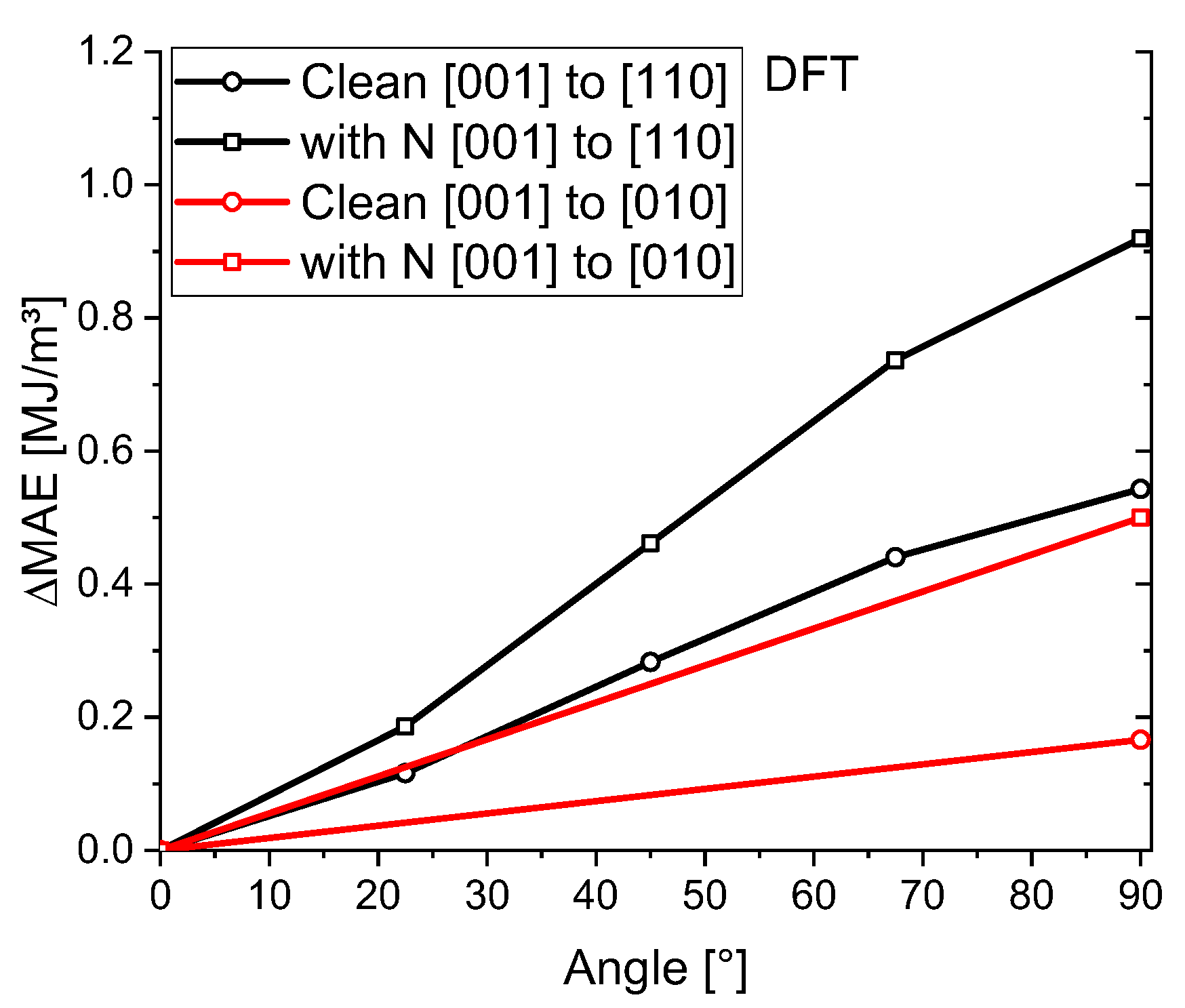}
\end{subfigure}
\begin{subfigure}{0.48\textwidth}
\includegraphics[width=\textwidth]{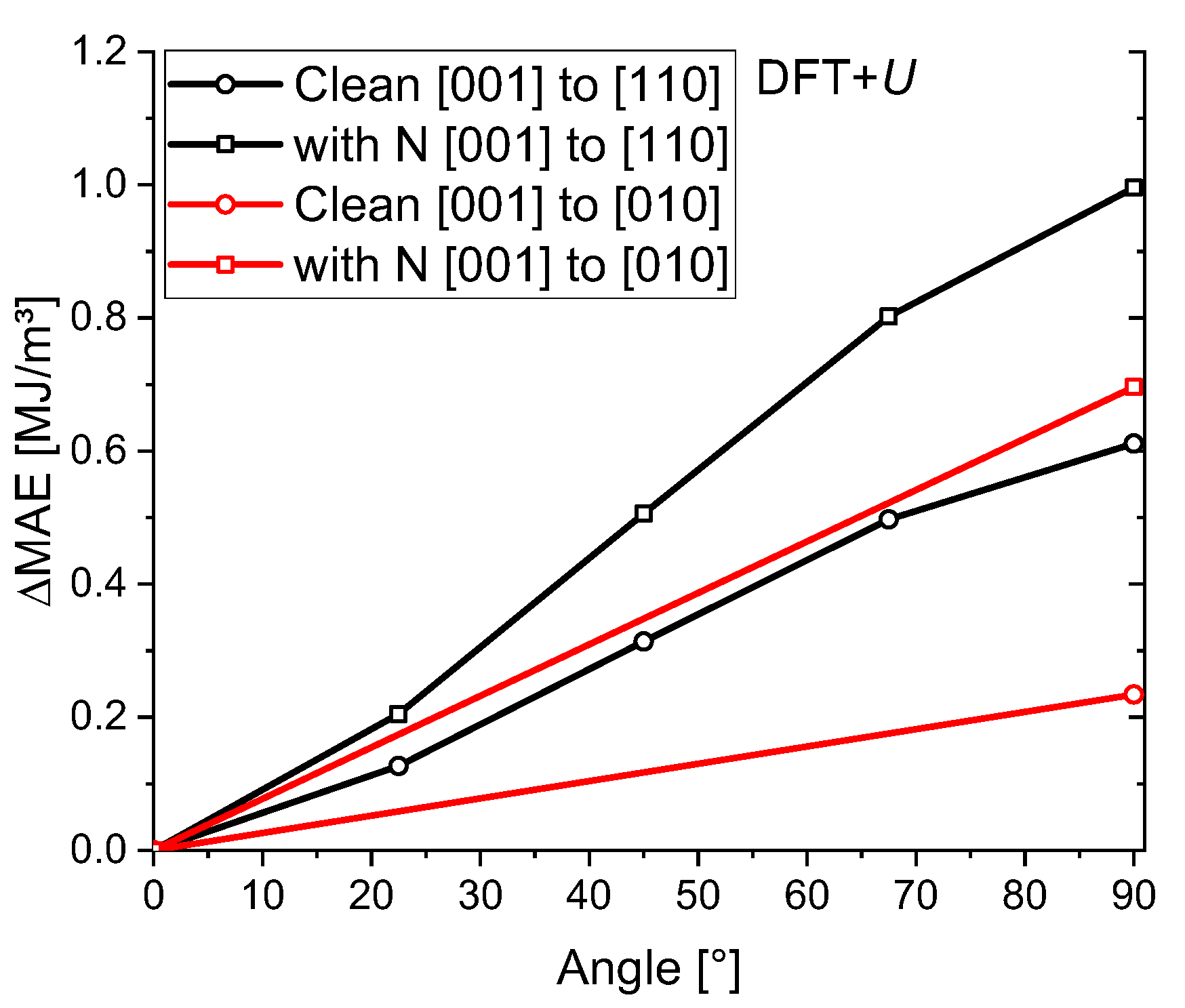}
\end{subfigure}
\caption{Calculated energy differences of the magnetocrystalline energy for the shift of the crystallographic direction from [001] to [110](black) and [100] to [010] (red) for the clean (circles) and nitrogenated (squares) NdFe$_{11}$Ti compound. On the left side are DFT and on the right side DFT+\textit{U} results shown. For the change from [001] to [110]  the angles 0, 22.5, 45, 67.5 and 90° are considered. For the change from [001] to [010] only the 0 and 90 ° angles are considered.\label{fig_mae_angles}}
\end{figure}

For MAE calculations, the orientation of the magnetization along the crystallographic directions plays an important role. In Fig.~\ref{fig_mae_angles}, the resulting energy differences on basis of DFT and DFT+\textit{U} calculations for changes from the easy magnetization axis [001] to the directions [110] and [010] are shown. DFT+\textit{U} treatment leads to an increase of the energies of $\sim 0.07$ MJ/m$^3$ for the clean and $\sim$ 0.1 to 0.2 MJ/m$^3$ for the nitrogenated case. The influence of interstitial nitrogenation is depicted with the change from circles to squares with an energy increase of $\sim$ 0.4 MJ/m$^3$. For the different direction changes from [001] to [110] and [001] and [010], shown in black and red, respectively, a significant energy difference of $\sim 0.4$ MJ/m$^3$ is visible. Due to the higher energy difference, the change to [110] is mainly focused on in this work.\\

Based on the calculated magnetocrystalline anisotropy energy one can find the anisotropy constant $K_1$ and the anisotropy field $H_a$. The anisotropy energy can be written as an expansion whose leading term is \cite{Coey2011}

\begin{equation}
\label{Eq_K1}
\Delta E_{\text{MAE}}=K_1 \sin(\phi)^2,
\end{equation}

where $\phi$ is the angle between the magnetization direction and the easy axis. Further terms in the expansion reflect the crystal symmetry, but for many purposes, it is enough to consider only the leading term. A positive (negative) $K_1$ value corresponds to uniaxial (planar) anisotropy. \\

The anisotropy field $H_a$ is the upper limit for the coercivity and can be estimated from \cite{Coey2011}

\begin{equation}
\label{Eq_Ha}
H_{a}=\frac{2 K_1}{\mu_{0}M_{S}}.
\end{equation}

$H_a$ is an important property for permanent magnets, and a high value is needed for hard magnetic applications. Nevertheless, it must be noted that the calculated $K_1$ and $M_S$ values correspond to $T=$0 K, therefore, the $T$ dependence of $H_a$ is excluded. \\

The calculated magnetocrystalline anisotropy constant $K_1$ and anisotropy field $H_a$ are given in Tab.~\ref{tab_MAE} for the clean compounds. As listed, all the considered 1:12 phases have uniaxial anisotropy ($K_1$>0). This linearity yields that higher-order anisotropy constants are negligible. We have examined our considered compounds and find that $K_2$ is almost 2 orders in magnitude smaller than the lowest order anisotropy constant, $K_1$, which makes it sufficient to describe the MCA (for more details regarding the anisotropy constants see Ref.~ \cite{Nikitin1998}).\\

\begin{table*}[h!]
\caption{Calculated theoretical magnetocrystalline anisotropy constant $K_1$ according to Eq.~\ref{Eq_K1}, anisotropy field $H_a$ according to Eq.~\ref{Eq_Ha} and magnetic hardness factor $\kappa$ from Eq.~\ref{Eq_kappa} for stable compounds with available experimental data. Values are given for both GGA and LSDA treatments for the exchange-correlation potential and DFT+\textit{U} (\textit{U}= 6 eV) results are reported in parenthesis Nd contained compounds. The experimental value of $K_1$ of well known hard magnets is also given for a comparison.
\label{tab_MAE}}
\centering
\scriptsize
\begin{tabular}{@{}cccccccccc@{}}
\toprule \hline
Alloy & \begin{tabular}[c]{@{}l@{}}$K_1$\\ GGA\\ (MJ/m$^3$)\end{tabular} & \begin{tabular}[c]{@{}l@{}}$K_1$\\ LSDA\\ (MJ/m$^3$)\end{tabular} & \begin{tabular}[c]{@{}l@{}} $K_1$\\experiment \\ clean \\ (MJ/m$^3$)\end{tabular} & \begin{tabular}[c]{@{}l@{}}$H_a$\\ GGA \\ (T)\end{tabular} & \begin{tabular}[c]{@{}l@{}}$H_a$\\ LSDA\\ (T)\end{tabular} & \begin{tabular}[c]{@{}l@{}} $H_a$\\ experiment \\ clean \\ (T)\end{tabular} \\ \midrule

Y$_2$Fe$_{23}$Ti & 0.81 & 1.65 &  & 0.98 & 2.28 &  \\
Y$_2$Fe$_{22}$Ti$_2$ & 0.92 & 1.73 & \begin{tabular}[c]{@{}l@{}}0.85$^a$ \cite{Nikitin1998}\\ 1.89$^b$ \cite{Nikitin1998} \\ 1.91$^c$ \cite{Tereshina2003} \\ 2.0$^c$ \cite{Qi1992} \\ 2.1$^d$ \cite{Herper2022} \end{tabular} & 1.22 & 2.76 & \begin{tabular}[c]{@{}l@{}}2.2$^e$ \cite{Wang2000} \\ 4.1$^f$ \cite{Wang2000} \\ 4.8$^a$ \cite{Obbade1997} \\5.7$^c$ \cite{Obbade1997} \end{tabular} \\ \midrule

Ce$_2$Fe$_{23}$Ti & 1.67 & 1.91 &  & 2.08 & 2.70 &  \\
Ce$_2$Fe$_{22}$Ti$_2$ & 2.08 & 2.12 & \begin{tabular}[c]{@{}l@{}}1.4$^e$ \cite{Isnard1998} \\ 1.78$^g$ \cite{Pan1994} \\ 1.9$^g$ \cite{Goll2014} \\ 4.3$^e$ \cite{Akayama1994}  \end{tabular} & 2.88 & 3.40 & \begin{tabular}[c]{@{}l@{}}1.5$^h$ \cite{Pan1994}\\ 1.7$^e$ \cite{Zhou2014} \\ 2.3$^e$ \cite{Akayama1994} \\ 3.5$^i$ \cite{Pan1994} \end{tabular}  \\ \midrule

Nd$_2$Fe$_{23}$Ti & \begin{tabular}[c]{@{}l@{}}0.37\\ (0.39)\end{tabular} & 1.32 &  & \begin{tabular}[c]{@{}l@{}}0.45\\ (0.54)\end{tabular} & 2.12 &   \\
Nd$_2$Fe$_{22}$Ti$_2$ & \begin{tabular}[c]{@{}l@{}}0.54\\ (0.61)\end{tabular} & 1.41 & \begin{tabular}[c]{@{}l@{}}0.94$^e$ \cite{Bouzidi2021} \\ 1.78$^d$ \cite{Herper2022}\end{tabular} & \begin{tabular}[c]{@{}l@{}}0.74\\ (0.96)\end{tabular} & 2.57 & \begin{tabular}[c]{@{}l@{}}1.9$^e$ \cite{Bouzidi2021} \\ 2.0$^e$ \cite{Akayama1994} \end{tabular}  \\ \midrule

NdYFe$_{23}$Ti & \begin{tabular}[c]{@{}l@{}}0.63\\ (0.66)\end{tabular} & 1.53 &  & \begin{tabular}[c]{@{}l@{}}0.77\\ (0.85)\end{tabular} & 2.28 &   \\
NdYFe$_{22}$Ti$_2$ & \begin{tabular}[c]{@{}l@{}}0.74\\ (0.91)\end{tabular} & 1.65 &  & \begin{tabular}[c]{@{}l@{}}1.00\\ (1.31)\end{tabular} & 2.75 &   \\ \midrule

NdCeFe$_{23}$Ti & \begin{tabular}[c]{@{}l@{}}0.87\\ (0.94)\end{tabular} & 1.43 &  & \begin{tabular}[c]{@{}l@{}}1.07\\ (1.23)\end{tabular} & 2.13 &   \\
NdCeFe$_{22}$Ti$_2$ & \begin{tabular}[c]{@{}l@{}}1.27\\ (1.35)\end{tabular} & 1.61 &  & \begin{tabular}[c]{@{}l@{}}1.75\\ (2.00)\end{tabular} & 2.71 &   \\ \midrule
Nd$_2$Fe$_{14}$B &  &  & 4.9$^j$ \cite{Skomski2016} &  &  &  \\
Sm$_2$Co$_{17}$ &  &  & 4.2$^j$ \cite{Skomski2016} &  &  &  \\
SmCo$_5$ &  &  & 17.0$^j$ \cite{Skomski2016} &  &  &  \\
Alnico 5 &  &  & 0.32$^j$ \cite{Skomski2016} &  &  &  \\ \hline \bottomrule
\end{tabular}
\begin{tablenotes}
    \item Theoretical references are represented by *
        \item $^a$ M\"ossbauer measurement at 300 K,
        $^b$ M\"ossbauer measurement at 77 K,
        $^c$ M\"ossbauer measurement at 4.2 K,
        $^d$ Single Crystal, Physical Property Measurement System (PPMS) measured at 10 K,
        $^e$ Derived from magnetization curve at 300 K,
        $^f$ Derived from magnetization curve at 4.2 K,
        $^g$ Determined by Kerr microscopy at 300 K,
        $^h$ X-ray diffraction measurement at 300 K,
        $^i$ X-ray diffraction measurement at 1.5 K,
        $^j$ Properties measured at 300 K.
\end{tablenotes}
\end{table*}


As given in Tab.~\ref{tab_MAE}, an increasing Ti concentration also increases the anisotropy constant. Since periodic boundary conditions are used, the presence of Ti can be an artifact in an unevenly distributed small supercell. The missing real random distribution of Fe/Ti at 8\textit{i} site also results in slightly non-equivalent basal directions [100] and [010].\\

In the case of YFe$_{11}$Ti, the DFT (GGA) approximation yields the anisotropy constant as 0.92 MJ/m$^3$. This is a clear underestimation of low temperature experimental values reported in the range of 1.89-2.1 MJ/m$^3$ \cite{Nikitin1999,Tereshina2003,Qi1992,Herper2022}. In addition, a DFT based underestimated MCA is also reported by Ke \textit{et al.} \cite{Ke2016} and Herper \textit{et al.} \cite{Herper2022}. Therefore, the MCA is further examined by using LSDA. Significant improvement is achieved by using LSDA with 1.73 MJ/m$^3$, which agrees better to experimental data. \\

The GGA results for CeFe$_{11}$Ti already yield a very promising $K_1$ against experimental data. Known reported experimental values are ranging from 1.4 to 4.3 MJ/m$^3$ \cite{Isnard1998,Akayama1994} based on the measurement technique and temperature. Nevertheless, the calculated $K_1$ with 2.08  MJ/m$^3$ fits well to these experiments. The LSDA treatment for Ce-based compounds does not have a significant impact, and the $K_1$ is calculated to be 2.12 MJ/m$^3$. \\

As mentioned earlier, 4\textit{f}-electrons treatment in general DFT is not straightforward and based on the choice of the 4\textit{f} localization MCA can yield different results. According to DFT (GGA) calculations in this study, where \textit{f}-electrons are treated in core and no hybridization takes place between 4\textit{f} and valence states, the anisotropy constant $K_1$ is calculated to be 0.54 MJ/m$^3$ for the NdFe$_{11}$Ti compound. Once we reduce the level of localization by applying the Hubbard \textit{U} correction to the on-site Coulomb interactions $K_1$ is predicted to be 0.61 MJ/m$^3$ for \textit{U=}6 eV. In comparison to the experimental value of 1.78 MJ/m$^3$ \cite{Herper2022}, both cases underestimates the MCA. As in the case of Y-based compounds, the LSDA treatment for NdFe$_{11}$Ti yields better MCA results with respect to an experimental value of 1.41 MJ/m$^3$.\\

According to the calculations in this work, both Y and Ce substitutions have improved the MCA compared to the NdFe$_{11}$Ti case. In case of Y substituted 1:12 phase (Nd,Y)Fe$_{11}$Ti, GGA (GGA+\textit{U}) has a $K_1$ value of 0.74 (0.91) MJ/m$^3$ and it is further improved for the Ce substituted case in (Nd,Ce)Fe$_{11}$Ti with 1.27 (1.35) MJ/m$^3$. Although, the GGA based calculations underestimate the MCA of NdFe$_{11}$Ti with respect to measurements, it is known that experimentally reported YFe$_{11}$Ti and CeFe$_{11}$Ti $K_1$ values (1.89 \cite{Nikitin1999} and 2 MJ/m$^3$ \cite{Goll2014}) are higher than the $K_1$ corresponding to NdFe$_{11}$Ti (1.78 MJ/m$^3$ \cite{Herper2022}). Therefore, such a theoretical increase is accurately captured for both GGA (GGA+\textit{U}) and LSDA calculations. In addition, for all considered compounds MCA is found to be uniaxial ($K_1$>0) independently from the choice of exchange-correlation functionals, although GGA calculations for Y and Nd-based compounds strongly underestimate MCA. This is a good agreement with experimental data. \\

As given in Eq. \ref{Eq_Ha}, the anisotropy field $H_a$ can be calculated in a combination of the anisotropy constant $K_1$ and the saturation magnetization $M_S$. In addition, as in the case of $K_1$ there is only available experimental data for the ternary compounds. In the case of the YFe$_{11}$Ti compound the GGA calculations yield 1.22 T, which underestimates the low-temperature experimental values, reported to be in the range of 4.1 to 5.7 T \cite{Wang2000,Obbade1997}, and room temperature experimental values ranging from 2.2 to 4.8 T \cite{Obbade1997, Wang2000}. The LSDA treatment leads to a significant improvement with a calculated value of 2.76 T, although there is still a deviation compared to low-temperature experimental data. \\

The anisotropy field $H_a$ of the CeFe$_{11}$Ti compound is already well described by the GGA calculations with a value of 2.88 T. The experimental findings are reported in the range of 1.5 to 3.5 T \cite{Pan1994,Zhou2014,Akayama1994} based on measurement technique and temperature. LSDA treatment yields a slightly higher value of 3.40 T, which also fits into the range of the reported experimental values and is near the value of 3.5 T measured by Pan \textit{et al.} \cite{Pan1994} with X-ray diffraction at 1.5 K. Therefore, the LSDA treatment does not have a significant impact as in the case of the Y and Nd-based ternaries.\\

In the NdFe$_{11}$Ti calculations, the DFT (GGA) approach, with 4\textit{f}-electrons treated as in core and no hybridization taking place, a value of 0.74 T is calculated. Applying the Hubbard \textit{U} correction to the on-site Coulomb interactions the result increases to 0.96 T for \textit{U=}6 eV. Comparing these results to the experimental findings of 1.9 T by Bouzidi \textit{et al.} \cite{Bouzidi2021} and 2.0 T by Akayama \textit{et al.} \cite{Akayama1994}, a clear underestimation of our calculations is observed. Nevertheless, the LSDA treatment leads to an significant increase with 2.57 T, which better agrees with the reported experimental values.\\

In the case of quaternaries, the partial Nd substitution leads to an improvement for $H_a$ compared to NdFe$_{11}$Ti ternary. For the Y substituted 1:12 compound NdYFe$_{22}$Ti$_2$ GGA (GGA+\textit{U}), calculations result in a $H_a$ value of 1.0 (1.31) T. In the case of Ce substitution NdCeFe$_{22}$Ti$_2$, the increase is even larger with a value of 1.75 (2.00) T. As shown before, the LSDA treatment yields larger $H_a$ values for Y and Ce substituted quaternaries with 2.75 and 2.71 T, respectively. Since there are no experimental values to compare the theoretical findings of this work with for the considered quaternaries, it can only be deduced that the  calculated $H_a$ values for the ternaries have a good agreement with experiments which is encouraging for the quaternary findings. \\

 \begin{figure}[h]
\centering
\includegraphics[width=0.75\textwidth]{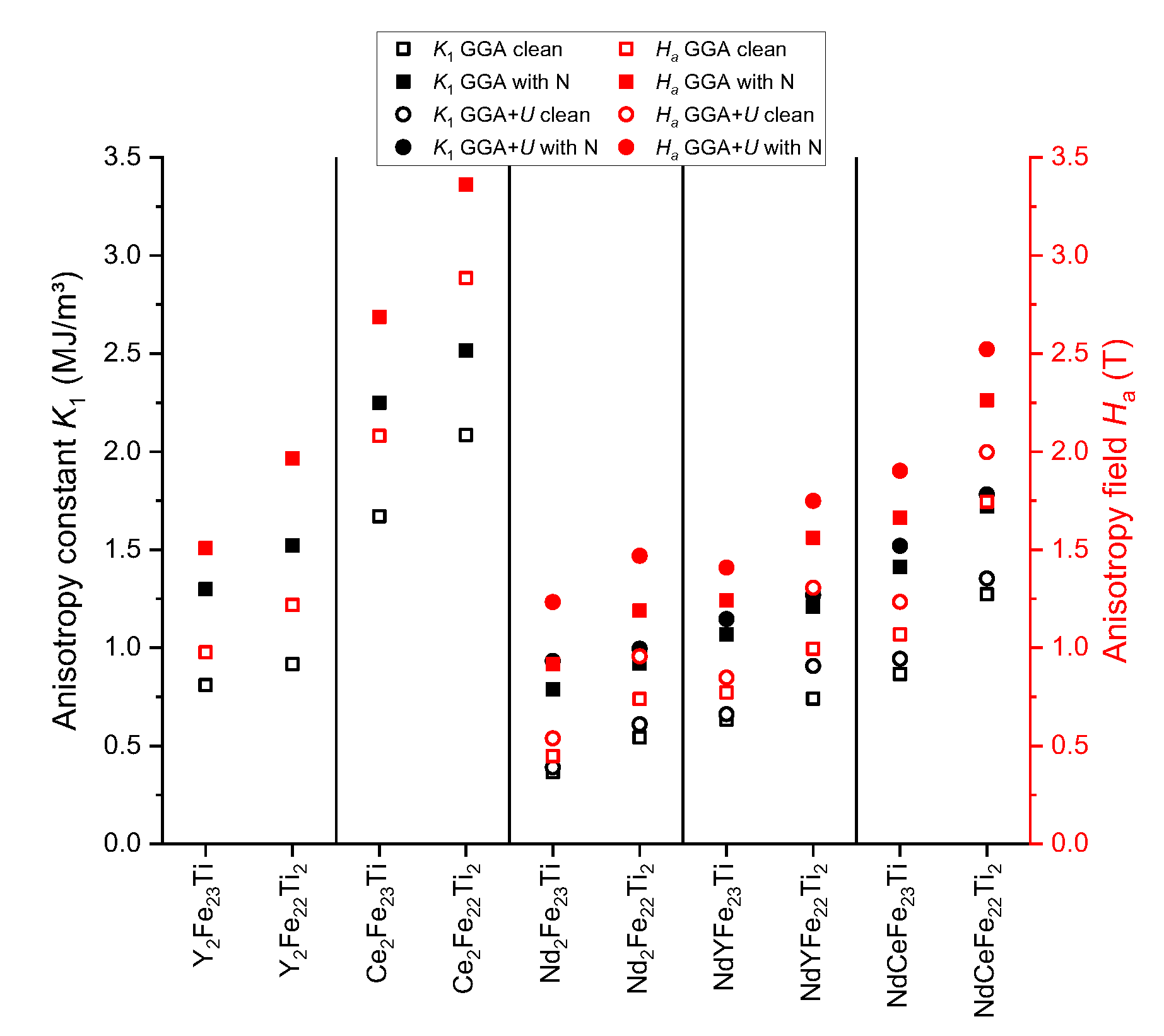}
\caption{Influence of interstitial nitrogenation on the anisotropy constant $K_1$ (black) and anisotropy field $H_a$ (red). Treatment with the GGA (GGA+\textit{U}) functional is depicted as squares (circles). Clean compounds are shown as empty and nitrogenated ones as filled symbols. \label{fig_n_doping_mae}}
\end{figure}

The calculated influence of interstitial nitrogenation on $K_1$ and $H_a$ for our GGA and GGA+\textit{U} calculations is shown in  Fig.~\ref{fig_n_doping_mae}. A general trend of an increase for $K_1$ and $H_a$ due to the addition of N is visible for all compounds. For instance, $K_1$ calculated via DFT for the Y-based ternary compound YFe$_{11}$Ti is improved by 0.60 MJ/m$^3$ to a value of 1.52 MJ/m$^3$.
In case of the CeFe$_{11}$Ti compound, the calculated value of 2.08 MJ/m$^3$ for the clean case is raised by N doping to a result of 2.52 MJ/m$^3$. The anisotropy field is consequently also increased due to nitrogenation to 3.36 T. 
For NdFe$_{11}$Ti the nitrogenated $K_1$ value is computed with DFT (DFT+\textit{U}) to be 0.92 (1.0) MJ/m$^3$. The according result for $H_a$  is calculated as 0.96 (1.47) T.\\

The impact of N on the quarternary compounds NdYFe$_{22}$Ti$_2$ NdCeFe$_{22}$Ti$_2$ is similar to the ternary ones. For $K_1$, an increase from 0.74 (0.91) to 1.21 (1.27) MJ/m$^3$ and 1.27 (1.35) to 1.72 (1.78) MJ/m$^3$ for the respective compounds is observed. Since the calculation of the anisotropy field is based on the anisotropy constant, the increasement of $K_1$ leads to an improvement of the anisotropy field up to values of 1.56 (1.75) and 2.26 (2.52) T, respectively.\\

For all considered compounds, an average increase of $\sim 0.5$ MJ/m$^3$ due to the additon of N atoms into the 2b site can be recorded. The biggest individual increase is seen for the above mentioned YFe$_{11}$Ti compound with an improvement of 0.60 MJ/m$^3$.


\section{Hardness Factor}

From the first-principles calculations, the magnetic hardness factor can be evaluated from the expression of saturation magnetisation $M_S$ and magnetocrystalline anisotropy constant $K_1$ by

\begin{equation}
\label{Eq_kappa}
\kappa = \sqrt{\frac{K_1}{\mu_{0}M_{S}^{2}}}.
\end{equation}

This parameter indicates the potential for a material to be developed into a permanent magnet independently of the shape when $\kappa$>1. In addition, a material can be considered as a semi-hard (0.1 <$\kappa$< 1) and soft magnet ($\kappa$<0.1) accordingly \cite{Skomski2016}. \\

\begin{figure}
\includegraphics[width=0.95\textwidth]{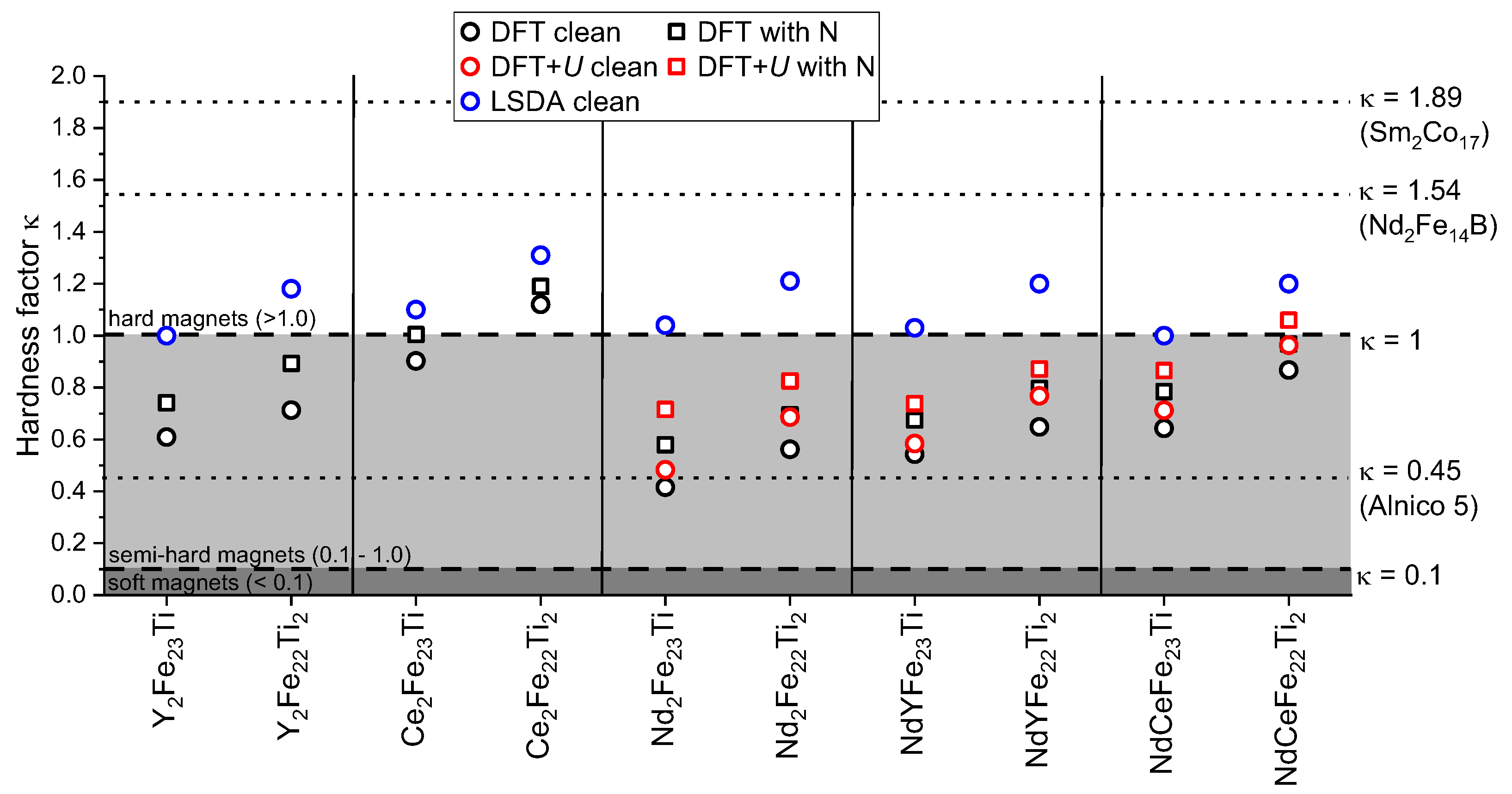}
\caption{Calculated hardness factors $\kappa$ for GGA (black), GGA+\textit{U} (red) and LSDA (blue) functionals. The clean compounds are represented by circles and the nitrogenated compounds considered in the GGA and GGA+\textit{U} approaches are depicted as squares. The region colored as dark grey represents soft magnets ($\kappa$ < 0.1), the light grey one semi-hard magnets (0.1 < $\kappa$ < 1.0) and the white background hard magnets ($\kappa$>1). The magnetic hardness factor of the most well-known magnets are also shown by horizontal dotted lines.
\label{fig_kappa}}
\end{figure}

The calculated magnetic hardness factors $\kappa$ are depicted in Fig.~\ref{fig_kappa} for both GGA (GGA+\textit{U}) and LSDA functionals. As mentioned above, for Y and Nd-based compounds, a better $K_1$ agreement with respect to experimental data by LSDA calculations was achieved. Therefore, LSDA based $\kappa$ values are more reliable, and it is noticeable that for all considered compounds $\kappa$>1. Herper \textit{et al.} \cite{Herper2022} calculated the magnetic hardness factor for YFe$_{11}$Ti for both GGA and LSDA and found $\kappa$=0.63 and 1.23, respectively. This agrees with the calculated values in this work, which are 0.71 (GGA) and 1.18 (LSDA), respectively. \\


For CeFe$_{11.5}$Ti$_{0.5}$ and CeFe$_{11}$Ti Snarsky-Adamsky \textit{et al.} \cite{Snarski-Adamski2022} reported theoretical values, calculated using the full-potential local-orbital code from 2018 (FPLO18) with the PBE functional, of 0.52 and 0.79 respectively. While these values are slightly smaller than the GGA results of 0.90 and 1.12 in this study, the trend of an increasing hardness factor with an increase of Ti concentration is in good agreement. \\

In case of Y and Ce substituted Nd-based 1:12 phases, promising $\kappa$ values for LSDA treatment are observed. For both NdYFe$_{22}$Ti$_2$ and NdCeFe$_{22}$Ti$_{2}$ cases the calculated value is $\kappa$=1.20. Note that the $\kappa$ value for the commonly used Nd$_2$Fe$_{14}$B magnet is reported to be 1.54 \cite{Skomski2016}. In the considered benchmark system of this thesis, by considering 50\% Nd-lean quaternaries, a very close magnetic hardness factor with respect to the market leading hard magnet was reached which is technologically encouraging.

\chapter{Conclusion and Outlook}
\label{Sec_Results_conclusion}

In conclusion, based on the promising finite temperature phase stability findings for Nd-lean permanent magnets \cite{Sozen2022}, systematic investigations on the intrinsic magnetic properties for (Nd,X)Fe$_{12-y}$Ti$_y$ compounds have been performed. As an alternative element to substitute critical Nd, abundant and inexpensive elements, X= Y and Ce, have been chosen. Since there is no experimental data for the considered quaternary compounds in the literature, only the theoretical results for ternary compounds could be compared and a promising agreement could be seen. This gives a strong confidence in the calculated magnetic properties of Nd-lean quaternaries. \\

An accurate description of RE 4\textit{f}-electrons has critical importance for a successful prediction of intrinsic magnetic properties. Due to the strong hybridization of Ce-4\textit{f} with TM-3\textit{d}-electrons, in addition to previous works by S\"ozen \textit{et al.} \cite{Sozen2019,Sozen2019a,Sozen2020,Sozen2022}, conventional DFT calculations are performed for Ce-based compounds and yield satisfactory results. However, Nd 4\textit{f}-electrons are strongly localized and available DFT exchange-correlation functionals have difficulties to accurately capture this property. Therefore, as a correction scheme, the DFT+\textit{U} approach with Hubbard \textit{U}=6 eV has been used.\\

The methodology of this work perfectly predicts the total magnetization trend of the 1:12 phases, that NdFe$_{11}$Ti has the highest magnetisation and is followed by YFe$_{11}$Ti and CeFe$_{11}$Ti. Experimentally reported saturation magnetisation $\mu_0M_S$ for NdFe$_{11}$Ti is deviating from 1.38 to 1.70 T \cite{Yang1991,Piquer2004,Herper2022,Akayama1994} based on the measurement temperature and method. Our DFT (DFT+\textit{U}) calculations yield 1.47 (1.28) T, which is a good agreement. In the case of Y and Ce substitution $\mu_0M_S$ is found to be 1.49 (1.39) T for NdYFe$_{22}$Ti$_2$ and 1.46 (1.35) T for NdCeFe$_{22}$Ti$_2$ based on DFT (DFT+\textit{U}). In addition, for all considered compounds further $\mu_0M_S$ improvement by $\sim$0.1 T is achieved after nitrogenation and it is found that the magnetovolume effect plays a major role rather than chemical effects. Based on the magnetisation calculations, very high |BH|$_{max}$ values for NdYFe$_{22}$Ti$_2$ and NdCeFe$_{22}$Ti$_2$ were achieved, which are 384 and 365 kJ/m$^3$, and they are higher than the |BH|$_{max}$ of Sm$_2$Co$_{17}$B, 294 kJ/m$^3$ \cite{Coey2011}. Note that, lower Ti contained quaternaries, NdYFe$_{23}$Ti (484 kJ/m$^3$) and NdCeFe$_{23}$Ti (466 kJ/m$^3$), have close |BH|$_{max}$ values to the Nd$_2$Fe$_{14}$B magnet that exhibits 512 kJ/m$^3$ \cite{Coey2011}. In addition, a further increase of |BH|$_{max}$ values around 40 kJ/m$^3$ is calculated with nitrogenation. \\

Curie temperatures $T_C$ are calculated by the mean-field approximation (MFA). Systematic overestimation of $T_C$ for ferromagnetic (FM) state calculations is observed. Nevertheless, quantitative agreement of decreasing $T_C$ for increasing Ti concentration and increasing $T_C$ after nitrogenation is captured with respect to experimental data. Miyake \textit{et al.} \cite{Miyake2021} reported that local moment disorder (LMD) calculations yield better performance for the $T_C$ of REFe$_{11}$Ti compounds. Our LMD calculations agree with this finding. In the case of YFe$_{11}$Ti, CeFe$_{11}$Ti and NdFe$_{11}$Ti, the LMD-based $T_C$ calculations overestimate the experimental data by 9, 12 and 20 \%, respectively. Note that, in the case of FM-based calculations, the margin of error is around 45\% for these ternaries. The LMD calculations yield 595 and 593 K $T_C$ values for the Nd-lean quaternaries NdYFe$_{22}$Ti$_{2}$ and NdCeFe$_{22}$Ti$_{2}$, respectively. For both cases, theoretical $T_C$ is slightly higher than the experimental $T_C$ of Nd$_2$Fe$_{14}$B, which is 588 K.\\

In the case of the magnetocrystalline anisotropy constant $K_1$, the generalized gradient approximation (GGA) based exchange-correlation functional failed for Y and Nd-based compounds. The shortcomings of GGA is also reported for such compounds by Herper \textit{et al.} and Ke \textit{et al.} \cite{Herper2022,Ke2016}. Therefore, the local spin density approximation (LSDA) was further considered and reached a better agreement with respect to the experiment. Knowing $K_1$ and $\mu_0M_S$ allows to calculate the magnetic hardness factor $\kappa$, which relates to the potential of a material to be developed into a permanent magnet and $\kappa$>1 is desired. Our 50\% Nd-lean quaternaries yield a 1.20 theoretical $\kappa$ value which is close to the magnetic hardness factor of 1.54 for Nd$_2$Fe$_{14}$B.\\

As a conclusion to this thesis, it was shown that the substitution of critical RE elements, such as Nd, in compounds of the ThMn$_{12}$ structure with more available elements like Ce or Y yields possible alternatives to the RE-TM type Nd-Fe-B magnets with comparable or only slightly smaller magnetic properties. These new quarternary (NdX)Fe$_{24-y}$Ti$_y$ (X=Y or Ce) compounds can consequently be considered as an achievable way to produce more affordable magnets with a smaller amount of critically labeled elements. As an outlook to this work, the further investigation of other elements, besides Ce and Y, with a combination of DFT and KKR calculations for the substitution with the Nd-atoms can be considered. Another possibility would be the expansion to other known magnetic structures such as the 2:14 or 2:17 phases and research the possible reduction of RE-elements in these compounds as well. Due to similar $T_C$s the investigated quarternary compounds are very appealing in a technological perspective as they could be used in almost the same field of applications as the Nd-Fe-B magnets when slightly lower magnetic properties are still acceptable.

\bibliographystyle{ieeetr}

\bibliography{masterthesis}

\chapter*{List of Abbreviations}
\begin{list}{}
\item RE - Rare-earth
\item TM - Transition metal
\item $m_{tot}$ - Total magnetic moment
\item $\mu _0$ - Magnetic constant
\item $\mu _B$ - Bohr magneton
\item $H$ - Magnetic field
\item $H_a$ - Anisotropy field
\item $H_c$ - Coercivity field
\item $H_d$ - Internal magnetic field
\item $M$ - Magnetization
\item $M_r$ - Remanence magnetization
\item $M_S$ - Magnetization saturation
\item $M(T)$ - Finite temperature magnetization
\item $m(t)$ - Reduced magnetization
\item $T_C$ - Curie temperature
\item |BH|$_{max}$ - Maximum energy product
\item MCA - Magnetocrystalline anisotropy
\item MAE - Magnetocrystalline anisotropy energy
\item $\Delta E_{MAE}$ - Anisotropy energy
\item $K_1$ - Anisotropy constant 1$^{\text{st}}$ order
\item $\kappa$ - Hardness factor
\item HF - Hartree-Fock
\item post-HF - post Hartree-Fock
\item DFT - Density functional theory
\item DFT+\textit{U} - Density functional theory with applied Hubbard \textit{U} correction
\item DMFT - Dynamical mean field theory
\item LDA - Local Density Approximation
\item LSDA - Local Spin Density Approximation
\item GGA - Generalized Gradient Approximation
\item PW91 - Functional by Perdew and Wang
\item PBE - Functional by Perdew, Burke and Ernzerhof
\item PAW - Projector Augmented Wave
\item USPP - Ultrasoft Pseudopotentials
\item VASP - Vienna \textit{ab initio} Simulation Package
\item KKR - Korringa-Kohn-Rostoker
\item SCF - self-consistent field
\item FM - Ferromagnetic
\item LMD - Local Moment Disorder
\item DLM - Disorder local moment
\item MFA - Mean-field approximation
\item ASA - Atomic sphere approximation
\item CPA - Coherent potential approximation
\item MJW - Moruzzi, Janak and Williams
\item TB-LMTO-ASA - Tight-binding linear-muffin-tin-orbital atomic-sphere-approximation
\item FPLO18 - Full potential local orbital code released in 2018
\item DOS - Density of states
\item XRD - X-Ray diffraction
\item ESM - Extraction sample magnetometer
\item PPMS - Physical property measurement system
\item VSM - Vibrating sample magnetometer
\item bcc - Body centered cubic

\end{list}

\chapter*{Erklärung}

Hiermit versichere ich an Eides statt, dass ich diese Arbeit selbstständig verfasst und keine anderen als die angegebenen Quellen und Hilfsmittel benutzt habe. Außerdem versichere ich, dass ich die allgemeinen Prinzipien wissenschaftlicher Arbeit und Veröffentlichungen, wie sie in den Leitlinien guter wissenschaftlicher Praxis der Carl von Ossietzky Universität Oldenburg festgehalten sind, befolgt habe.

\vspace{1cm}

\textbf{--------------------------------}

\vspace{0.15cm}
Stephan Erdmann

\end{document}